\documentclass[12pt]{iopart}
\begin{document}

\input{epsf}

\title[A primer on hierarchical galaxy formation]
{A primer on hierarchical galaxy formation: the semi-analytical approach}

\author{C.~M. Baugh}

\address{Institute for Computational Cosmology,  
Department of Physics, 
Durham University,
South Road
Durham, DH1 3LE, UK}

\begin{abstract}
Recent observational and theoretical breakthroughs make this an exciting 
time to be working towards understanding the physics of galaxy formation. 
The goal of this review is to make the principles behind the hierarchical 
paradigm accessible to a wide audience by providing a pedagogical 
introduction to modern theories of galaxy formation. I outline the 
ingredients of the powerful approach called semi-analytical modelling and 
contrast this method with numerical simulations of the gas dynamics 
of baryons. Semi-analytical models have enjoyed many successes, but 
it is the observations which the models struggle to match which mark 
out areas where future progress is most likely to be made; these are 
also reviewed.  
\end{abstract}



\maketitle

\section{Introduction}

These are exciting times to be studying how galaxies are made. 
After a decade of spectacular breakthroughs in physical cosmology, 
the focus is beginning to shift away from determining the values of 
the basic cosmological parameters towards attacking the problem of 
galaxy formation. A combination of factors is responsible for this 
change in emphasis. Firstly, the cold dark matter cosmological model 
has been placed on a much firmer footing by recent measurements of 
the cosmic microwave background radiation and galaxy clustering. Many of the 
fundamental cosmological parameters, such as the density of matter, 
are known to an uncertainty of around 10\%. Secondly, the 1990s saw 
the first detections of sizeable populations of galaxies at high redshifts, 
allowing evolutionary trends to be established. Finally, the increase in  
readily available computing power coupled with the development of 
powerful new techniques, such as the subject of this review, semi-analytical 
modelling of galaxy formation, means that we are in a position to 
generate accurate predictions for the properties of galaxies in 
hierarchical cosmologies. We will elaborate on each of the above points 
in turn in the introduction. Given these favourable conditions, there is 
now a genuine chance of making real progress in the advancement of our 
understanding of the process of galaxy formation. 

The aim of this review is to provide an introduction to the ideas 
and concepts that underpin modern ideas about galaxy formation in 
a universe in which cosmic structures build up hierarchically through 
gravitational instability. The main focus is the semi-analytical approach 
to modelling galaxy formation. As we shall see, this technique is 
currently the most developed theoretical tool available 
and can be used to make quite detailed predictions of galaxy properties. 
It has also enjoyed 
a large measure of success, though many problems and gaps remain as we 
will point out. The intention is to provide a pedagogical overview 
of this area, which will enable a reader with no prior knowledge of 
the subject to reach a position from which they are better placed 
to tackle the more advanced papers in the literature on the physics of galaxy 
formation. For this reason, I have sacrificed mathematical 
detail in favour of trying to provide a verbal description of the 
relevant physical processes, with the aim of providing a broad view 
of hierarchical galaxy formation. I have also deliberately avoided 
undertaking a detailed description of the implementations of various 
processes in the 
models, which would serve only to generate long lists of parameter 
definitions and confusion. Nor do I attempt to carry out a detailed 
comparison between the published results of different groups. 
There are many factors that conspire to make such an exercise largely futile. 
Papers presenting predictions from semi-analytical models stretch back over 
15 years. Until very recently, a whole slew of different cosmological 
models were considered in the literature. Furthermore, different groups 
have different aims 
and priorities and set their model parameters according to different 
criteria. The models are dynamic entities and as such are continually being 
improved, as one would expect in a subject in which our current knowledge can 
at best be described as rudimentary. Instead of making a comparison 
of the different semi-analytical models, I will concentrate on expounding 
the principles upon which the models are based, which will hopefully 
provide a more useful grounding in the physics of galaxy formation 
and of the philosophy behind the semi-analytical approach.  

There are many topics related to hierarchical galaxy formation which I do 
not have space to cover at anything other than a cursory level in this 
review. Fortunately, there are now many excellent textbooks which give a 
formal introduction to the growth of cosmological density perturbations 
(for example Peebles 1993, Padmanabhan 1993, Peacock 1999 and Coles \& 
Lucchin 2002). For those readers wanting to find out more about the cosmic 
microwave background radiation, it is hard to beat a visit to the web site 
of Wayne Hu (http://background.uchicago.edu/~whu/), which contains many 
lucid explanations, accompanied by neat animations. The recent review by 
Springel, Frenk \& White (2006) 
provides an introduction to the large scale structure of the Universe and 
the latest attempts to model this using computer simulations.  
There are relatively few reviews of the theory of galaxy formation available. 
The article by Kauffmann \& White (1994), the Les Houches lectures by 
White (1994) and the recent lecture notes by Avila-Reese (2006) cover many 
of the topics in this review (and more) often with a different emphasis 
and are valuable sources for researchers in galaxy formation.    

This review is laid out as follows. In the remainder of the introduction, 
a few general comments are made to set the scene, building upon the 
points made in the first paragraph. The next two sections describe 
the basic physic processes that underpin models of galaxy formation. 
Section 2 deals with the less controversial aspects of the modelling  
which relate to the dark matter component of the universe. 
Section 3 introduces the more complex physics of the baryonic component. 
Section 4 compares and contrasts the semi-analytical approach to galaxy 
formation with direct gas dynamical simulations. In Section 5, I review some 
of the successes and failures of the models, the latter pointing to the areas 
in which future developments are most likely to be made. 

\subsection{The hierarchical cosmology}
The cold dark matter (CDM) model has steadily gained acceptance since 
it was first mooted in the early 1980s (Peebles 1982; Blumenthal et al. 1984; 
Davis et al. 1985). The adoption of CDM as the theorist's model of choice 
can be attributed to three features. Firstly, there are many candidates 
for the cold dark matter particle predicted by extensions to the standard 
model of particle physics; perhaps the most promising of these is the lightest 
stable supersymmetric particle, the neutralino (e.g. Gondolo 2004). 
Secondly, the model has tremendous 
predictive power. Collisionless N-body simulations of the growth of structure 
in a CDM universe are essentially straightforward to do, given a suitably 
big and fast computer and an efficient algorithm to compute the gravitational 
forces between particles. As we will see later on in this article, the state 
of the art in numerical simulations allows incredibly detailed predictions 
to be made for a wide range of properties of the dark matter at all epochs. 
The simulation results in turn can be used to calibrate analytical and 
phenomenological models. Thirdly, and most importantly, many of these 
predictions have turned out to be impressively successful. 

Perhaps the most convincing support for the CDM model comes from the 
measurement of temperature anisotropies in the cosmic microwave background 
(CMB) radiation. The pattern of hot and cold spots in the radiation can be 
related to density fluctuations present in the Universe a mere few hundred 
thousand years after the Big Bang, when the radiation and baryon fluids 
stopped interacting with one another at the epoch of recombination. 
Since the initial detection of these anisotropies on large angular scales by 
the COBE satellite in 1992 (Smoot et~al. 1992), the power spectrum of 
the fluctuations has been gradually uncovered, culminating in the clear 
detection of three Doppler peaks due to acoustic oscillations in the 
photon-baryon fluid at the last scattering surface (de Bernardis et~al. 2000; 
Hanany et~al. 2000; Hinshaw et~al. 2003; Jones et~al. 2006; 
Hinshaw et~al. 2006). 

\begin{figure}
{\epsfxsize=14.truecm
\epsfbox[-150 0 590 360]{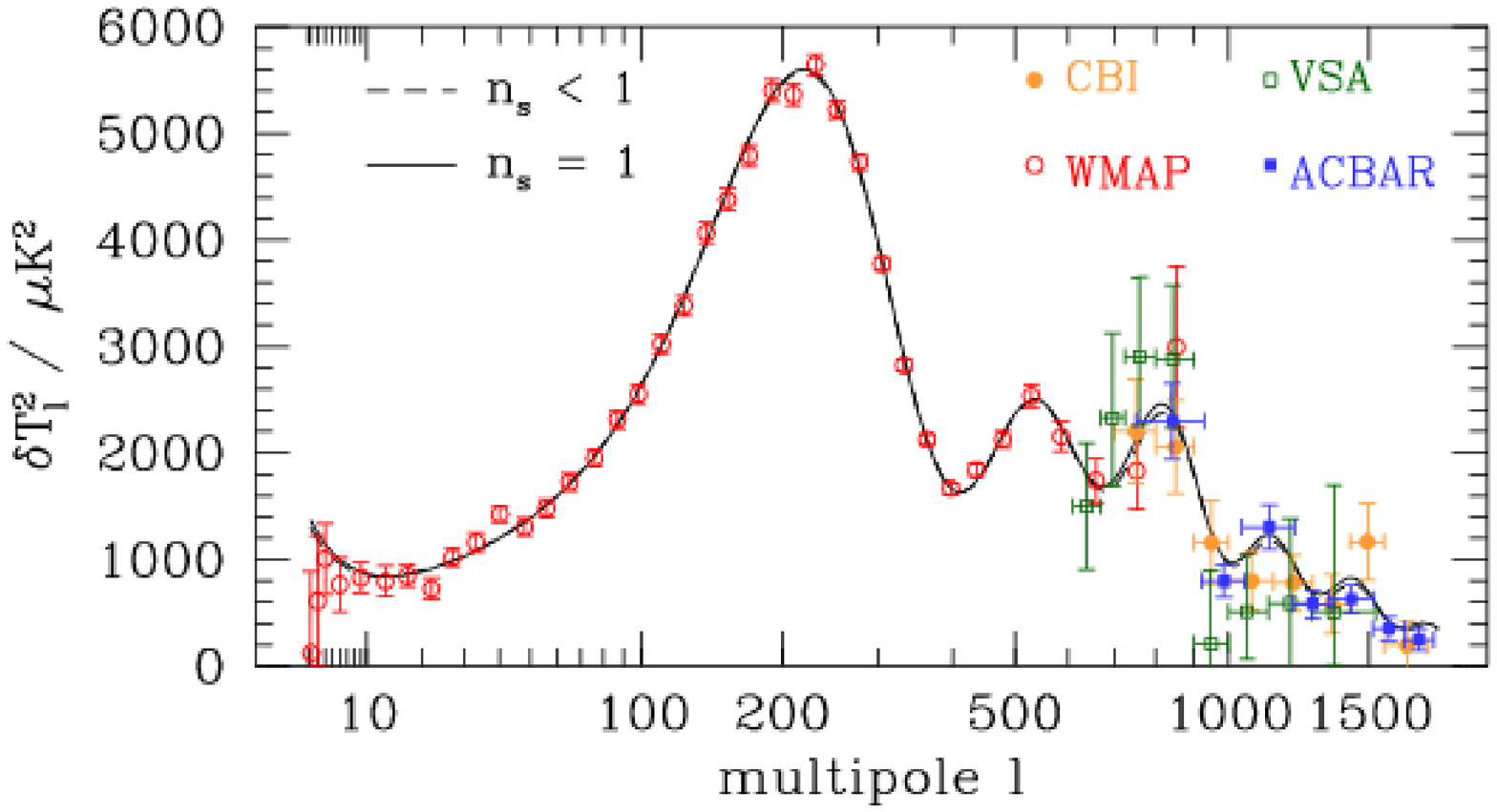}}
{\epsfxsize=14.truecm
\epsfbox[-150 150 590 720]{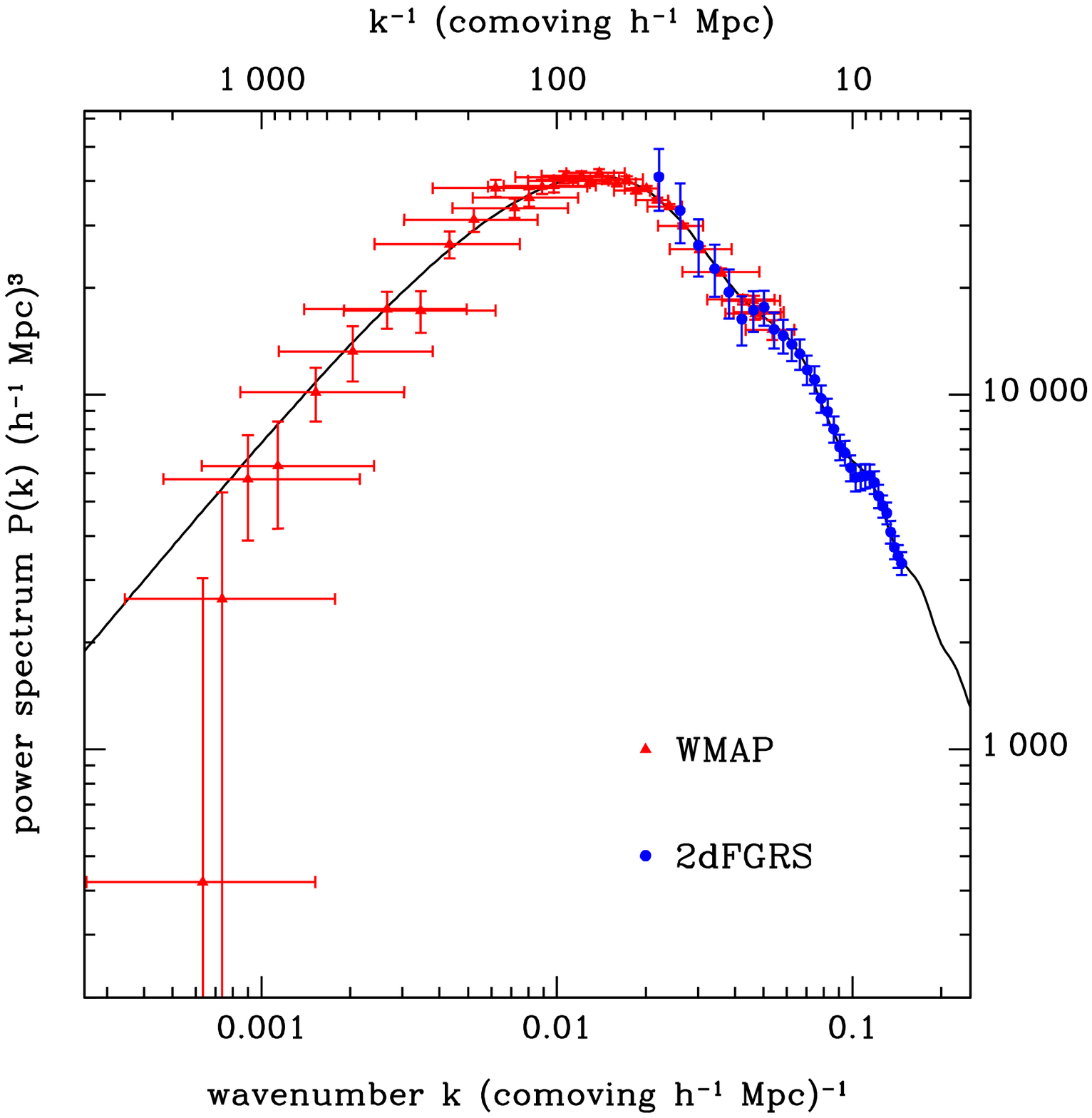}}
\caption{
Top panel: The power spectrum of temperature fluctuations in the CMB 
as shown by a compilation of recent data (WMAP - Hinshaw et~al. 2003; 
CBI -- Readhead et~al. 2004; VSA -- Dickinson et~al. 2004; 
ACBAR -- Kuo et~al. 2004). The solid and dashed lines show variants of 
the best fitting CDM model in which the spectral index of primordial 
fluctuations is held at $n=1$ (solid) or allowed to float ($n<1$, dashed 
line). 
Bottom panel: The power spectrum of density fluctions. The solid line shows 
the best fit CDM model with $n=1$. The circles show the galaxy power spectrum 
measured from the final 2dFGRS (Cole et~al. 2005). The triangles show the 
first year temperture power spectrum measured by WMAP plotted in the same 
units. Adapted from Sanchez et~al. (2006).   
}
\label{fig:cdm}
\end{figure}

Further compelling support for the CDM model has come recently from 
two galaxy surveys which have revolutionised our view of the local Universe, 
the two-degree Field Galaxy Redshift Survey (2dFGRS; Colless et~al. 2001) 
and the Sloan Digital Sky Survey (SDSS; York et~al. 2000). 
The unprecedented size of these maps of the galaxy distribution has permitted 
the most accurate measurements to date of the power spectrum of galaxy 
clustering (Percival et~al. 2001; Tegmark et~al. 2004a; Pope et~al. 2004; 
Cole et~al. 2005; Padmanabhan et~al. 2006; Percival et~al. 2006a; 
Tegmark et~al. 2006). As an illustration of the accuracy of these 
measurements, the imprint of acoustic oscillations on the matter power 
spectrum, a much weaker signal than the Doppler peaks in the CMB, has now 
been firmly detected (Cole et~al. 2005; Eisenstein et~al. 2005; 
Padmanabhan et~al. 2006; Percival et~al. 2006b). 
By confronting the theoretical models with the combined CMB and galaxy 
power spectrum data, the constraints on cosmological 
parameters can be tightened, through the breaking of certain degeneracies 
(Efstathiou et~al. 2002; Percival et~al. 2002; Spergel et~al. 2003; 
Tegmark et~al. 2004b; Seljak et~al. 2005; MacTavish et~al. 2006; 
Sanchez et~al. 2006; Spergel et~al. 2006). 
For the first time, many of the basic cosmological parameters can be 
constrained with accuracies approaching or better than $10\%$ 
(see Fig.~\ref{fig:cdm} for an illustration of how well the 
CDM model can reproduce the CMB and large-scale structure data). 
By combining the galaxy power spectrum measured from the 
final 2dFGRS by Cole et~al. (2005) with a compilation of the CMB 
data available shortly after the release of the first year results from WMAP, 
Sanchez et~al. found strong evidence in support of a tilt in the spectrum 
of primordial density fluctuations, away from the simple scale invariant 
model (i.e. a spectral index of scalar fluctuations described by $n=1$, 
where the primordial spectrum is described by a power law form, 
$P(k) \propto k^{n}$), a conclusion which was confirmed upon the 
analysis of the third year of data from WMAP (Spergel et~al. 2006). 

Despite the burgeoning circumstantial evidence in support of the CDM model, 
it should be borne in mind that candidate particles for the non-baryonic 
dark matter have yet to be detected in the laboratory (Bergstrom 2000).
In the current best fit CDM model, the universe is close to being 
spatially flat (e.g. Sanchez et~al. 2006). However, less than 30\% of 
the critical density required for this geometry is contributed by matter. 
The remainder is thought to be in some form of ``dark energy'', one limiting 
case of which is the cosmological constant (Carroll 2004). Compelling 
support for a dark energy component came from the deduction that the 
expansion of the universe is accelerating, based on the Hubble diagram of 
distant type-Ia supernovae (Riess et~al. 1998, 2004; Perlmutter et~al. 1999). 
Arguments in favour of a universe whose dynamics are currently dominated 
by a cosmological constant were made some time before the supernovae results, 
in order to reconcile the measured galaxy clustering with the predictions of 
CDM (Efstathiou, Sutherland \& Maddox 1990) and to match faint galaxy 
counts (Yoshii \& Takahara 1988; Fukugita et~al. 1990). 
Theoretically, the density parameter of the dark energy, at the level 
implied by astronomical tests is many orders of magnitude smaller than 
can be motivated from particle physics considerations using dimensional 
arguments (Carroll 2004).
This theoretical tension has led some authors to consider alternative 
models without any form of dark energy. 
Blanchard et~al. (2003) present a model without dark energy that can 
reproduce the CMB and galaxy power spectrum data, albeit with an 
uncomfortably low value for the Hubble constant, thus leaving only the 
supernova observations as evidence in favour of cosmological constant. 
Another possibility is a modification to the law of gravity on large scales 
(Deffayet, Dvali \& Gabadadze 2002; Carroll et~al. 2006). 
The methodology described in this review is not wedded 
to CDM; the calculations can be carried out in any cosmological model 
in which structure grows hierarchically. 

The conclusion is that, within the context of the CDM model, we now have a 
very good idea of the values of most of the fundamental cosmological 
parameters. The growth of structure in the dark matter is reasonably well 
constrained, as we shall see in Section~\ref{dissless}. This provides a 
tremendous fillip to our efforts to understand how galaxies form, as it 
removes a whole swathe of parameter space, allowing us to concentrate on 
the more difficult and interesting physics of galaxy formation. 

\begin{figure}
{\epsfxsize=14.truecm
\epsfbox[-150 150 590 720]{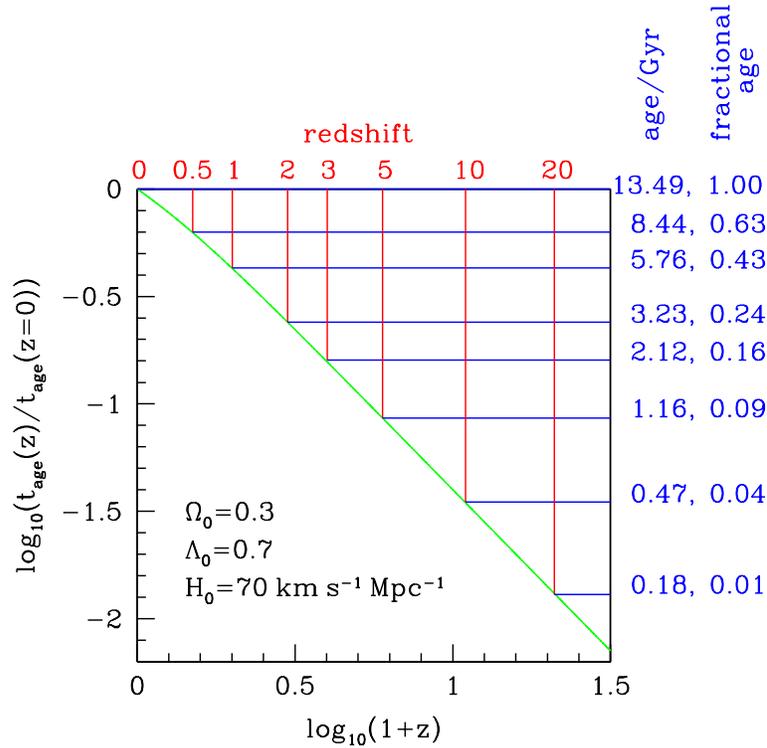}}
\caption{
The age of the `concordance' CDM universe, in units of the present day age, 
as a function of the log of $1+z$. The vertical lines mark a subset 
of redshifts, as indicated on the uper axis. The horizontal lines 
point to the corresponding age of the universe (first number right-hand 
axis, in Gyr) and the fraction of the present day age of the universe 
that this redshift corresponds to (second number); e.g. at $z=2$, the 
concordance universe is 3.23Gyr old, which is 24\% of its current age. 
}
\label{fig:lookback}
\end{figure}

\subsection{Observations of galaxies at high redshift}
The second key advance that makes progress in understanding galaxy 
formation possible is the unveiling of the high redshift universe 
in the second part of the 1990s. Observations of galaxies over a range 
of redshifts allow us to compare their properties at different 
epochs in the history of the universe. (For reference, 
Fig.~\ref{fig:lookback} shows the age of the canonical CDM model 
at different redshifts.) 
Such a comparison can be used to shed light on the formation and evolution 
of galaxies, and in particular can be used to establish whether galaxy 
formation is a steady process or if it took place much more vigorously 
at some earlier epoch. Perhaps the first major breakthrough in characterising 
the high redshift universe was the Hubble Deep Field (Williams et~al. 1996; 
Ferguson, Dickinson \& Williams 2000). 
The unprecedented faint imaging of galaxies, combined with the Lyman-break 
dropout technique to isolate high redshift ($z > 2$) galaxies 
(Steidel et~al. 1996) was essential in making possible the first 
determination of the cosmic star formation history over more than 
80\% of the age of the universe (Madau et~al. 1996; Ellis 1997; 
Steidel et~al. 1999). 
This approach uses measurements of the rest frame ultra-violet flux from 
a galaxy to infer the instantaneous star formation rate. (The UV flux 
is dominated by stars with masses in excess of several times the mass 
of the sun; these stars are also short lived, producing the UV flux for 
timescales on the order of 10\,Myr.)
One problem with conducting such a census at these wavelengths is that 
the rest-frame ultra-violet can be strongly attenuated by dust 
extinction. 
Technological advances have opened up the electromagnetic 
spectrum. Of particular relevance is the detection of the emission from 
galaxies at sub-millimetre wavelengths (Smail et~al. 1997; 
Barger et~al. 1998; Hughes et~al. 1998; for a review of the properties of 
sub-millimetre galaxies see Blain et~al. 2002). 
Submillimetre observations offer the chance of uncovering up heavily 
extincted galaxies which are too faint to appear in optical surveys. 
This emission can arise from starlight or emission from an active 
galactic nucleus (AGN) that is absorbed by dust and 
re-radiated at longer wavelengths. 
The presence and relative importance of an AGN 
can be constrained by combining sub-millimetre observations with X-ray 
imaging. Alexander et~al. (2005) report that practically all objects with 
sub-millimetre emission contain an AGN, but that the star formation in these 
objects accounts for the bulk of the emission at sub-millimetre wavelengths.  
 
\subsection{The theory of galaxy formation: a brief historical overview}

We now give a brief historical sketch of the some of the main ideas 
that underpin the current paradigm of galaxy formation and show how 
these led to the development of semi-analytical modelling. 
Many of these references will be revisited in later sections, 
where the ideas they put forward will be explained in more detail. 
However, it is instructive to point out the chain of key papers 
responsible for shaping our current understanding of galaxy formation.  

The idea that cosmic structures grow through the mechanism of 
gravitational instability is the oldest part of the paradigm. 
The application of perturbation theory and numerical simulations 
to understanding this process started in earnest in the early 
1970s (see Peebles 1980). Gunn \& Gott (1972) looked at the growth 
of clusters through the infall of material, using the spherical 
top-hat model to track the evolution of the cluster overdensity (e.g. 
Peacock 1999). 
Press \& Schechter (1974) used the top-hat model to compute the 
abundance of structures of different masses that form by gravitational 
condensation in a density field with a Gaussian distribution of 
fluctuations. The first calculation of the pattern of density 
fluctuations expected at early times in a cold dark matter universe 
was made by Peebles (1982; see also Bardeen et~al. 1986). The first numerical 
simulations of the hierarchical growth of structures in a CDM universe 
were carried out by Davis et~al. (1985). 

Two of the core ideas underpinning today's paradigm for galaxy formation 
can be traced back more than fifty years to Fred Hoyle (see Efstathiou 2003 
for a review of Hoyle's work in a modern context). Hoyle (1949) was the 
first to propose that the rotation of galaxies could be generated by the 
tidal torques which operate during their collapse. This idea was expanded 
upon by Peebles (1969) and White (1984), and tested with N-body 
simulations by Efstathiou \& Jones (1979); Hoyle also argued that the 
observed range of galaxy masses could be explained by considering the time 
taken for gas to cool and condense into galaxies (Hoyle 1953). This idea 
was later developed by 
Rees \& Ostriker (1977) and Silk (1977). 

White \& Rees (1978) presented a synthesis of the theory of Press \& 
Schechter, which describes the hierarchy of gravitationally bound structures, 
and the gas cooling arguments used to motivate the observed sizes of 
galaxies, to produce a model of galaxy formation that set the foundations  
for today's models. 
White \& Rees proposed that galaxy formation was a two stage process, with 
dark haloes forming in a dissipationless, gravitational collapse, 
with galaxies  
forming inside these structures, following the radiative cooling 
of baryons. The additional condensation of the gas through dissipative 
cooling stabilized galaxies against the disruption caused by the merging 
of the dark haloes. White \& Rees also argued that an additional process, 
feedback, was needed to make small galaxies more diffuse so that they 
would be less successful at surviving the merging process, thus avoiding 
the production of more faint galaxies than are observed. 

The first fully fledged semi-analytical model came over ten years after 
the work by White \& Rees (1978). White \& Frenk (1991) produced a 
galaxy formation model that included many of the ingredients of today's 
models: cold dark matter, gas cooling, star formation, feedback and 
stellar populations (see also Cole 1991 and Lacey \& Silk 1991). 
The first models to track the formation and evolution of galaxies 
in the setting of evolving dark matter haloes came a few years later 
(Kauffmann et~al. 1993; Cole et~al. 1994).

\subsection{The relation between semi-analytical modelling and gas dynamics simulations}

At some level numerical simulations and semi-analytical models have more 
in common than most people perhaps realise. There are always aspects of a 
numerical calculation for which either the resolution becomes inadequate or 
a complete physical model is simply not available (e.g. star formation); 
such phenomena are described as ``sub-resolution'' physics 
(Springel \& Hernquist 2003). These processes can only be dealt with 
using the same types of recipes employed in semi-analytical modelling. 

In the past, numerical simulations have tended to be more ``photogenic'' 
than semi-analytical models in the sense that you can see what is 
happening in a simulation through animations and snapshots. The seductive 
power of a movie that shows the evolution of dark matter and galaxies in 
a simulation cannot be overstated, even in the face of the traditionally 
better developed semi-analytical predictions (e.g. generating galaxy 
luminosities with dust extinction as well as star formation histories), 
which allow more direct comparison with observations. Now that semi-analytical 
models are routinely grafted onto N-body simulations, parity between the two 
approaches will no doubt ensue in the motion picture industry at conferences. 

Another problem with the perception of semi-analytical models probably 
lies with the name ``semi-analytical'', which some in the community 
have clearly taken to imply some half-baked witches' brew of ingredients, 
from which any result can be coaxed with a suitable incantation, as and 
when required to fit new observational data.  

There are three points one can make to dispel this misguided notion. 
Firstly, criticising the models on the grounds that they appear to contain 
too many free parameters seems unfair. The models contain parameters simply 
because of our lack of understanding of the physics underpining galaxy 
formation. 
Secondly, it is essential to be clear that the parameters are physical 
and not statistical. For instance, the galaxy luminosity function is 
reasonably well described by the three parameter fit proposed by Schechter 
(1976). In this case, the parameter values are set by demanding a good 
fit to the data. 
There is simply not the same freedom to adjust the values 
of the parameters in a physical recipe, because in this case changing 
the parameter value has consequences which extend beyond the prediction 
for the luminosity function. 
For example, if the efficiency of gas heating by supernova explosions 
is increased with the aim of reducing the abundance of faint galaxies, 
then gas tends to cool more effectively in more massive haloes, 
with the consequence that the disks of bright spirals are predicted to 
be larger (see Fig.~8 of Cole et~al. 2000). A more powerful model  
with a wider range of galaxy properties for which predictions can 
be made has a smaller parameter space available to it than a more naive model. 
Model parameters are set to match a range of observations as well as 
possible. The parameter values required to meet these targets 
can then be assessed critically, e.g. to see if the amount of energy 
injection required for a feedback prescription to work is consistent with 
the number of supernova explosions that have taken place.  
Thirdly, the recipes used in semi-analytic modelling can be tested against 
numerical simulations and improved as required. The results of such 
comparisons are the subject of Section 4. In that section, 
we will discuss an example, the calculation of galaxy sizes and the 
conservation of angular momentum, where results from semi-analytic 
modelling have influenced the astrophysics implemented in numerical 
simulations. 

The level of sophistication attained by many current semi-analytical 
models is a double-edged sword. One person's admiration at the power 
of the models and their ability to make a wide range of predictions is 
tempered by another's impression of unnecessary complexity. Galaxy formation 
{\it is} complex, involving complicated, nonlinear physics, much of which, 
as we shall see in later sections, we are only just beginning to get 
to grips with. In the face of the scepticism the models naturally faced 
when they were first introduced, there was an onus to show how well the models 
reproduced different observations. Now that the semi-analytical approach 
is slowly gaining acceptance, the focus can shift to improving everyone's 
understanding of how galaxies are made. One of the great advantages of 
semi-analytical modelling is that facets of the model can easily be varied 
or switched on and off to gain a better appreciation of which ingredients 
have the most bearing on a particular observation. 

In the final section of this review, we will list 
some of the successes and failures of semi-analytical modelling. 
The fact that some observations stubbornly resist reproduction 
by the models should welcomed, not as an excuse to through away the 
whole machinery developed to model galaxy formation, but rather 
as an indication that we may need to include a new physical ingredient 
in the models. One such example which was uncovered by semi-analytical 
models is the difficulty in matching simultaneously the normalisation 
of the galaxy luminosity function and the zero-point of 
the luminosity-rotation speed scaling relation for spirals 
(Kauffmann, White \& Guiderdoni 1993; Cole et~al. 1994); this tension lead 
to variants of the cold dark matter model being explored and new physics being 
incorporated into the models, such as the calculation of the rotation curve 
of the model galaxies and the effects of dust extinction.

\subsection{Galaxy properties: clues and challenges for a model of galaxy formation}

Theories of galaxy formation are driven by observations. Any successful theory 
needs to explain certain basic measured properties of the galaxy population; 
these observations therefore set a challenge to the theorists but also contain 
important clues about the nature of the galaxy formation process. To set 
the scene, we now give a brief list of some fundamental observed properties 
of galaxies. These observations are returned to and discussed in more detail 
in different parts of the review. 

\begin{itemize}

\item Why is there a characteristic mass for galaxies? The most fundamental 
statistic describing the galaxy population is the luminosity function, 
a census of the number of galaxies per unit volume as a function of their 
luminosity. The luminosity function has a sharp break, brightwards of which 
the abundance of galaxies falls off exponentially (Norberg et~al. 2002b; 
Blanton et~al. 2003). The main process driving the growth of cosmic 
structures, gravitational instability, has no preferred scale, so processes 
in addition to gravity are responsible for the break. 

\item Why is star formation such an inefficient process? Only a small 
fraction of the baryons in the universe (on the order of $10\%$) 
is locked up in stars (Cole et~al. 2001). Where are the remaining 
baryons (Fukugita, Hogan \& Peebles 1998)?  

\item Why are there remarkably tight correlations between certain galaxy 
properties? Spiral and elliptical galaxies exhibit tight correlations between 
characteristic speeds of internal motion, a structural property, and 
luminosity, which depends upon the star formation history (Faber \& 
Jackson 1976; Tully \& Fisher 1977; Kormendy 1977; Djorgovski \& 
Davis 1987; Dressler et~al. 1987). 

\item Another correlation is perhaps fundamental enough to merit its 
own bullet point: the correlation between the mass of the central 
supermassive black hole in a galaxy and the mass of the spheroidal 
component (Magorrian et~al. 1998). Why is this relation so tight when 
there is such a huge difference in the spatial scale of these components? 
Does this correlation mean that bulges and 
black holes share a common formation mechanism? Did the energy released by 
the accretion of material onto the black hole play a role in the formation 
of the galaxy? 

\item What role does the environment play in galaxy formation? The mix of 
morphological types is strongly dependent on local density, with elliptical 
galaxies more prevalent than spirals in the cores of clusters (Dressler 1980). 
The fraction of galaxies contained in groups is expected to grow with time 
in hierarchical models. Are the physical processes which operate within 
groups, such as ``strangulation'' or ram pressure stripping, which acts 
to remove the supply of cold gas in a satellite galaxy, or dynamical 
effects such as tidal disruption or harassment, responsible for 
switching off the star 
formation these galaxies (Gunn \& Gott 1972; Moore et~al. 1996; Balogh et~al. 2004b; Wilman et~al. 2005a; Mayer et~al. 2006).

\item Why do we see significant changes in galaxy properties below a particular galaxy mass (Kauffmann et~al. 2003)? Why are there distinct populations or a 
bimodality in properties such as colour (e.g. Baldry et~al. 2004)? 

\item How can we reconcile observations of seemingly massive galaxies at 
high redshift, some of which are forming stars at prodigious rates, with 
a universe in which structures grow hierarchically? What do the galaxies 
seen at high redshift turn into by the present day? Are we seeing the 
formation of today's elliptical galaxies? When did the first galaxies 
begin to form?

\end{itemize}

\section{The basic ingredients -- Part 1: The dissipationless universe}
\label{dissless}

In this section we review the set of ingredients in the recipe for 
hierarchical galaxy formation that are on the firmest footing and 
upon which the majority of modellers would agree. The physics 
behind the topics discussed in this section is the growth 
of fluctuations in the dark matter, due to gravitational instability. 
In the cold dark matter model, the process of perturbation growth is 
dissipationless. This means that the total kinetic and potential 
energy of a system of dark matter is retained, although energy can be 
converted from potential to kinetic. The candidates for cold dark matter 
experience only the weak and gravitational forces. Therefore, they cannot 
lose energy through electromagnetic interactions which generate radiation. 
Our treatment in this section is brief as there are many sources to which 
the reader can turn for a more rigorous exposition (e.g. the textbooks 
by Padmanabhan (1993) and Peacock (1999) give an excellent overview of  
cosmic structure formation, covering perturbation theory, the spherical 
top-hat model and Press-Schechter theory). 

\subsection{The cosmological model}

Our starting point is to specify the background cosmology. 
The current ``standard'' model is a cold dark matter universe 
with a cosmological constant ($\Lambda$CDM). The initial fluctuations 
are assumed to follow a Gaussian random distribution. Once the 
values of the basic cosmological parameters have been set, such 
as the density parameter of matter ($\Omega_{\rm M}$), the density 
parameter of baryons ($\Omega_{\rm b}$) and the current amplitude of density 
fluctuations on some reference scale, the pattern of primordial 
density fluctuations is put in place, as described by the linear 
perturbation theory power spectrum (see Fig.~\ref{fig:cdm}) 
and the timetable for their collapse into gravitationally bound 
structures is set. 

\subsection{Dark matter haloes}

Dark matter haloes are the cradles of galaxy formation. Hierarchical 
galaxy formation models require three basic pieces of information about 
dark matter haloes: (i) The abundance of haloes of different masses. 
(ii) The formation history of each halo, commonly called the merger tree. 
(iii) The internal structure of the halo, in terms of the radial density 
and their angular momentum. 

These fundamental properties of the dark matter distribution are now 
well established, thanks mainly to the tremendous advances made possible 
by N-body simulations. 
The current state of the art in simulations of large scale structure 
is the Virgo Consortium's Millennium Simulation (Springel et~al. 2005).  
Driven on by the spectacular increase in the available computing 
power and developments in the algorithms used to compute the gravitational 
forces between particles, the Millennium simulation is a landmark in 
computational cosmology. 
Coming twenty years after the first calculations of 
hierarchical clustering in a CDM universe which employed 32,768 particles in a 
box of side $32.5h^{-2}$Mpc (Davis et al. 1985), the Millennium simulation 
volume is $500h^{-1}$Mpc on a side and uses in excess of ten billion 
particles to represent the dark matter. The smallest haloes that can be 
identified have a mass around $10^{10}h^{-1}M_{\odot}$, much smaller than the 
expected mass of the Milky Way's halo. 

\subsubsection{The abundance of dark matter haloes}

\begin{figure}
{\epsfxsize=12.truecm
\epsfbox[-200 180 550 700]{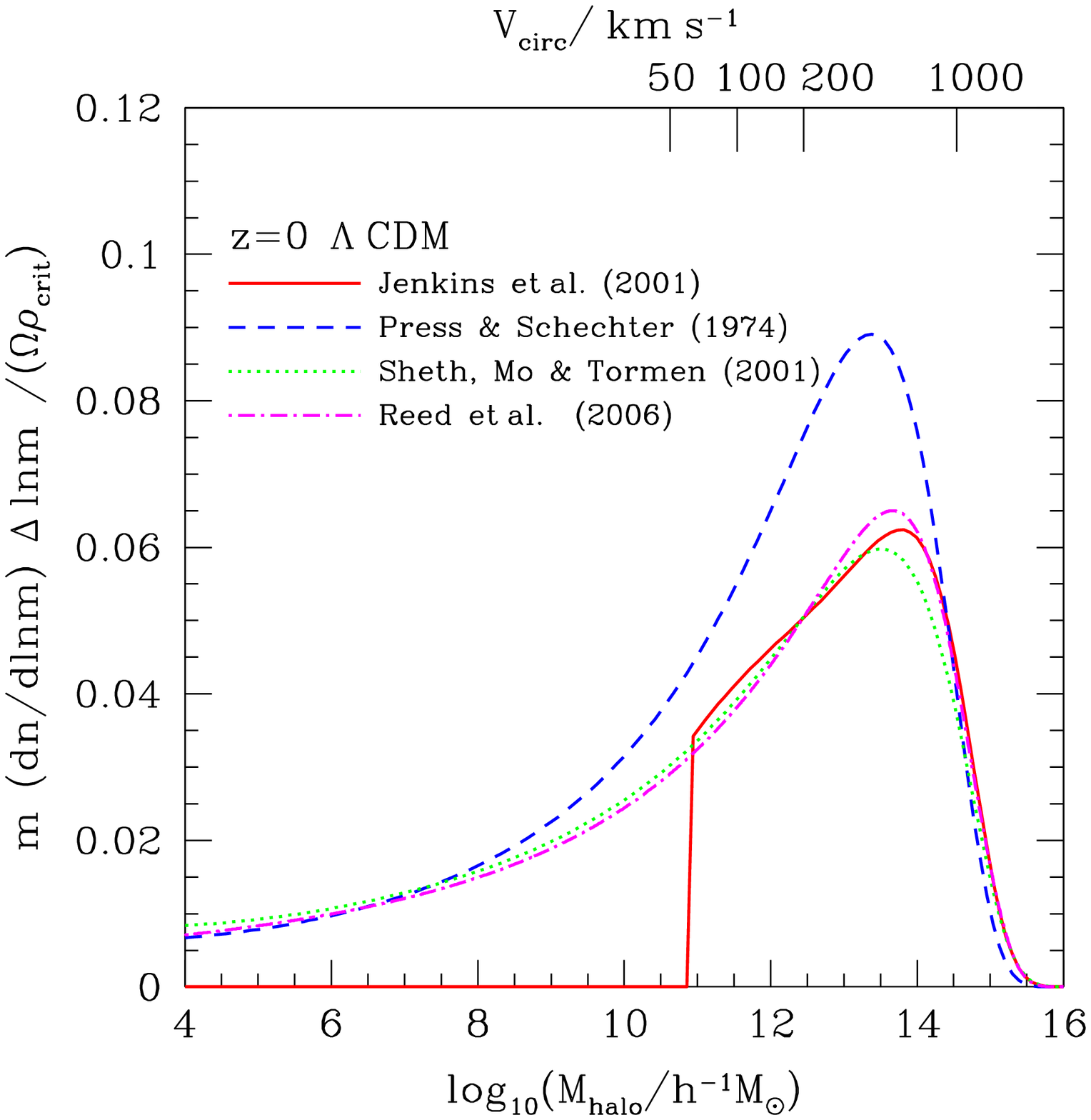}}
{\epsfxsize=12.truecm
\epsfbox[-200 180 550 700]{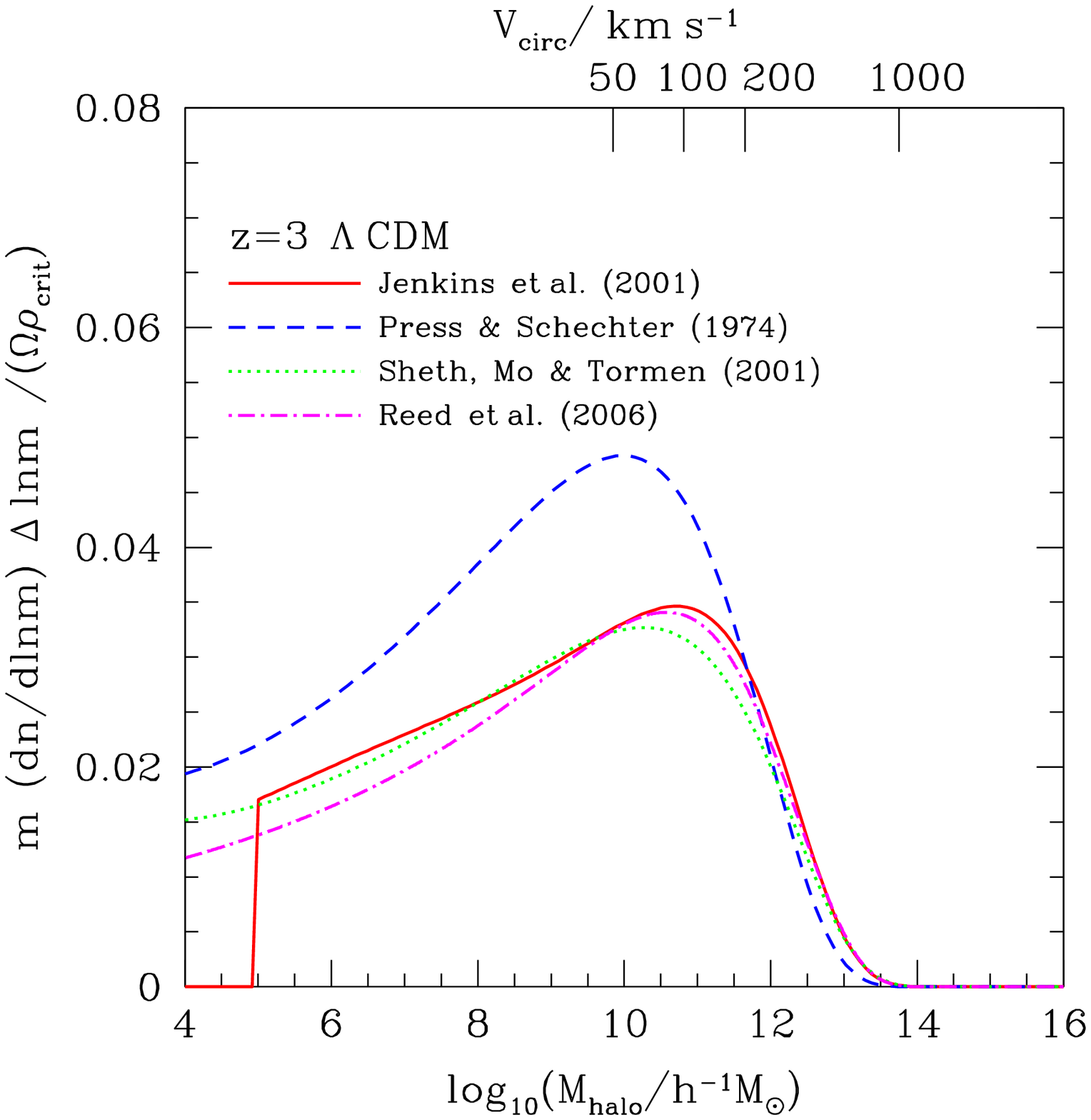}}
\caption{
The fraction of mass contained in haloes of different masses, at z=0 
(top panel) and z=3 (bottom panel). The upper axes in each panel give 
the equivalent circular velocity for selected halo masses. The curves 
show various theoretical predictions for the mass function of dark matter 
haloes, as indicated by the key. The same range of halo masses is plotted 
in each panel to emphasize the hierarchical growth of the mass function 
between $z=3$ and $z=0$. Note that the Jenkins et al. fit is only plotted 
over the range of masses available in the suite of N-body simulations these 
authors used to determine their fit; this curve goes to zero outside 
the range in mass over which it is applicable. 
}
\label{fig:massfn}
\end{figure}

The first attempt to calculate the abundance of gravitationally bound 
structures was made by Press \& Schechter (1974), long before the CDM 
model was introduced. Press \& Schechter assumed a Gaussian density 
field and smoothed the field on different scales. By varying the radius, 
$R$, of the spherical top-hat smoothing window, structures of different 
mass, $M$, could be considered, where $M=4/3 \pi \rho R^{3}$. The 
abundance of haloes above a given mass simply depends upon the fraction of 
spheres put down in the density field for which the linear theory 
overdensity or density contrast ($\delta = \rho(x,t)/\bar{\rho}(t)-1$) 
exceeds some critical value, $\delta_{\rm c}$. Press \& Schechter used the 
spherical top-hat collapse model to derive an appropriate value for  
$\delta_{c}$ (e.g. Peacock 1999). Thus, the fraction of the total mass 
that is contained within haloes of mass $M$ is obtained by integrating 
over the tail (i.e. $\delta > \delta_{\rm c}$) of a Gaussian with zero 
mean and a variance appropriate for smoothing the field on a radius 
defined by $M$. The Press-Schechter derivation neglects underdense parts 
of the universe, and so omits half the mass. Press \& Schechter adopted a 
pragmatic approach and multiplied their expression for the halo mass 
function by a factor of two. More convincing arguments have been put forward  
for the missing factor of 2, which led to the development of 
extended Press-Schechter theory (Peacock \& Heavens 1990; Bond et~al. 1991; 
Bower 1991; Lacey \& Cole 1993; Jedamzik 1995; Yano, Nagashima \& Gouda 1996, Nagashima 2001).  
A lucid exposition of the excursion set formalism behind extended 
Press-Schechter theory is given by White (1994). 
The extended version of the theory gives the distribution of 
masses of the progenitors of a halo at some earlier epoch, called the 
conditional mass function; this will be discussed in more detail in 
Section~\ref{sec:mergertree1}.  

The mass function predicted by this simple calculation agrees 
surprisingly well with the results obtained from N-body simulations 
(e.g. Efstathiou et~al. 1988; Lacey \& Cole 1994; Gross et~al. 1998; 
Governato et~al. 1999; Somerville et~al. 2000). 
The accuracy with which the mass function of haloes can now be predicted  
using N-body simulations has led to refinements in the Press Schechter 
ansatz. Sheth, Mo \& Tormen (2001) presented a model in which the collapse 
of a fluctuation is allowed to proceed more quickly along one axis, 
replacing the spherical collapse model with an ellipsoidal collapse.  
Jenkins et~al. (2001) established the mass function of haloes over four 
decades in mass using a suite of N-body simulations and proposed a fitting 
formula that encapsulates the numerical results. An extension of this 
work to five orders of magnitude in mass was carried out by 
Warren et~al. (2006) (see also Reed et~al. 2006a).  
The fraction of mass locked up in dark matter haloes is shown as a 
function of halo mass in Fig.~\ref{fig:massfn}. The upper panel shows 
$z=0$ and the lower panel $z=3$. The shift in the positions of the peaks 
of the curves between the panels shows the hierarchical nature of structure 
formation in a CDM universe. Fig.~\ref{fig:massfn} compares the Jenkins 
et~al. fit for the mass function with the analytical predictions of 
Press-Schechter and Sheth, Mo \& Tormen.
One further point to note from Fig.~\ref{fig:massfn} is that while 
clusters may be a useful laboratory for studying galaxy evolution, 
they are unrepresentative of the mass of the universe. 
Reed et~al. (2006a) extended the range of applicability of the Jenkins 
et~al. fit to lower masses (see also Yahagi, Nagashima \& Yoshii 2004). 
The mass function can be probed at low masses by resimulating a region 
from a larger cosmological volume at ultra-high resolution. The high 
resolution region is surrounded by a volume in which the particle mass 
used is much larger; this ensures that the correct tidal torques act 
on the high resolution volume. Diemand et~al. (2005) applied this 
multiscale technique to follow the collapse of structures in a CDM 
universe. Diemand et~al. considered initial fluctuation power spectra 
for the cases in which the dark matter is made up of axions or a 
supersymmetric particle with a rest mass of $\sim 100$GeV; for the 
latter, the power spectrum is truncated at a scale equivalent 
to a mass of $10^{-6}$ times the mass of the Sun. Diemand et~al. argue 
that the first structures to form in the dark matter have a mass 
similar to that of the Earth.  

\begin{figure}
{\epsfxsize=24.truecm
\epsfbox[-100 50 750 800]{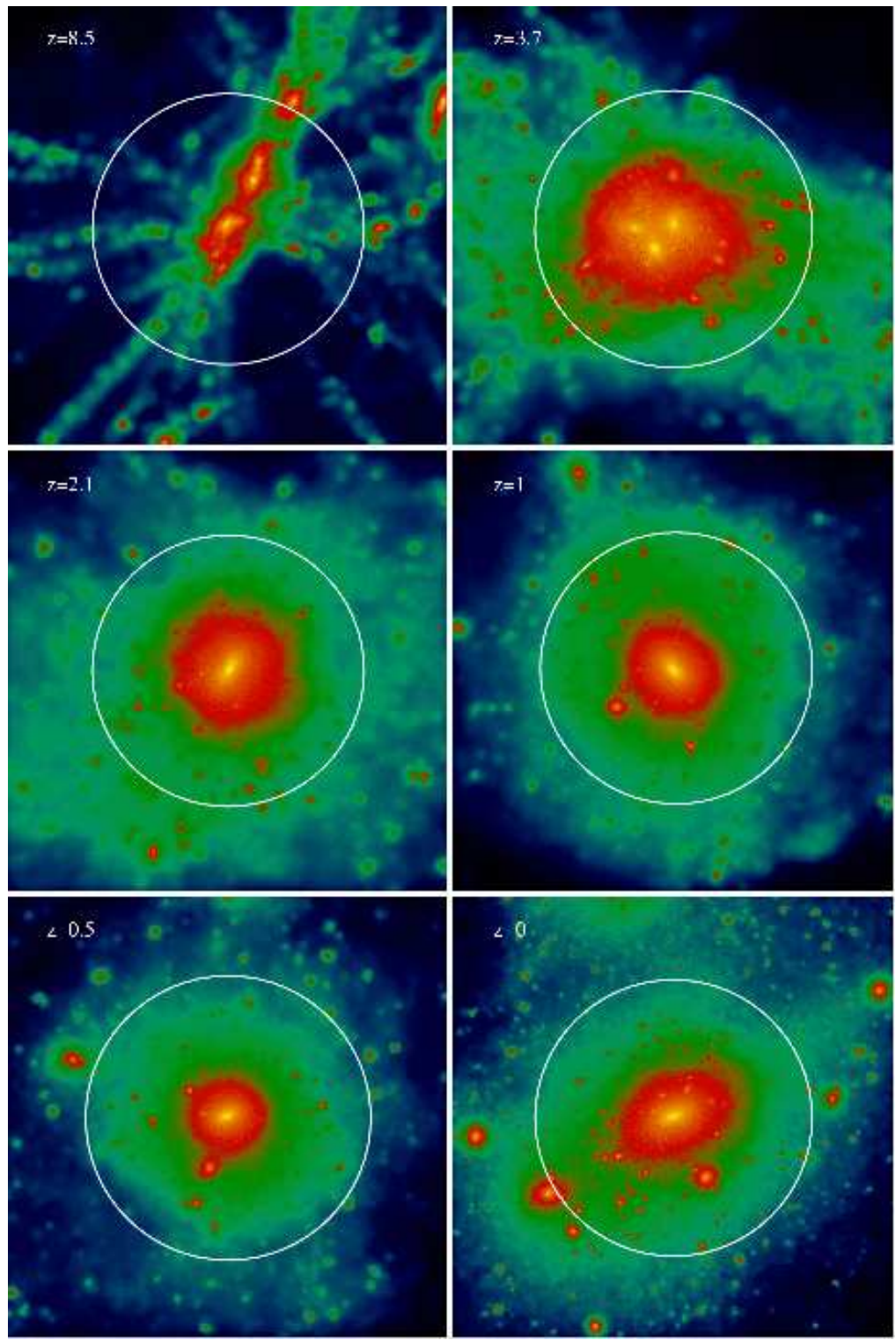}}
\caption{
The formation of a dark matter halo in a high resolution N-body simulation. 
The present day mass of the halo is $3 \times 10^{11}h^{-1}M_{\odot}$; 
the circle marks the present day virial radius $145h^{-1}$Kpc. The panels are 
fixed in comoving size and show snapshots at redshifts in the interval $z=8.5$ 
to $z=0$, as indicated by the labels. The colours reflect the density of dark 
matter, with ``warmer'' or redder colours indicating higher density. 
Figure courtesy of Chris Power. 
}
\label{fig:dmhalo}
\end{figure}

Until the late 1990s, numerical simulations of the growth of structure 
through gravitational instability suggested that dark matter haloes 
were smooth and featureless (e.g. Summers et~al. 1995; Frenk et~al. 1996). 
Subsequent advances in computing power and the development of techniques 
allowing the resimulation of selected volumes at greatly improved resolution 
showed that this phenomenon, dubbed `over-merging', was in fact a numerical 
artefact (Ghigna et~al. 1998; Moore et~al. 1998; Moore et~al. 1999a; 
Klypin et~al. 1999; Colin et~al. 2000). With the improved spatial and 
mass resolution afforded by packing more and more particles within 
the virial radius of the final resimulated halo (typically in excess of 
a few million particles), the gravitational potential of the progenitor 
haloes is better defined and these haloes are less diffuse. 
Fig.~\ref{fig:dmhalo} shows the formation of a dark matter halo in 
a high resolution N-body simulation (courtesy of Chris Power). 
Once a halo enters within the virial radius of a more massive halo 
it is referred to as a satellite halo or substructure within the larger halo. 
There is a wealth of substructure apparent within the virial radius of 
the halo at the present day in Fig.~\ref{fig:dmhalo}.
As the substructure halo orbits within the more massive halo, its mass 
is reduced as the more diffuse outer parts are stripped off by tidal 
effects and interactions with other substructures. 
Typically, around 15\% of the total mass of a dark matter halo is in the 
form of identifiable substructures, with the bulk of this mass accounted for 
by a small number of substructures (Ghigna et~al. 2000). The mass of the 
substructures can be reduced substantially from their original mass before 
infall into the larger halo. The circular velocity of the substructure 
is also affected by tidal effects, but to a lesser extent 
(Hayashi et~al. 2003; Kazantzidis et~al. 2004; Kravtsov, 
Gnedin \& Klypin 2004). The cores of the substructure haloes survive 
due to their high density compared with the outer parts of the haloes. 
 
\subsubsection{The assembly of dark matter haloes} 
\label{sec:mergertree1}

\begin{figure}
{\epsfxsize=14.truecm
\epsfbox[-120 150 570 680]{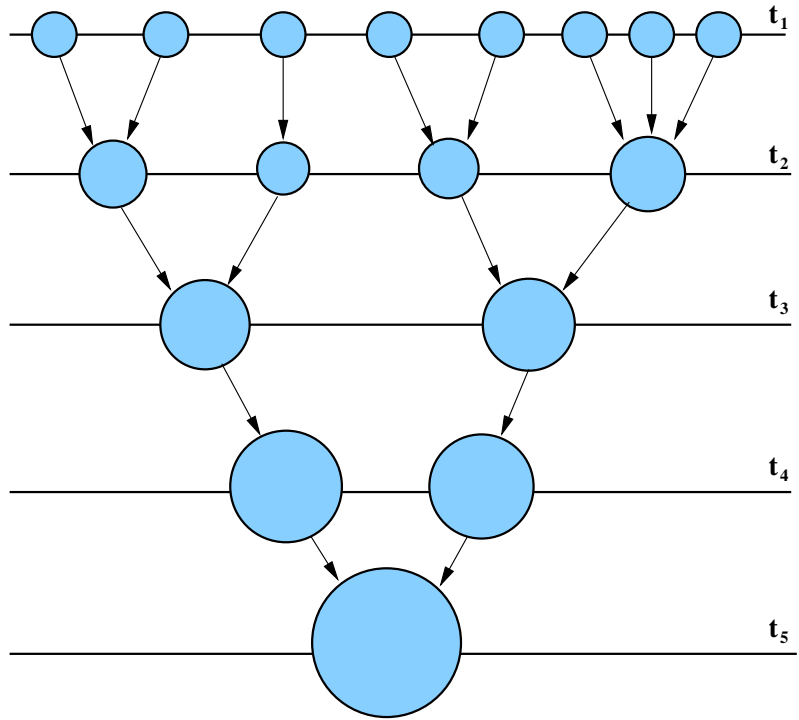}}
\caption{
A schematic merger tree for a dark matter halo. The horizontal lines 
represent snapshots in the evolution of the history of the halo, 
corresponding to timesteps in an N-body simulation or Monte-Carlo 
realization of the merger tree ($t_{1}<t_{2}$). The size of the circle indicates the 
mass of the halo. The haloes grow through merger events between haloes 
and by accretion of objects below the (halo) mass resolution 
(e.g. as depicted between steps $t_{3}$ and $t_{4}$). The final 
halo is shown at $t_{5}$. 
}
\label{fig:tree}
\end{figure}

The merger histories of dark matter haloes can be extracted from N-body 
simulations which have sufficiently frequent outputs 
(see Fig.~\ref{fig:tree} for a schematic merger tree). This requires  
typically around 50 outputs over a redshift interval of approximately 
z=20 to z=0. Haloes are identified in a given output using a percolation 
algorithm, such as friends-of-friends (Davis et al. 1985) or some 
other prescription designed to find a local overdensity (e.g. DENMAX, Gelb \& 
Bertschinger 1994; Spherical-overdensity, Cole \& Lacey 1996; SKID, Governato 
et~al. 1997; Bound-density-maximum, Klypin, Nolthenius \& Primack 1997; HOP, 
Eisenstein \& Hut 1998).  
The percolation algorithm links together all particles that are within some 
specified distance of one another. The linking length is quoted 
as some fraction of the mean interparticle separation and is set 
to return objects of a particular overdensity (see White 2002). 
The indices of the particles that belong to a particular halo can 
then be tracked in the halo list generated from 
the preceding (in expansion factor) output. 
Merger trees can also be generated using a Monte-Carlo approach by 
sampling the distribution of progenitor masses predicted using extended 
Press-Schechter theory (Lacey \& Cole 1994; Somerville \& Kolatt 1999; 
Cole et al. 2000; for a critique of extended Press-Schechter theory, 
see Benson, Kamionkowski \& Hassani 2005). The Monte-Carlo approach generally gives a less faithful representation of the merger trees than those 
extracted from N-body simulations, particularly as the difference 
in expansion factor increases between the parent halo and the progenitor 
branches (e.g. Somerville et~al. 2000). 
For example, if one generates merger trees for a representative sample of 
haloes at z=0, and we then attempt to construct the mass function of haloes 
at high redshift by combining the branches of the merger trees, with an 
appropriate weighting based on the abundance of the parent haloes, 
then the result will not agree with the mass function extracted 
from an N-body simulation at this epoch. 
The level of this discrepancy can be reduced by empirically tuning the 
progenitor distributions, though no theoretical justification exists for 
the form of such a correction (see e.g. Benson et al. 2001). 

A fundamental assumption that underpins the Monte-Carlo approach to 
growing merger trees is that the formation history of the halo does 
not depend upon its environment. If this assumption holds, then one 
can generate a merger history for a dark matter halo based upon its 
mass alone. Early simulation results appeared to validate this assumption 
(Lemson \& Kauffmann 1999; Percival et~al. 2003). However, some of these 
results have recently been reanalysed and found to show evidence in 
{\it favour} of an environmental dependence of halo properties 
(Sheth \& Tormen 2004). This effect was confirmed by recent analyses 
of the properties of dark matter haloes 
in large volume, high resolution simulations 
(Gao et~al. 2005b; Harker et~al. 2006; Wechsler et~al. 2005; Reed et~al. 2006b; 
Zhu et~al. 2006). 
These authors found a dependence of the clustering amplitude of galactic 
mass haloes on their formation history, as quantified by the formation 
redshift, which is defined as the redshift at which one progenitor first 
contains half the mass of the final object. For haloes at the extremes 
of the distribution of formation times (the 10\% which have either the 
highest or lowest formation redshifts), there is a strong change in 
clustering amplitude. Strictly speaking, this result invalidates the 
Monte-Carlo approach, at least in terms of using a one parameter model, 
halo mass, to assign a tree to a halo. On the other hand, the Monte-Carlo 
approach is correct 80\% of the time for such haloes (and 100\% of the time 
for higher mass haloes, for which no dependence of clustering signal on 
formation redshift is evident). Therefore, unless the galaxy population in 
question has extreme properties which lead to a strong correlation with the 
formation redshift of the halo, there will be little difference between 
results obtained from a Monte-Carlo approach and from the merger trees drawn 
directly from an N-body simulation. 
  
Both approaches, extracting the trees directly from an N-body simulation 
and growing Montre-Carlo trees, have their pros and cons. The N-body 
trees allow one to connect galaxies directly between outputs in the 
simulation and give predictions for galaxy positions within haloes. 
On the whole, they are more accurate and incorporate the 
environmental effects discussed in the previous paragraph. However, 
the N-body trees are not without their problems. Objects that a 
group finder has identified as one structure in a given output may 
actually fly part in a later output; spatial proximity is no guarantee 
that particles belong to a bound, self-gravitating structure. 
Furthermore, when objects do merge together, the mass of the remnant may 
not always equal the mass of the progenitors. 
These two effects mean that the mass of a halo in a merger tree extracted 
from an N-body simulation may not always increase monotonically with time. 
The smallest haloes that can be reliably identified a self-gravitating 
structures, do not, by definition, have a merger history that can be 
extracted from the simulation. The main drawback of N-body merger trees 
is their finite resolution 
(see Section~\ref{sec:mergertree2} and Helly et~al. 2003a). 
Monte-Carlo trees, on the other hand, can, in principle have arbitrarily 
high resolution, because the whole of the computer memory can be devoted to 
one tree at a time, rather than to all the haloes within some cosmological 
volume. 

\subsubsection{The structure of dark matter haloes}

The internal structure of dark matter haloes is important for determining 
the rate at which gas can cool (Section~\ref{sec:cooling}) and the size and 
dynamics of galaxies (Section~\ref{sec:size}). 

The structure of dark matter haloes has been studied extensively over the past 
decade and a half using computer simulations (Dubinski \& Carlberg 1991; 
Navarro, Frenk \& White 1996, 1997; Moore et~al. 1998; 
Fukushige \& Makino 1997, 2001; 
Klypin et~al. 2001; Power et~al. 2003; Hayashi et~al. 2004; 
Navarro et~al. 2004). 
One can either study well resolved (more massive) haloes within a 
standard N-body volume, or track selected objects with better resolution using 
either an adaptive scheme or a resimulation technique. 
In the case of resimulations, haloes are first 
identified in the output of a simulation of a cosmologically 
representative volume. The region containing the chosen halo is then 
resimulated using a much larger number of particles than was employed 
in the original calculation. At the same time, the input power spectrum 
is extended to higher wavenumbers to include the small 
scale power appropriate to the new, improved mass resolution.  
The remainder of the original simulation volume is represented using a 
number of high mass particles, so that the tidal torques which operate 
on the halo during its formation are reproduced. This high 
resolution resimulation technique now permits the structure of dark matter 
haloes to be resolved down to $0.5\%$ of the virial radius 
(Power et~al. 2003). The density profile of the dark matter varies with 
radius within the halo. Navarro, Frenk \& White (1996) reported a 
density profile that is significantly shallower than 
$\rho \propto r^{-2}$ near the centre but which tends to $r^{-3}$ 
as the virial radius is approached. Over much of the radius, 
an isothermal halo density profile, $\rho \propto r^{-2}$, is a reasonable 
description. This work was extended to show that the density profile of 
dark matter haloes could be described by a simple, universal formula, 
with a inner scale or concentration parameter 
which depends upon halo mass (Navarro, Frenk \& White 1997; see also 
Merritt et~al. 2005a, 2005b). 
The concentration parameter displays appreciable scatter as a function of 
halo mass (Jing 2000; Bullock et~al. 2001b; Eke, Navarro \& Steinmetz 2001; 
Wechsler et~al. 2002).
The most recent resimulations follow the hierarchical 
formation of haloes with several million particles within the virial radius 
(e.g. Moore et~al. 1999a,1999b; Springel et~al. 2001; Power et~al. 2003).

\subsection{A simple model: Is this all we need?}

\begin{figure}
{\epsfxsize=9.5truecm
\epsfbox[-160 180 550 710]{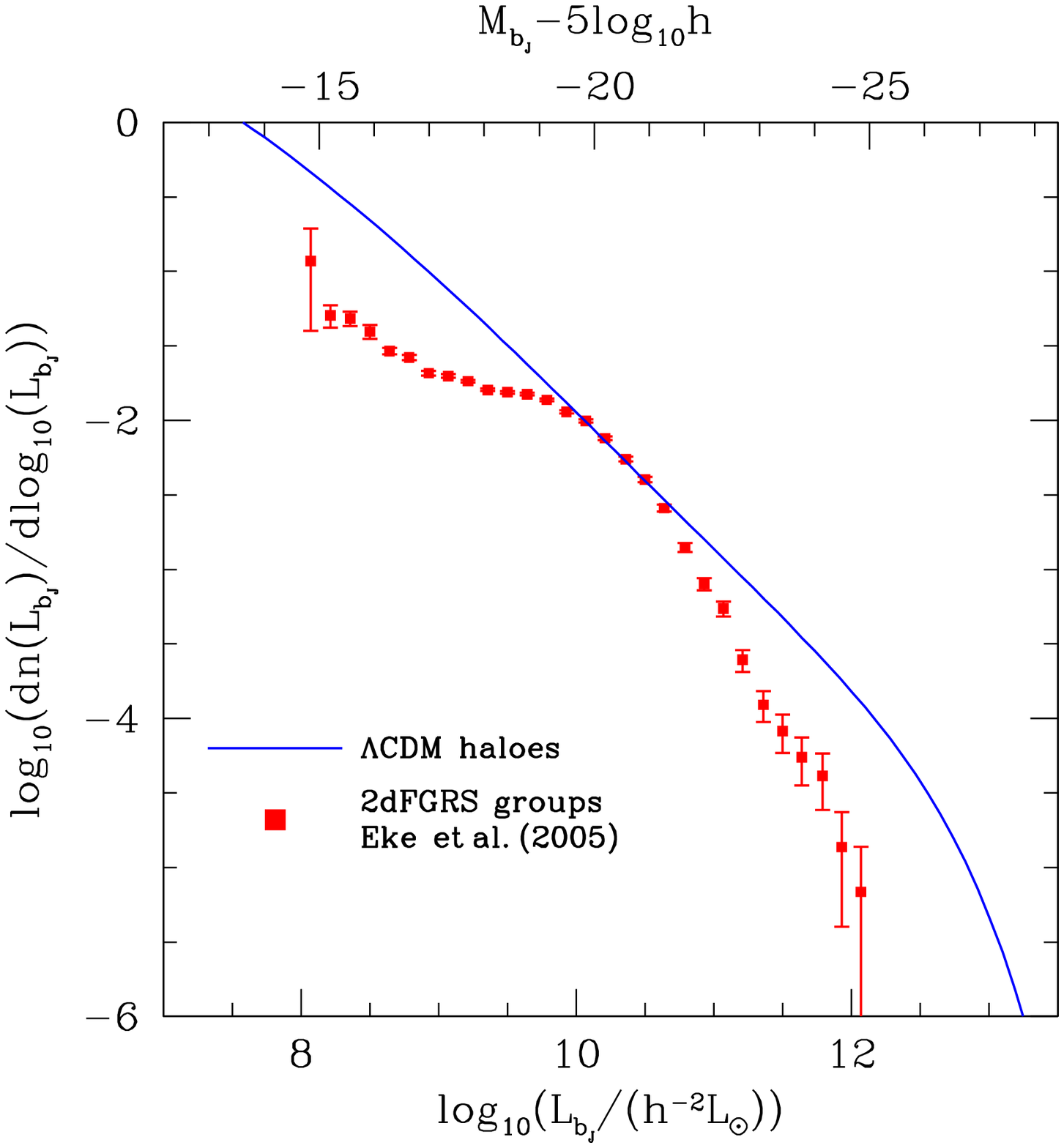}}
{\epsfxsize=7.5truecm
\epsfbox[0 180 550 710]{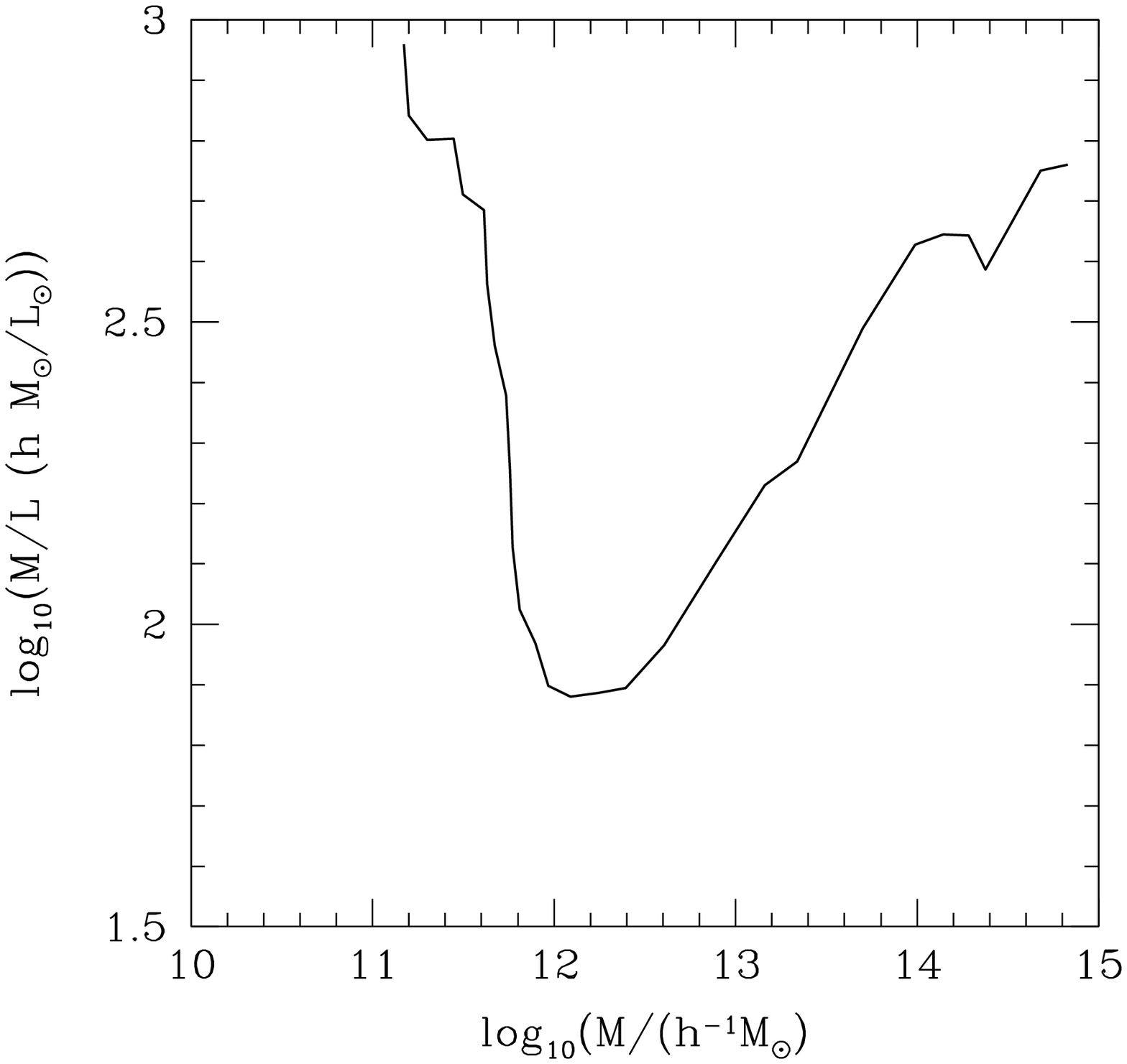}}
\caption{
Top: A simple prediction for the total luminosity function of 
galactic systems (solid line) compared with the group luminosity 
function estimated from the {\tt 2PIGG} catalogue by 
Eke et~al. (2006). The halo mass function of Jenkins et~al. 
(2001) has been converted into a group luminosity function 
by assuming a constant mass to light ratio for each halo. 
Bottom: The mass-to-light ratio required to match the observed group 
luminosity function is plotted in the right hand panel. 
Note that the strength of the up-turn below 
$M \sim 10^{12}h^{-1}M_{\odot}$ is affected by systematic 
errors in the determination of the total luminosity of 
groups in the 2dFGRS. 
}
\label{fig:haloml}
\end{figure}

Now that we have specified a cosmological model and can compute the 
abundance of dark matter haloes, we are in a position to make a very 
simple model of galaxy formation. This naive calculation will serve to 
reveal some basic facts about how the efficiency of galaxy formation 
must depend upon the mass of dark matter halo. The shortcomings of 
this toy model will motivate the more physical (and complicated) modelling 
that is the focus of this review. 

The first calculation that we can do is to take each dark matter halo 
and assign to it a luminosity that scales linearly with the mass of the halo. 
Thus, each halo is given a fixed mass to light ratio. Note that we have 
not made any assumption about how this light is distributed between 
galaxies within the halo. We can compare this prediction with the 
abundance of galaxy groups as a function of their total luminosity. 
This quantity was measured recently for galaxy groups extracted from 
the two-degree field galaxy redshift survey by Eke et al. (2004a,b). 
The comparison is shown in Fig.~\ref{fig:haloml}. 
A fixed mass-to-light ratio ($\sim 80 h M_{\odot}/L_{\odot}$) 
was chosen such that haloes of mass $\approx 10^{12}h^{-1}M_{\odot}$ 
match the break in the observed group luminosity function. 
We can see that this simple prediction gives a poor match to the observed 
luminosity function of groups. The predicted group luminosity function simply 
has the wrong shape, with too many faint groups and too many bright groups. 
Thus, if we are to retain the otherwise highly successful background 
$\Lambda$CDM cosmology, our assumption of a mass to light ratio which 
does not vary with halo mass is seriously flawed. 

We can of course choose the mass to light ratio of each dark matter halo 
more carefully. If we match the observed galaxy groups to dark matter 
haloes that are predicted to have the same space density, we can derive 
a mass to light ratio that guarantees a match between the theoretical 
prediction and the observed group luminosity function. The mass to light 
ratio obtained by this procedure is a strong function of halo mass, 
as shown by the right-hand panel of Fig.~\ref{fig:haloml}. The mass to light 
ratio is lowest for haloes of mass $\approx 10^{12}h^{-1}M_{\odot}$, 
and rises by a factor of $\approx 6$ to lower and higher mass haloes 
(Yang, Mo \& van den Bosch 2003; Eke et~al. 2004a, 2004b). 
(Note the sharpness of the increase in the mass to light ratio for 
haloes with masses below $10^{12}h^{-1}M_{\odot}$ is exaggerated 
somewhat by errors in the determination of the total group luminosity; 
see Eke et~al.~2006 for a discussion.) 
Thus, galaxy formation is expected to be most efficient in 
haloes of mass $\sim 10^{12}h^{-1}M_{\odot}$ (Eke et~al. 2006); 
we expect that these haloes should produce the most luminosity 
per unit mass. For some reasons, which may differ depending upon 
the mass scale, the efficiency with which galaxies 
form drops as haloes of mass lower or higher than 
$\sim 10^{12}h^{-1}M_{\odot}$ are considered. 

This simple exercise reveals two key facts about galaxy formation. 
Firstly, the efficiency of galaxy formation is low. Most baryons do not 
end up as stars. Audits of the distribution of baryons in the Universe 
suggest that galaxy formation is not particularly efficient at 
turning hot gas into cold gas and stars (Persic \& Salucci 1992; Fukugita, 
Hogan \& Peebles 1998; Balogh, Pearce, Bower \& Kay 2001). 
Cole et~al. (2001) used their measurement of the K-band galaxy luminosity 
function and simple stellar population synthesis models (see later) 
to construct the local stellar mass function of galaxies 
(see also Kochanek et~al. 2001 and Bell et~al. 2003b). Integrating 
over the stellar mass function gives the density parameter in stars today. 
Cole et~al. found that only a small fraction of baryons, around 10\% depending 
upon the choice of stellar initial mass function, have been turned into stars. 
An even smaller fraction of baryons, around 1\%, are in the form of cold gas 
in galaxies today (Zwaan et~al. 2003). 
Secondly, the efficiency of galaxy formation is not the same in haloes 
of different mass. The mass of the dark matter halo plays an important 
role in shaping the galaxies that it contains. Direct observational evidence 
for this has been obtained using group catalogues derived from 
the two-degree field galaxy redshift survey 
(Eke et~al. 2006; Yang et~al. 2005).

\section{The basic ingredients -- Part 2: ``gastrophysics''}
\label{sec:ingred}

\begin{figure}
{\epsfxsize=18.truecm
\epsfbox[-50 80 590 750]{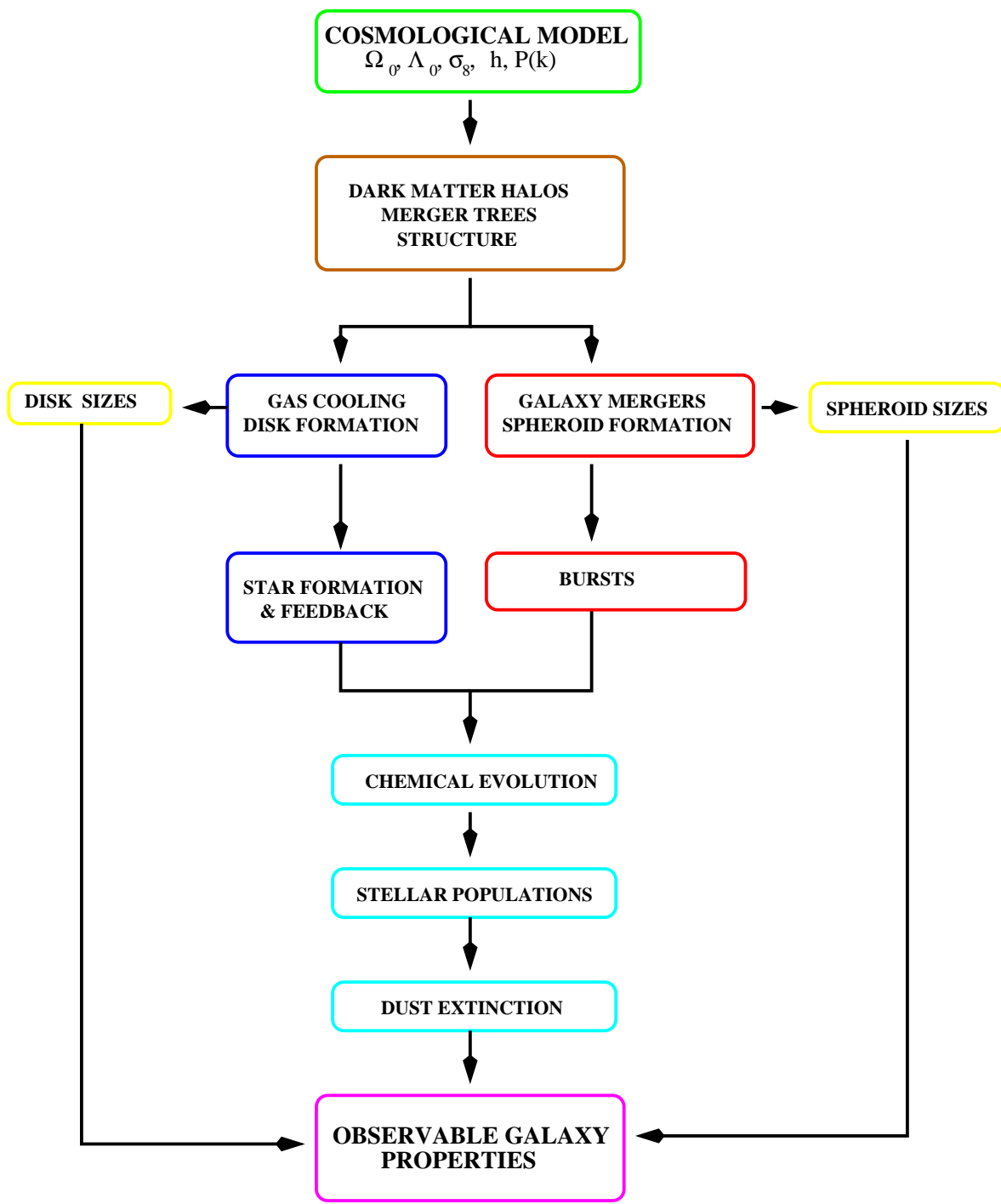}}
\caption{
A schematic overview of the ingredients of a hierarchical 
galaxy formation model. Adapted from Cole et~al. (2000). 
}
\label{fig:physics}
\end{figure}

In this section, we outline the more complicated elements of 
hierarchical galaxy formation. These processes are far more difficult 
to deal with than gravitational instability, and are often dissipative 
and nonlinear, which led Dick Bond to coin the apt umbrella label 
of ``gastrophysics''. The physics behind the phenomena that are 
described in this section are in general poorly understood. To 
counter this, recipes or prescriptions which contain parameters 
are employed. The values of the parameters are set by demanding 
that the model reproduces a subset of the available observations, typically 
low redshift data. The form of the rule adopted to describe a 
process is motivated by a result from a more detailed numerical 
simulation or from observations. 
We give a generic description of how phenomena are modelled, rather 
than providing a detailed comparison between the implementations 
used in different models; such a comparison would be tedious and 
would soon be out of date, since the models are continually being 
improved and developed. An overview of the processes typically incorporated 
in semi-analyical models is shown in Fig.~\ref{fig:physics}.

\subsection{The cooling of gas}
\label{sec:cooling}

\begin{figure}
{\epsfxsize=10.truecm
\epsfbox[-220 23 490 830]{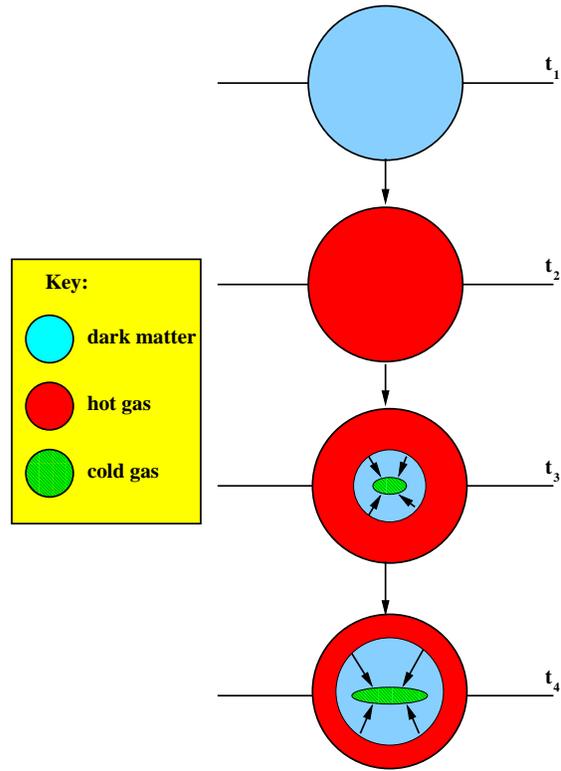}}
\caption{
A schematic of the basic cooling model used in semi-analytical models. 
Each line represents a stage in the cooling process. In the first 
step ($t_{1}$), baryons fall into the gravitational potential well of the dark 
matter halo. The presence of a photo-ionising background may reduce the 
fraction of baryons that fall into low mass haloes, as described in the 
text. This gas is assumed to be heated by shocks as it falls into the 
potential well, attaining the virial temperature associated with the 
halo ($t_{2}$). In the third step ($t_{3}$), the inner parts of 
the hot gas halo cool, forming a rotationally supported disc. 
At a later stage ($t_{4}$), the radius within which gas has had time to 
cool advances outwards towards the virial radius of the halo and the 
cold gas disc grows in size. 
}
\label{fig:cooling}
\end{figure}

The cooling of gas is central to the process of galaxy formation, 
as it sets the rate at which the raw material for star formation 
becomes available (Blumenthal et~al. 1984). The basic model of how 
gas cools inside dark matter haloes was set out in detail by 
White \& Frenk (1991; see also Cole 1991 and Lacey \& Silk 1991); 
White \& Frenk based their framework on the arguments about gas 
cooling set out in Rees \& Ostriker (1977) and Silk (1977). 
White \& Rees (1978) were the first to postulate that gas cooling 
took place within dark haloes.  

A schematic picture of the standard gas cooling model used in 
semi-analytic models is presented in Fig.~\ref{fig:cooling}. 
The gas initially has the same spatial distribution as the 
dark matter ($t_{1}$). As fluctuations in the dark matter separate from 
the Hubble expansion, turn around and collapse, the gas is assumed to be 
heated by shocks as it falls into the gravitational potential well of 
the dark halo, producing a hot gas halo that is supported against 
further collapse by the pressure of the gas (step $t_{2}$). 
The gas attains the virial temperature of the halo, 
which depends upon the mass of the halo: 
\begin{equation}
T_{\rm vir} = \frac{1}{2} \frac{\mu m_{\rm H}}{k} V^{2}_{\rm H},
\label{eq:vir}
\end{equation}
where $\mu=1/1.71$ is the mean molecular mass of the gas, $m_{\rm H}$ is 
the mass of a hydrogen atom and $k$ is Boltzmann's constant. 
Dark matter haloes are supported against further gravitational 
collapse by a pressure created by the thermalized velocities of the 
dark matter particles. However, it is common practice to quote an 
equivalent circular velocity for the halo at the virial radius, 
$r_{\rm vir}$, using $V_{\rm H} = \sqrt{G M /r_{\rm vir}}$; the halo has 
an angular momentum, but this should not be thought of as rotation with 
velocity $V_{\rm H}$. 
The relation between the circular velocity at the virial radius and 
halo mass is a function of redshift and cosmology
\footnote{
We caution the reader that the literature contains several definitions 
of the virial mass of a dark matter halo, quoted in terms of the mean 
overdensity contained within the virial radius times some reference density 
as calculated using the spherical collapse model (see Padmanabhan 1993).  
For a universe with the critical density in matter, the mean overdensity 
of mass within the virial radius is 178 times the background density. 
This value of 178 is often rounded up to 200. In a low density universe, 
haloes have a lower mean overdensity when expressed in units of the critical 
density e.g. for a flat, $\Omega=0.3$ universe, the mean overdensity of a 
dark matter halo is around 100 times the {\it critical density}; see Fig.~1 of 
Eke, Cole \& Frenk (1996).} (e.g. Mo \& White 2002).  
Gas can subsequently cool from the hot halo, through the processes 
outlined in the next paragraph. The rate at which the gas 
can cool depends upon the temperature of the gas, which determines its 
ionisation state, the chemical composition of the gas and the density 
of the gas, which determines the rate at which collisions between electrons 
and ions take place. 
As the gas cools, the pressure of the gas drops and the removal of 
pressure support means that the gas sinks to the centre of the 
dark halo on the free-fall or dynamical timescale in the halo (step $t_3$). 
If the angular momentum of the cooling gas is conserved 
(see Section~\ref{sec:disk}), the cold gas forms a rotationally 
supported disk. 
Hence, the rate at which cold gas is added to the galactic disc depends 
upon (i) how quickly the gas can cool (i.e. the cooling time) 
and (ii) how quickly the cooled gas can move from the halo to the disk 
(i.e. the free-fall time). 
In this simple model, the rate at which gas can cool is used to compute 
a cooling radius, $r_{\rm cool}$. The gas enclosed 
within $r_{\rm cool}$ has had sufficient time to cool since the 
formation of the dark matter halo. The cooling radius continues to 
propagate outwards (step $t_4$) until either all of the hot gas halo has 
cooled, or a merger with another halo results in the formation of a new halo. 

Gas can cool via a number of mechanisms (see, for example, the discussion 
in Kauffmann \& White 1994). The relative importance of the various 
mechanisms depends upon the conditions in the universe at the time 
the gas is cooling and the temperature of the gas. The cooling channels are:  
(i) Inverse Compton scattering of CMB photons by 
electrons in the hot halo gas. The time for gas to cool via inverse 
compton scattering exceeds the age of the universe for redshifts $z < 10$ 
(Rees \& Ostriker 1977), so this process is only important in the very early 
universe (see Fig~\ref{fig:lookback}).    
(ii) The excitation of rotational or vibrational energy levels in 
molecular hydrogen through collisions. The subsequent decay of these 
levels removes energy from the gas, allowing it to cool. This channel 
is important in haloes with virial temperatures (see Eq.~\ref{eq:vir}) 
below $T \sim 10^{4}$K (see e.g. Figure 12 in Barkana \& Loeb 2001).  
(iii) Emission of photons following transitions between energy 
levels. Collisions between partially ionised atoms and electrons 
excite the atoms to higher energy levels. The gas cools when the 
excited level decays radiatively. This process is important for haloes 
with intermediate virial temperatures (i.e. $10^{4} {\rm K} < T < 10^{6}$K).
(iv) Bremsstrahlung radiation as electrons are accelerated in an ionized 
plasma. This the dominant emission mechanism in massive clusters 
($ T \sim 10^{7}$K). 

\begin{figure}
{\epsfxsize=14.truecm
\epsfbox[-200 173 557 700]{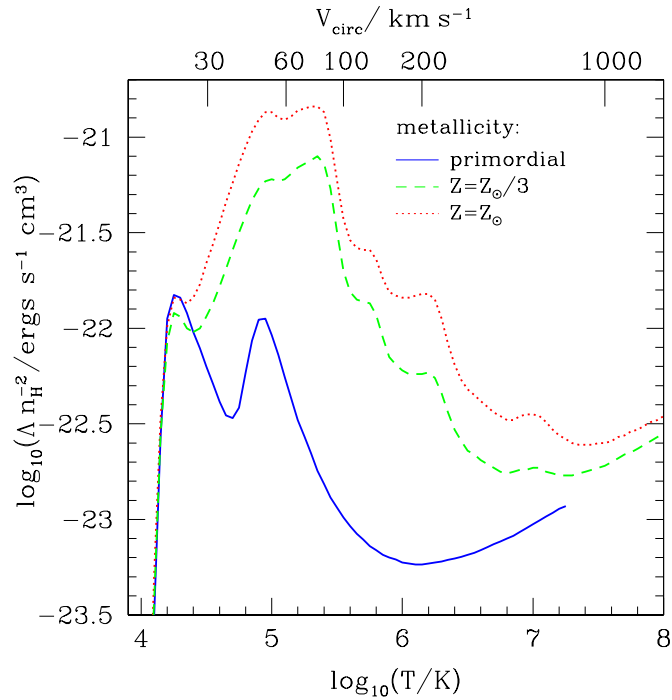}}
\caption{
The cooling rate plotted as a function of the virial temperature of 
the hot halo gas. The equivalent circular velocity of the halo is 
indicated on the top axis. The different curves show how the cooling 
rate depends upon the metallicity of the gas, as indicated by the key.
}
\label{fig:coolrate}
\end{figure}

The primary cooling processes relevant to the formation of galaxies 
are (iii) and (iv) in the list above (Kauffmann \& White 1994). 
A cooling time can be specified by dividing the thermal 
energy density of the gas by the cooling rate per unit volume: 
\begin{equation}
t_{\rm cool}(r) = 
\left(\frac{3}{2} \frac{\rho_{\rm gas}kT_{\rm vir}}{ \mu m_{\rm H}}\right) 
/ 
\left(\rho^2_{\rm gas} \Lambda ({T_{\rm vir},Z_{\rm gas}}) \right).
\end{equation}  
Here, $\rho_{\rm gas}$ is the gas density. The function $\Lambda$ is 
a function of the temperature and metallicity of the 
gas, $Z_{\rm gas}$. 
Primordial gas is partially ionised at $T\sim 10^{4}$K and fully ionised 
at a temperature of around $10^{6}$K; enriched gas becomes ionised when 
the temperature exceeds $10^{7}$K. For a plasma, the dominant cooling 
mechanism is Bremsstrahlung radiation by electrons 
experiencing acceleration in the electric field of ions, with an associated 
cooling rate $\propto T^{1/2}$. This explains the form of the cooling 
function at high temperatures in Fig.~\ref{fig:coolrate}, which shows  
the model results from Sutherland \& Dopita (1993).
At intermediate temperatures, the dependence of the cooling rate on 
temperature is more complicated, particularly once the gas becomes 
enriched. 
In this temperature regime, electrons can recombine with ions emitting a 
photon or partially ionized atoms can be excited by a collision with 
an electron, and then emit radiation as they decay to the ground state. 
Decays following the excitation of ionised atoms by collisions 
dominate in primordial gas, causing the peaks seen in the cooling 
function at $\,\,15\,000$ K (Hydrogen) and $100\, 000$ K (singly-ionised Helium). 
Gas with a solar abundance of metals has a much stronger peak at 100 000 K 
due to oxygen; other elements enhance the cooling rate around 
$10^{6}$K (Kauffmann \& White 1994).

The standard cooling model can be modified in a number of ways which 
affect the rate at which gas cools in either low or high mass haloes. 
These changes to the cooling rate are the result of phenomena which  
depend upon gas cooling that took place in the past, and as such can 
be considered as feedback processes that regulate the rate at which 
star formation can occur. 

The first of these phenomena is the suppression of cooling in low mass 
haloes due to the presence of a background of photo-ionising radiation 
(Couchman \& Rees 1986; Efstathiou 1992; Babul \& Rees 1992; 
Thoul \& Weinberg 1996; Nagashima, Gouda \& Sugiura 1999; Gnedin 2000). 
The background of high energy UV photons could be generated by quasars or 
by massive stars made in primitive galaxies. 
Two effects combine to reduce the rate at which gas can cool in low mass 
haloes. Firstly, the radiation heats the 
intergalactic medium to a temperature of $\sim 10^{4}$K, increasing the 
pressure of the baryons. This restricts the infall of baryons into haloes 
with virial temperatures below $10^{4}$K, with the consequence that these 
haloes accrete less than the universal mass fraction of baryons. 
Secondly, the radiation increases the ionisation of the hot gas, 
thereby removing channels for cooling following the excitation of atoms 
and ions in collisions. The low temperature peak in the cooling curve is 
effectively deleted by the radiation background, thus dramatically 
increasing the cooling time in low mass haloes. Benson et~al. (2002a,b) 
produced a fully coupled model of the intergalactic medium and the formation 
of galaxies, in which the contribution to the photo-ionising background 
from star formation is treated self-consistently and the 
background from quasars in included empirically (Haardt \& Madau 1996). 
Benson et~al. also discussed a simple model which encapsulates some of 
the main features of their more detailed calculation; in this simple scheme, 
haloes with a circular velocity below some threshold 
($\approx 60 {\rm kms}^{-1}$) are not allowed to cool gas below the 
redshift at which the universe is reionised (see also Somerville 2002; 
Tully et~al. 2002). 

The cooling rate in massive haloes can be reduced by heating the hot halo 
gas. Three mechanisms have been discussed in the literature. In the 
first of these, the energy released by supernova explosions is injected 
into the hot gas halo, causing it to expand and become more diffuse, 
thus lowering the cooling rate (Bower et~al. 2001). Bower et~al. 
estimated that an energy of around $\sim 1$KeV per particle in the hot 
halo is required to suppress cooling sufficiently 
in order to reproduce the break in the galaxy luminosity function 
(see also Wu, Fabian \& Nulsen 1998, 2000). 
Benson et~al. (2003) discuss the second mechanism which can diminish  
cooling flows in massive haloes, the thermal conduction of energy from the 
outer parts to the inner parts of the hot gas halo. Spectra of clusters 
taken by the Chandra and XMM-Newton satellites indicate a lack of cold 
gas (i.e. below 1-2 KeV) in the intra-cluster medium, suggesting that 
some form of heating mechanism may be preventing the gas from attaining 
these temperatures (Fabian et~al. 2001; Kaastra et~al. 2001; 
Peterson et~al. 2001). If the ionised plasma in the halo has a conductivity 
of the order calculated by Spitzer (1962), this could explain 
the suppression of the cooling flow (Fabian et~al. 2001, 2005). 
Benson et~al. modelled the impact of thermal conduction on the cooling rate 
by preventing cooling in haloes with a rotation speed above a set value; 
note that the cut-off velocity scales with redshift 
as $(1+z)^{3/4}$ (see Nagashima et~al. 2005a). 
The value of the cut-off velocity required to produce a good match to the 
bright end of the luminosity function implies a thermal conductivity that is 
greatly in excess of the Spitzer value. 
The third mechanism is heating of the hot halo by energy released from the 
accretion of material onto a central black hole. Croton et~al. (2006) and 
Bower et~al. (2006) describe models in which the cooling rate is gradually 
reduced with halo mass. This model is described in more detail in 
Section~\ref{sec:feedback}. 

A critical assessment of the standard cooling model is given in 
Section~\ref{sec:simcool}.

\subsection{Star formation}
\label{sec:starformation}

One fairly safe prediction that can be made in writing this review is the 
following: we still have some considerable time to wait until a theory of star formation 
is developed such that the local properties of the interstellar medium 
can be fed into a subroutine which will return the star formation rate as a 
function of position within a model galaxy, without any free parameters. 
Despite this, there has been a huge amount of activity on several fronts: 
(i) Understanding the properties of the first generation of stars, such as 
when and where these objects are likely to have formed and what mass they 
had. (ii) The distribution of stellar masses produced in episodes of star 
formation as quantified by the stellar initial mass function (IMF). (iii) 
The conditions for star formation in galactic disks and starbursts.  

The formation of the first stars from gas with a primordial composition 
is a problem with several features which combine to make it 
more amenable to simulation than tracking subsequent generations 
of stars (Abel, Bryan \& Norman 2002; see also Bromm \& Larson 2004; 
Yoshida et~al. 2003; Ciardi \& Ferrara 2005): 
(i) The gravitational potential wells most likely to host the 
first sites of star formation can be identified given 
a power spectrum of density fluctuations. 
(ii) The chemistry of the unenriched gas is much simpler than is 
the case for subsequent generations of stars that form from gas 
polluted with metals from supernova explosions (Tegmark et~al. 1997). 
(iii) Magnetic fields are unimportant in primordial gas clouds 
(Abel et~al. 2002). 
(iv) By definition, the first star forms from gas that has not been 
ionised by radiation or winds from other stars. 
Abel et~al. were able to follow the formation of a proto-stellar core 
in a primordial gas cloud, up until the point at which radiation from 
the proto-star has an impact on the rate at which material can be accreted 
onto the growing star, since their calculation did not treat the radiative 
transfer of the photons from the proto-star. They argued that the masses 
of the first generation of stars could be much higher than expected 
from a standard IMF. Gao et~al. (2005a) and Reed et~al. (2005) extracted 
the most massive halo from a $\Lambda$CDM simulation at the present day and, 
using a resimulation technique, traced the progenitors of this structure back 
to $z \sim 50$. They postulated that the potential wells already in place 
by $z \sim 50$ could host the first stars, which is a considerably higher 
redshift than that proposed by Abel et~al.

There are a wide range of theories of star formation from gas clouds which  
attempt to explain the slope of the IMF and the rate of star formation in 
galaxies. One issue that does not seem to have been resolved between the 
modellers is the nature of the dominant process for determining the masses 
of stars. Starting from a molecular gas cloud, there are top-down and 
bottom-up scenarios for star formation. In the bottom-up case, 
low mass stellar cores acquire gas from the cloud in a competitive accretion 
process (Bonnell et~al. 1997). On the other hand, in the top-down 
model, the gas cloud simply fragments and the sub-clouds collapse to 
form stars (Krumholz, McKee \& Klein 2005). 
Other physical processes considered include collisions between clouds 
(Tan 2000), the escape of magnetic fields by ambipolar diffusion 
(Tassis \& Mouschovias 2004) and supersonic turbulence 
(Klessen, Heitsch \& Mac Low 2000; Li et~al. 2004). We do not attempt 
to go into the details of these processes here; instead we refer 
the interested reader 
to the review articles by Elmegreen \& Scalo (2004) and 
Mac-Low \& Klessen (2004).

The lack of a theory of star formation, may, at first sight, appear to 
thwart any attempt to produce a theory of galaxy formation. 
Semi-analytical modellers have instead be forced to take a more pragmatic,  
top-down approach. A simple estimate of the global rate of star formation 
in a model galaxy can be made on dimensional grounds: 
\begin{equation} 
\dot{M}_{*} \propto \frac{M_{\rm cold}}{\tau},
\label{eq:schmidt}
\end{equation} 
where the star formation rate, $\dot{M}_{*}$, depends upon the amount of 
cold gas available, $M_{\rm cold}$, and a characteristic timescale 
$\tau$. The timescale could be chosen to be proportional to the dynamical 
time within the galaxy, $\tau_{\rm dyn} = r_{\rm gal}/ v_{\rm gal}$, or to 
be some fixed value. Typically, some additional dependence on the circular 
velocity is incorporated into the definition of $\tau$, which is important 
when attempting to reproduce the observed gas fractions in spirals as a 
function of luminosity (e.g. Cole et~al. 1994; Cole et~al. 2000). 
The {\it effective} star formation timescale is 
in practice a modified version of $\tau$, due to feedback processes, 
which deplete the reservoir of cold gas, and the replenishment of 
the cold gas supply by material that is recycled by stars. 

Schmidt (1959) proposed a model in which the star formation rate per unit 
area of a galaxy ($\dot{\Sigma}_{*}$) scales with a power of the surface 
density of the cold gas, $\Sigma_{\rm g}$: $\dot{\Sigma}_{*}  \propto 
\Sigma^{n}_{\rm g}$. Kennicutt (1998a) verified this form for a large sample 
of spiral and starburst galaxies, finding $n \sim 1.4$ 
(see also Kennicutt 1998b). The Schmidt law can be rewritten 
in the form of Eq.~\ref{eq:schmidt}, with $\tau$ replaced by the 
dynamical time of the galaxy (see Kennicutt 1998b, Bell et~al. 2003a).

\subsection{Feedback processes}
\label{sec:feedback}

\begin{figure}
{\epsfxsize=14.truecm
\epsfbox[-200 145 557 700]{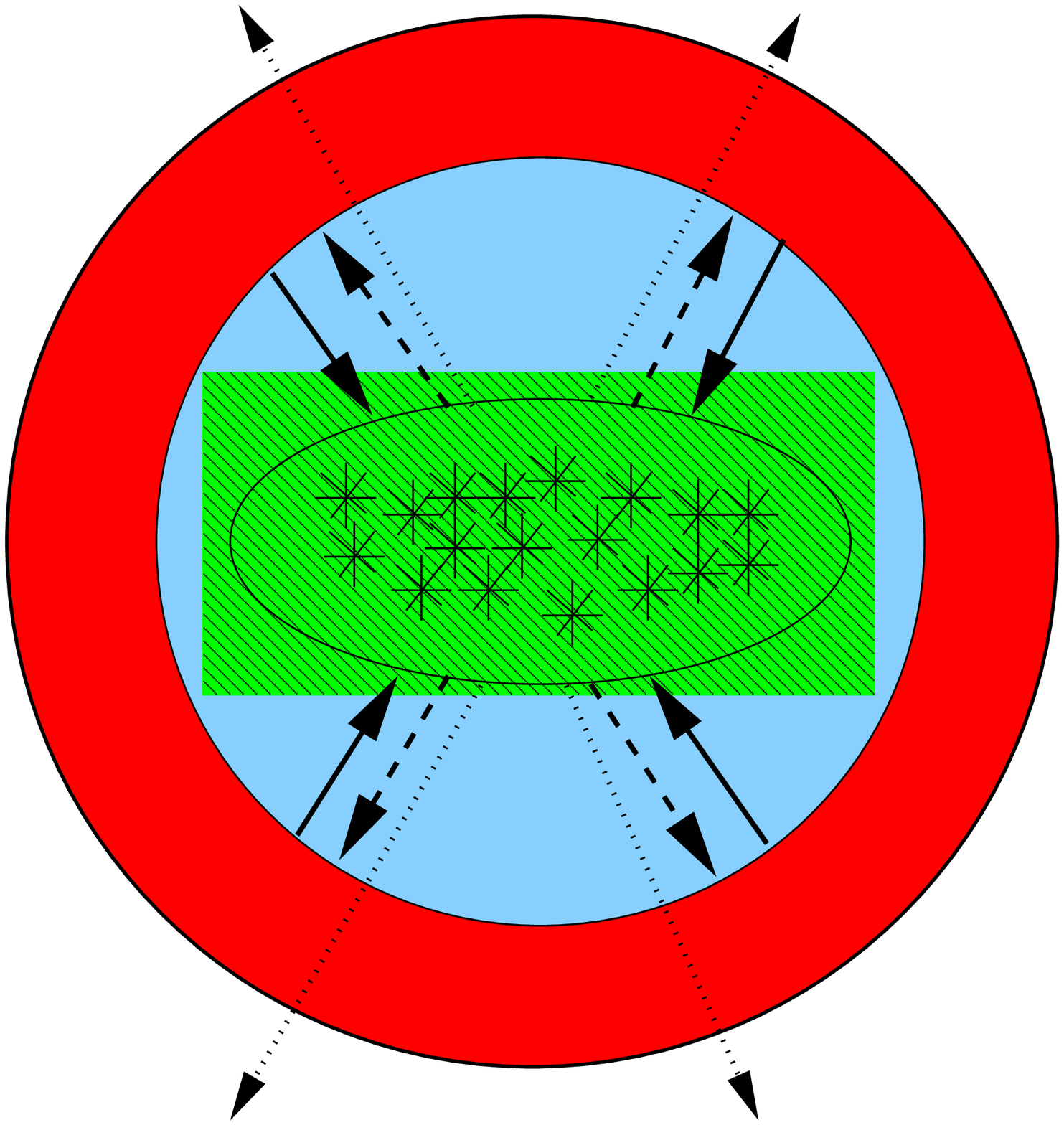}}
\caption{
A schematic figure showing gas cooling from the hot halo (solid lines) and 
building up the reservoir of cold gas in the galactic disc. The cooled 
gas is turned into stars on a timescale set by the parameters 
of the model. Supernova explosions can reheat a fraction of the cooled 
gas and return it to the hot phase (dashed lines) or eject material 
from the halo altogether (dotted lines).
}
\label{fig:stars}
\end{figure}

The need for physical mechanisms that are able to 
modulate the efficiency of galaxy formation as a function of 
halo mass, over and above the variation in the cooling time of the hot gas 
with halo mass, was recognised from the first calculations of 
the galaxy luminosity function in hierarchical clustering 
cosmologies. White \& Rees (1978) found that their prediction for the 
faint end of the luminosity function was steeper than the observational 
estimates available at the time, leading them to speculate that 
this discrepancy could be resolved if there was a process that would 
make ``low-mass galaxies relatively more vulnerable to disruption''. 

Such processes are included in modern models under the blanket heading 
of `feedback'. Feedback processes arguably have the largest impact on 
the form of the theoretical predictions for galaxy properties, whilst at 
the same time being amongst the most difficult and controversial phenomena 
to model; a detailed treatment should include a multiphase interstellar 
medium, with hot, cold and possibly warm gas components, tracking 
collisions between cold clouds and their evaporation by supernova heating 
(McKee \& Ostriker 1977; Efstathiou 2000; Monaco 2004). 
Broadly speaking, two forms of feedback are considered in galaxy formation 
models: in the first, cold gas is heated and removed from a galactic disk 
and in the second, the rate at which gas cools from the hot halo is 
suppressed. Both modes of feedback diminish the reservoir of cold gas 
available to be turned into stars (see Fig~\ref{fig:stars}).

The most common form of feedback used in hierarchical models is the 
ejection of cold gas from a galactic disk by a supernova driven wind 
(e.g. Larson 1974; Dekel \& Silk 1986). The reheated cold gas  
could be blown out to the hot gas halo, from which it may subsequently 
recool (sometimes called ``retention'' feedback), or it may even be 
ejected from the halo altogether (which is naturally enough called 
``ejection feedback''), and left unable to cool until it is incorporated 
into a more massive halo at a later stage in the merger hierarchy. 
The distinction between the ``ejection'' and ``retention'' modes of feedback 
can have a significant impact on the form of the galaxy luminosity function 
(e.g. Kauffmann et~al. 1999; Somerville \& Primack 1999; de Lucia, Kauffmann 
\& White 2004). 
There is now convincing observational evidence for the existence of 
supernova driven winds in dwarf galaxies (Martin 1997, 1998, 1999; 
Ott, Walter \& Brinks 2005). 
Other forms of feedback act to modify the rate at 
which gas cools, either by altering the density profile or entropy of 
the hot gas halo (following the injection of energy into the hot gas 
halo) or by reducing the fraction of baryons that fall into dark matter haloes 
and changing the cooling rate (i.e. photo-ionization suppression 
of cooling in low mass haloes) or by stifling the cooling flow by 
injecting energy  (Wu, Fabian \& Nulsen 2000; Bower et~al. 2001; 
Benson et~al. 2003; Granato et~al. 2004; Croton et~al. 2006; 
Bower et~al. 2006). 

Initially, as remarked upon above, the motivation for invoking 
feedback was to reduce the efficiency of star formation in low mass 
haloes, in order to flatten the slope of the faint end of the predicted 
galaxy luminosity function, thus bringing it in line with the extant 
observations (Cole 1991; White \& Frenk 1991). Cole et~al. (1994) 
appealed to a feedback model in which the rate of ejection of reheated gas 
was a very strong function of the circular velocity of the host 
dark matter halo ($\propto v^{5.5}_{\rm c}$) in order to bring their model 
predictions close to the flat faint end of the luminosity function 
estimated by Loveday et~al. (1992). The tension between the faint end 
of the observed galaxy luminosity function and the predictions of 
hierarchical models has been greatly reduced with the advent of much larger 
and deeper redshift surveys, which allow a more robust estimate of the faint 
end slope. Surveys such as the 2dFGRS and SDSS allow 
the measurement of the galaxy luminosity function with, for the most part, 
random errors (arising from the volume covered and the number of 
galaxies) that are smaller than systematic effects, such as the choice of 
band-shifting and evolutionary corrections (Norberg et~al. 2002b; 
Blanton et~al. 2003). 
Currently, the semi-analytic models do a good job of reproducing the 
faint end of the optical luminosity function as determined by recent 
measurements, with much more modest amount of supernova feedback 
(with a velocity dependence $\propto v^{2}_{\rm c}$). This is due in 
part to the shift in the favoured cold dark matter from a universe with 
the critical density in matter to a low density universe (Heyl et~al. 1995; 
Somerville \& Primack 1999). 
The inclusion of a photo-ionising background, along with an unexceptional 
amount of supernova energy injection into the ISM results in the model 
predictions matching the observations (e.g. Benson et~al. 2002b, 2003; 
Croton et~al. 2006). There is still some debate 
over the agreement between models and observations in the K-band.  
The Cole et~al. (2001) estimate of the near-IR luminosity function 
yields a relatively flat faint-end slope (see also Kochanek et~al. 2001). 
These estimates are derived from relatively shallow 2MASS photometry 
(Jarrett et~al. 2000). Huang et~al. (2003) find a significantly steeper 
faint-end slope, using deeper photometry, but over a much smaller solid 
angle than was considered in the estimates made from the 2MASS catalogue. 

In recent years, the focus has shifted to reproducing the break at the 
bright end of 
the luminosity function. The overproduction of bright galaxies is a problem 
that has dogged hierarchical galaxy formation models for more than a decade. 
We saw in Fig.~\ref{fig:coolrate} that the rate at which gas can cool 
peaks at the present day in haloes expected to host Milky Way like galaxies, 
and drops with increasing halo mass.
However, by itself, the corresponding increase in the cooling time of the 
hot gas is not sufficient to account for the observed sharpness of the 
break in the luminosity function. Various fixes have been 
proposed to this problem. Kauffmann et~al. (1993) suppressed the number 
of bright galaxies by simply turning off star formation by hand in the 
cooling flows present in high circular velocity haloes (see also 
Kang et~al. 2005). 
Cole et~al. (2000) proposed that the hot gas follows a different 
density profile 
from that exhibited by the dark matter. In particular, they proposed that 
the hot gas has a constant density core. Therefore, the gas in the central 
regions of the halo has a lower density than it would have had if it tracked 
the dark matter. A lower gas density means a longer cooling time. 
Furthermore, in the Cole et~al. model, the radius of the constant density 
core grows with time, as low entropy gas cools from the central regions. 
This model produced a better match to the observed abundance of bright 
galaxies because the cooling radius for the dark matter haloes which  
host such galaxies is typically within the constant density core, 
thereby resulting in the suppression of the rate at which gas can cool. 

Cole et~al. (2000) were also helped by using a baryon density parameter which, 
by comparison with the constraints available today, would be considered 
too low. The current best fitting value for the cosmological density in 
baryons ($\Omega_{b} \approx 0.04 $) is twice the value adopted 
by Cole et~al. in their fiducial model
(e.g. Sanchez et~al. 2006). Increasing the baryon density from the value 
used by Cole et~al. exacerbates the problem of matching the bright end 
of the luminosity function. 

An additional feedback mechanism is required that is effective in more 
massive haloes, acting either to prevent gas from cooling in the first 
place or to eject cold gas before it forms stars. Benson et~al. (2003) 
carried out a systematic study of the impact of various feedback 
mechanisms on the form of the predictions for the galaxy luminosity 
function. The standard supernova driven winds, whilst helping to reduce 
the number of faint galaxies to match observations, were found to 
have little effect at the bright end. The strength of this mode of 
feedback cannot be increased with impunity; excessive amounts of 
supernova feedback result in galactic disks bigger than are observed 
(Cole et~al. 2000; de Jong \& Lacey 2000) and introduce curvature into 
the predicted Tully-Fisher relation for spirals (Cole et~al. 1994; 
Somerville \& Primack 1999).   
Moreover, such a change would also tend to weaken the break in the predicted 
luminosity function rather than enhance it; stronger supernova feedback 
would wipe out galaxies around $L_{*}$ and instead  cause more gas to cool 
in more massive haloes (unless the heated gas is expelled altogether 
and is not allowed to recool), thus boosting the luminosity of the galaxies in 
these haloes. The resulting luminosity function would be closer to a  
power-law rather than a Schechter function form\footnote{The Schechter function is three parameter function which gives a resonable fit to the observed 
galaxy luminosity function. The parameters are the normalisation, 
$\phi_{\star}$, the slope of the power law $\alpha$ and the characteristic 
luminosity at which the function changes from a power law to an exponential, 
$L_{*}$ (Schechter 1976). The abundance of faint galaxies is described by 
a power law, whereas the number of galaxies brighter than $L_{*}$ drops exponentially with increasing luminosity.}.

Benson et~al. considered two more promising mechanisms: thermal conduction 
in the hot halo and superwinds (the ``ejection'' mode of supernova feedback). 
The thermal conduction model was discussed in Section~\ref{sec:cooling}; 
this model results in a break in the luminosity 
function but requires an unphysically high conductivity in the halo gas. 
The superwind model is motivated by the observations of outflows of 
gas from large, star forming galaxies at high and low redshift (Pettini et~al. 
2001, 2002; Dawson et~al. 2002; Adelberger et~al. 2003; Shapley et~al. 2003; 
Martin 2005; Wilman et~al. 2005b). The analysis of the profile of spectral lines in the 
spectra of Lyman-break galaxies supports mass ejection rates comparable to 
the star formation rate 
in the galaxy, with the material moving at speeds on the order of several 
hundreds of kilometres per second. In the Benson et~al. scheme, the superwind 
is most effective at removing cold gas from intermediate mass haloes. This is 
the case because the galaxies in such haloes have appreciable star formation rates, 
but do not have the high escape velocities of more massive haloes. The ejected 
gas can be recaptured at a later stage in the merger hierarchy; the circular 
velocity threshold for a halo to be able to entrain the gas removed by 
superwinds from its progenitors is treated as a parameter of the model.
It turns out that, in order to produce a good match to the exponential 
break in the luminosity function, the superwind is required to be 
extremely efficient (perhaps implausibly so) at coupling the energy 
released by supernovae into driving cold gas from the disk. Benson et~al. 
remarked that such  a superwind may be feasible if it is driven by the 
energy released by the accretion of material onto a black hole at the 
centre of the galaxy. 

Building upon previous work which examined the impact of AGN on aspects 
of galaxy formation (e.g. Granato et~al. 2004; Monaco \& Fontanot 2005; 
Cattaneo et~al. 2005; Di Matteo et~al. 2005), Croton et~al. (2006) and 
Bower et~al. (2006) have implemented simple AGN feedback schemes 
into the Munich and Durham semi-analytical galaxy formation models 
respectively. The first enhancement required to the standard galaxy 
formation model is to track the formation and evolution of black holes. 
This is done using the model introduced by Kauffmann \& Haehnelt (2000). 
Motivated by the observed correlation between black hole mass and bulge mass 
(Magorrian et~al. 1998; Ferrarese \& Merritt 2000; Gebhardt et~al. 2000), 
Kauffmann \& Haehnelt tied the formation of black holes to the same 
process, galaxy mergers, which is responsible for building spheroids 
and bulges. 

It is instructive to briefly compare and contrast the implementations 
of ``AGN feedback'' in the models of Croton et~al. (2006) and 
Bower et~al. (2006). Croton et~al. refer to the build up of black hole 
mass due to mergers of existing black holes and the accretion of cold gas 
during starbursts as the ``quasar mode''. They also consider a 
new mode, the ``radio mode'', during which the black hole accretes 
gas directly from the hot halo. In their model, the quasar mode is the more 
important channel for building up the mass of the black hole.  
However, the accretion of mass in the radio mode releases energy into 
the hot halo and is therefore responsible for suppressing the cooling 
flow in more massive haloes. Croton et~al. introduce a parametric form 
for the radio mode suppression of cooling flows, which depends upon the 
virial temperature of the halo and the mass of the central black hole, 
and then present arguments to motivate this recipe. Bower et~al. use the 
model of black hole growth described by Malbon et~al (2006). Bower et~al. 
use the mass of the black hole to compute the Eddington limit, i.e. 
the maximum luminosity at which the black hole can radiate energy. 
They argue that if the rate at which energy is released by gas cooling 
is less than some fraction of the Eddington luminosity (where the fraction 
is a model parameter), then cooling is suppressed. In both cases, AGN act 
to suppress gas cooling only in those haloes in which a quasi-static hot halo 
has formed, i.e. where the cooling time of the gas exceeds 
the free-fall time. 

The new models reproduce the observed break in the present day luminosity and 
the bimodality of the colour distribution of local galaxies (as measured 
by Kauffmann et~al. 2003; Baldry et~al. 2004; Balogh et~al. 2004a). 
Croton et~al. 
also find that their model displays the luminosity dependent clustering 
seen in the 2dFGRS and SDSS (Norberg et~al. 2001, 2002a; Zehavi et~al. 2002). 
De Lucia et~al. (2006) find a clear increase in the luminosity weighted age 
of elliptical galaxies with velocity dispersion in the Croton et~al. model. 
Bower et~al. examined the predictions of their model with AGN feedback at 
high redshift. They find remarkably good agreement with observational 
estimates of the stellar mass function to $z\sim 5$ (Fontana et~al. 2004; 
Drory et~al. 2005). There is also evidence for the ``downsizing'' of 
star formation, insofar as galaxies with high stellar masses were 
forming stars more vigorously in the past than they are today; we shall return 
to this point in the last section.  Other groups have also developed 
semi-analytical models in which AGN act to suppress cooling 
(Cattaneo et~al. 2006; Fontanot et~al. 2006; Kang et~al. 2006; 
Menci et~al. 2006; Monaco et~al. 2006a; Monaco et~al. 2006b).

\subsection{Chemical evolution}

The formation of stars changes the metal content of the interstellar 
medium (ISM) of a galaxy. The act of forming stars removes cold gas 
and associated metals from the ISM. Also, as stars evolve, they return 
material to the ISM with an enhanced metallicity. The return mechanism 
can take the form of stellar winds or supernova explosions. Mass loss 
through stellar winds becomes more important as the star evolves away 
from the main sequence. The amount of gas returned to the ISM per 
unit mass of stars formed, called the recycled fraction, is therefore 
dependent on the form of the IMF. The chemical evolution of the gas and 
stars in a galaxy is important for a number of reasons: (i) The rate at 
which gas cools from the hot halo depends upon the metallicity of the 
gas; a higher metallicity results in a shorter cooling time 
(Fig~\ref{fig:coolrate}). 
(ii) The metallicity with which stars are born has an impact on the 
luminosity and colour of the stellar population. 
(iii) The optical depth of a galaxy, which determines the extinction 
of starlight due to dust,  scales linearly with the metallicity of its 
cold gas. 

There are two broad categories of supernova explosions, type I and type II, 
characterized by their spectra (see Binney \& Merrifield 1998). The spectra 
of type I supernova do not contain any lines signifying the presence of 
hydrogen, whereas type II supernova spectra do display such lines.   
Further sub-division of the type I class is made according to the 
presence (type Ia) or absence (type Ib) of absorption lines in the spectra 
from silicon ions. The most pertinent difference between the various types 
of supernova from the semi-analytic modeller's viewpoint is the timescale 
on which the supernova explosion, and hence the metal enrichment, take 
place. Both type II and type Ib supernovae occur on a short timescale 
(on the order of 10 Myr) after an episode of star formation has taken place; 
type Ia supernovae happen on a much longer timescale (on the order of 1Gyr). 
Type II and type Ib supernovae occur when massive stars experience 
core-collapse; type Ib supernovae are believed to mark the end point of 
the evolution of more massive stars than those leading to type II supernovae, 
and so are less numerous than type IIs for most choices of IMF. Type Ia 
supernovae are thought to originate in the explosion of white dwarfs upon 
the accretion of material from a binary companion. The frequency of both 
types of supernovae depends upon the form of IMF adopted. 
Stars with masses in excess of $5$-$8M_{\odot}$ are thought to end 
in core-collapse, whereas type Ia are sensitive to the form of the IMF 
around one solar mass. 

Type 1a supernovae dominate the production of iron (Fe), whereas 
type II supernovae are primarily responsible for the production 
of nuclei formed by $\alpha$ particles (the $\alpha$ elements: 
O, Ne, Mg, Si, S, Ar, Ca, Ti) and also nitrogen and sodium. The relative 
proportions of these metal species, quantified by the $\alpha$/Fe abundance 
ratio, therefore tells us about the relative importance of type II and 
type Ia supernovae in a galaxy and the timescale over which star 
formation took place. Two examples of how abundance ratios can contain 
clues about galaxy formation are worthy of mention at this point. 
Elliptical galaxies display metal abundances which increase with galaxy 
luminosity and velocity disperion (Faber 1973; Bender et~al. 1993). 
Their super solar total metallicities and Mg/Fe ratios are described 
as $\alpha$-enhancement. The observed trend in Mg line strength and 
velocity dispersion can be reproduced with a standard IMF in a single 
burst model, if the timescale of the burst is an appropriate function 
of galaxy mass (Thomas, Greggio \& Bender 1998). In this simple picture, 
all of the stars in the elliptical form in a single burst at high redshift 
which is terminated by the ejection of gas in a wind (Larson 1974, 1975).
Also, the intracluster medium (ICM) shows $\alpha$-enhancement, with 
comparable amounts of $\alpha$ elements to those found in the solar 
neighbourhood, but with only 30\% of the iron content (Mushotzky et~al. 1996). 
The metal content of the ICM  depends on the star formation histories of 
the cluster galaxies and the ejection of metals from them into the 
cluster gas. The $\alpha$-enhancement of the ICM in hot X-ray emitting 
clusters could be explained if the IMF of star formation in the cluster 
galaxies is biased towards high mass stars (e.g. Renzini et~al. 1993).

The first attempts to follow the chemical evolution of galaxies in 
semi-analytical models considered type II supernovae, using the instantaneous 
recycling approximation introduced by Tinsley (1980). In this case, as 
stars are formed in a given timestep, a quantity of metals, determined by 
the yield for the chosen IMF, is generated instantaneously. 
The effective yield of metals depends on a number of factors and is 
different from that expected in a simple closed-box chemical evolution 
model (see Cole et~al. 2000 for a discussion): (i) Halo mergers mix hot 
gas reservoirs of differing metallicities. (ii) Gas cooling adds gas of 
one metallicity to a galactic disk which could have a different metallicity. 
(iii) Feedback processes can remove gas from galactic disks, depleting the 
metal content of the disk. 

The inclusion of type Ia supernovae is a much more recent development 
of semi-analytical models. Thomas (1999) was the first to consider the 
delayed enrichment due to type Ia supernovae in conjunction with 
semi-analytical models, using star formation histories extracted from 
the models of Kauffmann et~al. (1999), but neglecting any inflow or 
outflow of gas and metals (see also Thomas \& Kauffmann 1999). 
Nagashima \& Okamoto (2006) produced the first semi-analytical model 
which genuinely integrated the impact of type Ia. These authors used the 
simplification of assuming a fixed time delay for all type Ia explosions. 
Nagashima et~al. (2005a,b) carried out the first fully consistent calculation 
including both type II and type Ia supernovae in the semi-analytical model 
of Cole et~al. (2000). 

\subsection{Galaxy Mergers}
\label{sec:galmerge}

\begin{figure}
{\epsfxsize=14.truecm
\epsfbox[-200 85 557 740]{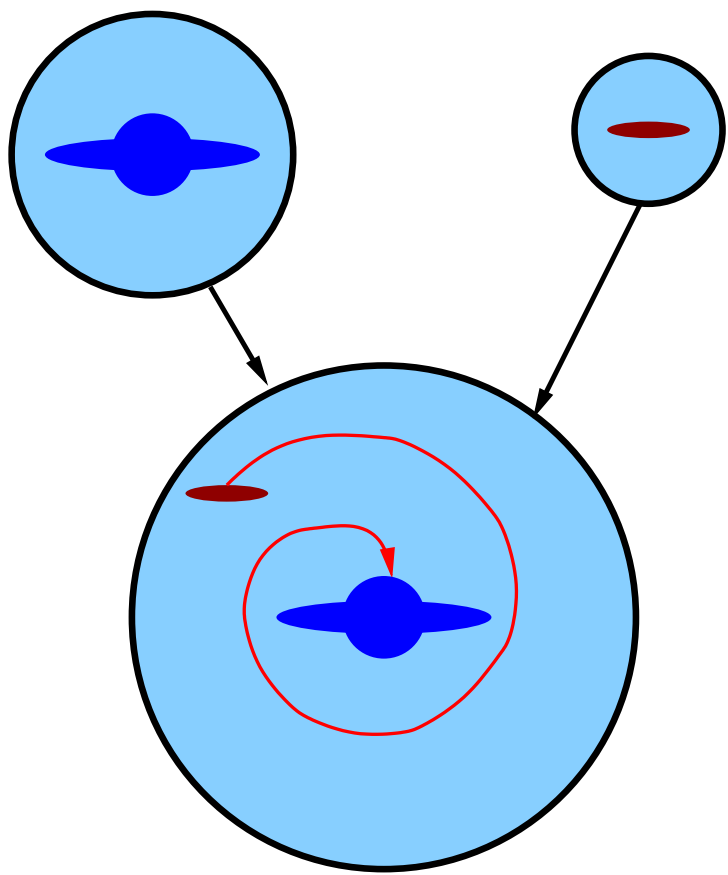}}
\caption{
A schematic of a merger between two dark matter haloes. The progenitors 
of the final halo each contain a galaxy. After the haloes merge, the more 
massive galaxy is placed at the centre of the newly formed halo. Any hot 
gas that cools would be directed onto the central galaxy (for simplicity,  
in this illustration, the haloes have exhausted their supply of hot gas). 
The smaller galaxy becomes a satellite of the central galaxy. The orbit 
of the satellite galaxy decays due to dynamical friction. The satellite 
may eventually merge with the central galaxy. 
}
\label{fig:merger}
\end{figure}

In the two-stage model of galaxy formation proposed by White \& Rees (1978), 
dark haloes are assumed to grow through mergers and accretion, with dynamical  
relaxation effects erasing any trace of the progenitor haloes at each 
stage of the merging hierarchy (see Press \& Schechter 1974). The halo 
resulting from a merger or accretion event is assumed to be smooth 
and devoid of any substructure. White \& Rees argued that galaxies 
survive the merger of their parent haloes as a result of them being 
more concentrated than the dark matter, due to the dissipative cooling 
of gas. 

The White \& Rees picture of galaxy formation leads naturally to a scenario  
in which a dark halo contains a massive central galaxy surrounded by smaller 
satellite galaxies. These satellites were formerly central galaxies in 
the progenitors of the current halo which were present in the 
earlier stages of the merger hierarchy. The satellite galaxies retain their 
identity after their parent halo merges with a more massive object due 
to their high concentration. However, as the satellites orbit the central 
galaxy in their common dark halo, they gradually lose energy through 
dynamical friction, an effect originally calculated for star clusters by 
Chandrasekhar (1943). The gravitational attraction exerted by the mass of 
the satellite galaxy on its surroundings draws the material in the halo 
towards it. This produces a wake of higher density material along the 
path of the satellite. The satellite therefore feels a stronger gravitational 
pull from the region of the halo that it has just passed through compared with 
the region is it about to travel through, which acts as a break on its 
motion. The orbital energy of the satellite decays as a result and it 
spirals in towards the central galaxy (Binney \& Tremaine 1987). A timescale can be 
computed for the dynamical friction process to remove the orbital energy 
of the satellite completely. If this timescale is shorter than the lifetime 
of the dark halo, then the satellite merges with the central galaxy 
(see Fig~\ref{fig:merger}).

In addition to providing an alternative mechanism to gas cooling for 
increasing the mass and luminosity of the central galaxy, the accretion 
of a satellite galaxy can have more dramatic consequences. 
The impact of a galaxy merger is usually quantified by the ratio of the mass 
of the accreted satellite galaxy to the mass of the central galaxy. 
Some numerical work exists in which satellites of different mass and 
gas content have been fired at central galaxies (e.g. Barnes \& Hernquist 
1991, 1992; Mihos \& Hernquist 1994, 1996; Walker, Mihos \& Hernquist 1996). However, the range of possible orbits 
and the parameter space of mass ratios and gas fractions in both the satellite 
and central galaxies is enormous; such calculations are computationally 
expensive and much numerical work remains to be done.
Typically semi-analytical modellers treat the mass ratio that sets the 
threshold for a merger to be termed a violent or major merger as a 
parameter in their models, using the extant numerical simulation results 
as a guide (Baugh et~al. 1996a,b; Kauffman 1996; Somerville, 
Faber \& Primack 2001). In a violent merger, the disk of the central 
galaxy is assumed to be destroyed and all of the stars involved 
in the merger event form a spheroidal remnant. 
In some models, a major merger can also trigger a burst of 
star formation, in addition to changing the morphology of the stars 
(Baugh et~al. 1996b; Somerville, Faber \& Primack 2001; 
Baugh et~al. 2005). More sophisticated analytic schemes have been devised 
that can track the loss of mass from the satellites as they move through the 
dark halo (Taylor \& Babul 2001; Benson et~al. 2002a; Zentner \& Bullock 2003).
Benson et~al. (2002a) implemented a similar scheme to that developed 
by Taylor \& Babul in the Cole et~al. (2000) model and used this to 
study the impact of satellites 
on the disk of Milky Way like galaxies (Benson et~al. 2004). 
In variations on the simple galaxy merging scheme set out above, 
some models also consider collisions between 
satellite galaxies, in addition to mergers of satellites onto the 
central galaxy due to dynamical friction (Somerville \& Primack 1999; 
Van Kampen, Jimenez \& Peacock 1999; 
Somerville, Primack \& Faber 2001; Menci et~al. 2002; Enoki et~al. 
2003; Menci et~al. 2004). 

\subsection{Galaxy sizes} 
\label{sec:size} 
An estimate of the size of the galactic disk or bulge is an important input 
into the model for star formation if the prescription used depends 
upon the dynamical time of the disk 
($t_{\rm dyn}= r_{\rm disk}/v_{\rm disk}$). The disk scale length is 
also required to compute its optical depth in order to calculate the 
extinction of starlight due to dust. Comparison of the model predictions 
for the sizes of the disks and bulges of galaxies with 
observations can constrain the models of gas cooling and merging that 
build up these components. 

The starting point for a calculation of the radius of a galaxy is the 
angular momentum of the host dark matter halo. If the halo is asymmetric and 
surrounded by a lumpy distribution of matter, then it can acquire an 
angular momentum or spin through a net tidal torque which acts to spin up the 
halo as it forms (Hoyle 1949). The spin of the halo can be quantified in 
a dimensionless number $\lambda$: 
\begin{equation}
\lambda = \frac{J |E|^{1/2}}{G M^{5/2}},
\end{equation}
where $J$, $E$ and $M$ are the total angular momentum, energy and mass 
of the dark matter halo. 
The halo gas is assumed to have the same specific angular momentum as the 
dark mater. If the further assumption is made that this angular momentum 
is conserved in the dissipative collapse of the gas (this assumption will 
be discussed further in Section~\ref{sec:disk}), then the factor by which the 
radius of the gas is reduced is $\approx 1 / 2 \lambda_{\rm H}$ 
(Fall \& Efstathiou 1980). N-body simulations indicate $\lambda \sim 0.04$ 
(Efstathiou \& Jones 1979; Barnes \& Efstathiou 1987; Frenk et~al. 1988; 
Warren et~al. 1992; Cole \& Lacey 1996; Bullock et~al. 2001a), which means 
that the gas collapses to a rotationally supported disk with a radius 
over an order of 
magnitude smaller than the virial radius of the dark halo. This simple 
argument has been used to estimate the sizes of galaxies in many semi-analytic 
models (Lacey et~al. 1993; Kauffman \& Charlot 1994; Poli et~al. 1999; Somerville \& Primack 1999; Firmani \& Avila-Reese 2000; Hatton et~al. 2003). 
Mo, Mao \& White (1998) presented a more physical calculation in with the 
following improvements: (i) realistic density profiles for the dark 
matter and the gas (ii) a distribution of spin-parameter values ($\lambda$), 
motivated by the results of N-body simulations (iii) the gravity of the 
disk and bulge (iv) the reaction of the halo to the gravity of the disk and 
bulge, which causes a contraction of the dark matter (see also 
Dalcanton, Spergel \& Summers 1997). Mo, Mao \& White (1998) 
treated the mass and angular momentum of the baryonic components and the 
mass-to-light ratio of the disk as input parameters to their model. 
Cole et~al. (2000) used a similar model to Mo, Mao \& White to compute 
galaxy sizes, but with the replacement of the parameters in the model of 
Mo et~al. with the predictions of the semi-analytical model of galaxy formation. 
Cole et~al. also gave a model for the size of the bulge component, considering 
the conservation of energy and the virial theorem to compute the size of the 
merger remnant. 

\subsection{
The generation of a spectral energy distribution for model galaxies
}

\begin{figure}
{\epsfxsize=17.truecm
\epsfbox[0 50 590 720]{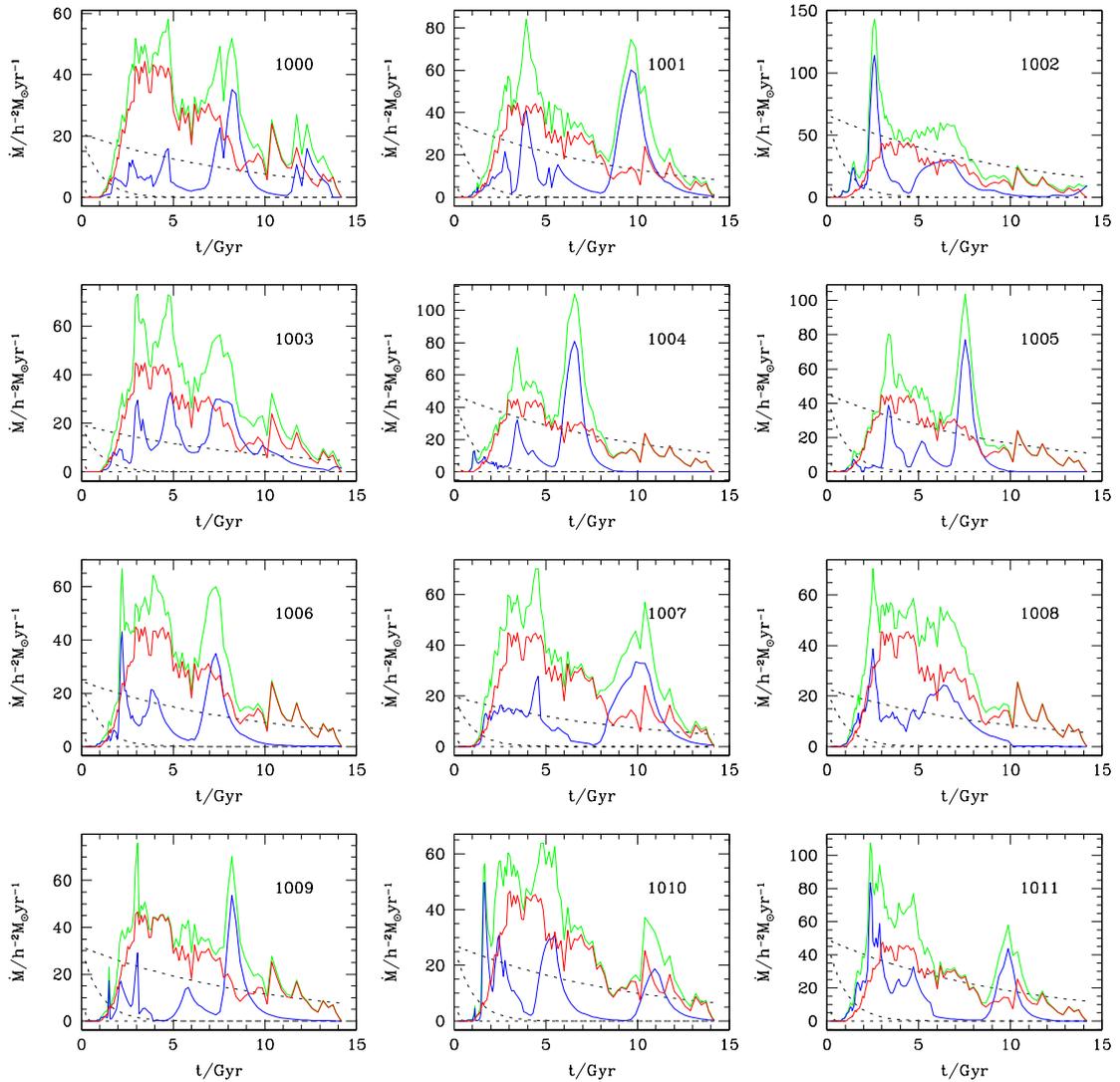}}
\caption{
The star formation history of a selection of massive galaxies, as 
predicted by the model of Baugh et~al. (2005). The horizontal axis 
gives the age of the universe. The star formation formation 
history is divided into quiescent star formation in galactic disks (blue) 
and starbursts triggered by galaxy mergers (red). The total star formation 
rate is show by the green curve. These curves are calculated by summing over 
all of the progenitors of the present day galaxy at each redshift. The black 
dashed lines show exponential curves, a typical {\it assumption} for 
the star formation history of galaxies used in other approaches, for reference.
}
\label{fig:sfr}
\end{figure}

The final step required to connect the theoretical predictions of a model 
of galaxy formation to observations is the production of a synthetic spectral 
energy distribution for each model galaxy, i.e. the amount of energy emitted 
by the galaxy as a function of wavelength or frequency. Semi-analytical models 
predict the complete star formation history of a galaxy, taking into 
account all mergers between the progenitors of the galaxy, 
star formation in bursts triggered by mergers and quiescent star formation 
in galactic disks (some examples of the model predictions are given in 
Fig.~\ref{fig:sfr}). The star formation history of a 
single galaxy is stored in a table that records how many stars of a 
given metallicity were formed in each timestep of the calculation, taking 
into account all the branches of the galaxy merger tree. This information 
is then combined with a stellar population synthesis model to compute a 
composite stellar population for the whole galaxy 
(Bruzual \& Charlot 1993; Worthey 1994; Devriendt, Guiderdoni \& Sadat 1999; 
Fioc \& Rocca-Volmerange 1999; Leitherer et~al. 1999; 
Girardi et~al. 2000; Bruzual \& Charlot 2003). 
The stellar population 
model provides a look-up table of the spectral energy distribution of a 
single-age population of stars as a function of the time elapsed since the 
stars were made; the stars are born with a distribution of masses set 
by an assumed initial mass function (IMF) and have a given metallicity. 
As a simple stellar population ages, hot, massive stars evolve off the 
main sequence most rapidly, with the result that the flux of ultra-violet 
photons declines with increasing age.  

Stellar population synthesis models are traditionally treated  
as trusted black boxes by semi-analytical modellers. If a particular 
prediction does not match an observation, the suspicion generally 
falls on the details of the galaxy formation model rather than on the 
accuracy of the population synthesis model. Charlot, Worthey \& Bressan 
(1996) carried out a comparison of their respective stellar population 
synthesis models and reached the following cautionary conclusion, 
quoting directly from the abstract of their paper: ``There appear to 
be persistent problems in virtually every ingredient of population 
synthesis models''.  
Whilst the accuracy of population synthesis models 
has improved since the mid-1990s, there are still areas 
where there are discrepancies, particularly in the near-infrared; 
this can be traced to the way in which the models attempt to 
follow the thermal pulsations of stars which have left the main sequence 
for the asymptotic giant branch -- AGB stars, (Maraston 1998; 2005; 
Bruzual \& Charlot 2003). Bruzual \& Charlot (2003) produced an 
update of their earlier models with higher spectral resolution ($3\AA$), 
the ability to incorporate some absorption lines and a treatment of 
AGB stars. Vazdekis (1999) 
produced models with even higher resolution over a limited 
range of wavelengths (see also Le Borgne et~al. 2004). 

The majority of the early semi-analytical models did not explicitly take 
into account the impact of dust on the spectral energy distribution of 
stars. By allowing themselves the freedom to rescale the luminosity of 
the model galaxy by a fixed factor in order to match the observed 
luminosity function at $L_*$, some of the resulting reduction in 
luminosity could be blamed on dust extinction, although the attenuation 
factor would necessarily be the same at all wavelengths (e.g. 
Cole et~al. 1994; Baugh et~al. 1998). Lacey et~al. (1993) were the first 
to incorporate extinction, invoking a screen of dust sitting between the 
galaxy and the observer, with the optical depth of the slab scaling with the 
metallicity of the cold gas, as computed self-consistently from a 
chemical evolution model (see also Guiderdoni et~al. 1998). 
In other models, the optical depth of the slab is modelled empirically 
(e.g. Kauffmann et~al. 1999; Somerville \& Primack 1999).
In reality, the stars and dust are mixed together. The propagation of 
starlight through the interstellar medium of a galaxy therefore requires 
a radiative transfer calculation, which takes into account the geometry 
and size of the galaxy (Silva et~al. 1998; Ferrara et~al. 1999).  
The Cole et~al. (2000) model uses the results of the radiative transfer 
calculations carried out by Ferrara et~al (1999) to compute a 
self-consistent optical depth for each galaxy, based on the model 
predictions for the metallicity of the cold gas and the size of the disk 
and bulge components. 

The energy absorbed by the dust heats the dust grains, resulting in 
emission at longer wavelengths, in the far-infra red and sub-millimetre 
ranges of the electromagnetic spectrum. 
A simple estimate of the luminosity at these wavelengths can be made 
by assuming a dust mass and temperature for each galaxy. 
Kaviani, Haehnelt \& Kauffman (2003) took this approach to generate 
number counts of sub-millimetre selected galaxies using the star 
formation histories predicted by the semi-analytic galaxy formation 
model of Kauffmann et~al. (1999). However, the resulting luminosity 
is extremely sensitive to the assumed dust temperature, scaling as 
the sixth power of the dust temperature for a standard choice of dust 
emissivity. 
Guiderdoni et~al. (1998) were the first to compute the dust emission in 
a semi-analytical model, though their model did not follow galaxy 
mergers explicitly.  
Devriendt \& Guiderdoni (2000) combined their semi-analytic 
model with empirical templates for the spectral energy distribution in 
the far-infrared. 
Granato et~al. (2000) combined the semi-analytic model of Cole et~al. (2000) 
with the spectro-photometric model developed by Silva et~al. (1998).  
Baugh et~al. (2005) applied the machinery developed by Granato et~al. to 
devise the only model at the time of writing which has been able to 
provide a reasonable match to the 
present day galaxy luminosity function in the optical and near 
and far-infrared at the same time as matching the observed number 
of sub-millimetre sources at high redshift, along with the luminosity 
function of Lyman-break galaxies.

\section{Semi-analytical modelling or direct simulation?} 

We now compare the ways in which semi-analytical models and direct numerical 
simulations of gas and dark matter treat the key ingredients of galaxy 
formation as set out in Section \ref{sec:ingred}, highlighting both 
the common ground and the differences between the two approaches.
For completeness, we first give a brief overview of the techniques 
used in gas dynamics simulations and explain how they are applied to 
address various problems (section~\ref{sec:gassim}). 
We then discuss two aspects of galaxy formation modelling that are carried 
out in different ways if a semi-analytical model is incorporated into a 
high resolution N-body simulation: the construction of dark matter halo 
merger trees (section~\ref{sec:mergertree2}) and the merging of 
galaxies (section~\ref{sec:simmerge}). After this, we turn our 
attention to the physics of the baryonic component and assess how 
the prescriptions 
used in semi-analytic models compare with the results of direct simulations, 
considering gas cooling (section~\ref{sec:simcool}), star formation 
(section~\ref{sec:stars}), feedback processes (section~\ref{sec:simfeedback}) 
and the angular momentum of galactic disks (section~\ref{sec:disk}).

\subsection{Gas dynamics techniques}
\label{sec:gassim}

There are two principle algorithms in common use to follow the 
hydrodynamics of gas in an expanding universe: particle based, Lagrangian 
schemes, which employ a technique called smoothed particle hydrodynamics 
(SPH, Monaghan 1992; Couchman, Thomas \& Pearce 1995; 
Gnedin 1995; Springel \& Hernquist 2003; 
Wadsley, Stadel \& Quinn 2004) and grid based, 
Eulerian schemes (e.g. Ryu et~al. 1993; Cen \& Ostriker 1999). 

In an SPH simulation, two sets of particles are used, one to trace the 
dark matter and one to represent the baryonic component of the universe.  
The dark matter particles are collisionless responding only to the 
gravitational force exerted by the other particles, whereas the baryonic 
particles can also feel pressure and dissipate energy through cooling. 
The local thermodynamic properties of the gas particles are computed by 
averaging over a number of neighbours, typically around 32 particles 
(see Springel \& Hernquist 2003 for a recent discussion of an SPH scheme). 

The SPH technique has traditionally achieved superior resolution compared 
with fixed grid schemes, due to its Lagrangian nature. 
The SPH particles move to the regions of interest, giving improved 
spatial resolution in regions where it is required, and, consequently, 
poorer resolution in voids. 
The spatial resolution attained using fixed-grid schemes typically lags 
more than ten years behind the standards of SPH simulations 
(Pearce et~al. 1999). 
The SPH method does have a number of drawbacks, however. In addition to 
the limited description of low density regions, SPH algorithms require 
numerical aids such as artificial viscosity to improve the handling 
of shock waves and strong density gradients, because the smoothing inherent in 
the approach smears out these features; the scheme may also violate 
conservation of various quantities (Okamoto et~al. 2003). 
Grid based codes deal much better with shocks and discontinuities 
(e.g. Ryu et~al. 1993; Quilis, Ibanez \& Saez 1994).  
Springel \& Hernquist (2002) introduced a new formulation of SPH in which 
they integrated the entropy as a function of time rather than the thermal 
energy. This new implementation enjoys improved conservation of energy 
and entropy compared with the traditional SPH scheme and also deals better 
with cooling flows in less well resolved haloes. 

To circumvent the spatial resolution problem faced by fixed-grid codes, 
much work has been done to develop adaptive-mesh refinement (AMR) 
codes (Bryan \& Norman 1997; Teyssier 2002; Kravtsov 2003; Quilis 2004; 
Nagai \& Kravtsov 2005). The mesh used to 
solve the hydrodynamic equations is refined in regions where better 
resolution is desirable, e.g. within dark matter haloes. Several levels 
of refinement can be used as required. 

Gas dynamics simulations are run in two regimes. In the first, a large 
representative volume of the universe is simulated, with the goal of 
following the properties of a population of galaxies 
(e.g. Blanton et~al. 1999; Pearce et~al. 1999). In the second, a single 
halo is extracted from a large volume, dark matter only simulation and 
resimulated at much higher resolution with gas (e.g. Frenk et~al. 1996). 
The region of interest is simulated with much 
lower mass particles than used in the original simulation. In the 
resimulation, the region surrounding the high resolution volume is  
represented using higher mass particles, so that the tidal forces exerted 
on the high resolution structure are properly included. Early work on the 
formation of single galaxies used static haloes without cosmological initial 
conditions. 

Various tests have been conducted of the gas dynamic codes on the 
market. Frenk et~al. (1999) supplied a standard set of initial conditions, 
the so-called ``Santa-Barbara'' cluster, to the writers of a wide 
range of SPH and grid-based codes in order to compare their performance 
in modelling the adiabatic evolution of the cluster mass and gas. 
The best agreement was found between the predictions for the dark 
matter in the cluster and the worst agreement for its X-ray luminosity. 
Quantities such as the gas temperature and the mass fraction of gas 
within the virial radius were found to agree to within $10\%$. 
Thacker et~al. (2000) applied standard gas simulation tests (e.g. the 
handling of shocks) to twelve different implementations of the SPH 
technique and concluded that the implementation of artificial viscosity 
was the main factor responsible for producing different results. 
O'Shea et~al. (2005) compared two specific SPH and AMR codes; the 
entropy-conserving formulation of SPH in the GADGET code (the latest release, 
GADGET II is described in Springel 2005) 
and the adaptive mesh code ENZO (O'Shea et al. 2004). 
As in the case of the Santa Barbara project, good agreement was found for 
the dark matter, provided that a fine resolution grid was used in ENZO. 
The gas was allowed to evolve adiabatically. A good match was obtained 
between the predictions of the two approaches for the temperature, entropy and 
pressure of the gas in regions of high density; significant differences were 
reported, however, in low density regions. 
Kay et~al. (2002) compared different implementations of recipes for star 
formation and feedback using the same N-body/SPH code. In the absence of 
any feedback, they found that the various prescriptions for star formation 
produced galaxies with similar stellar masses. However, as expected, these 
stellar masses were too high. This problem was diminished by invoking a 
kinematic form of supernova feedback, or by using a thermal feedback in 
which the reheated gas is not allowed to recool immediately.

\subsection{Dark halo merger trees} 
\label{sec:mergertree2}

As we discussed in Section~\ref{sec:mergertree1}, merger trees 
describing the assembly of dark matter haloes can either be 
extracted from an N-body simulation or grown using a theory for 
the distribution of progenitor masses and a Monte Carlo algorithm. 
N-body merger trees are generally regarded as the 
benchmark. As we remarked in Section~\ref{sec:mergertree1}, 
Monte Carlo trees based on extended Press-Schechter theory tend 
to become progressively more inaccurate as they are followed over a 
longer interval in time. However, it turns out that the construction 
of a merger tree from the outputs of an N-body simulation is not trivial. 
The mass of a halo can decrease with time, as haloes that overlap spatially 
at one output time (and which could therefore be identified as a single 
object by a halo finding algorithm) are not necessarily gravitational bound, 
and so can move apart again by a subsequent output. N-body merger trees, 
due to disk-space limitations, typically have poorer time resolution 
than a tree grown with a Monte-Carlo scheme. This means that additional 
care is needed when applying recipes for gas cooling and star formation, 
to ensure that the model predictions are insensitive to the number of 
timesteps. (Typically, N-body merger trees have around 50 outputs, 
whereas semi-analytic calculations typically use 150-300 timesteps.)   
The main deficiency of N-body merger trees is, however, limited mass 
resolution. The Millennium Simulation, despite being by far 
the best available in terms of providing high resolution merger trees 
over a wide range of masses within a single computational box, is still 
only able to yield trees whose mass resolution is a factor of three 
poorer than the standard Monte Carlo trees used, for example, 
in the Cole et~al. model. 

Several groups have used merger trees drawn from N-body 
simulations in their semi-analytical models, primarily to add 
information about the spatial distribution of galaxies to 
the predictions made by the galaxy formation model 
(Roukema et~al. 1997; Kauffmann et~al. 1999; 
Okamoto \& Nagashima 2001; 
Hatton et~al. 2003; Helly et~al. 2003a;  
Croton et~al. 2006; de Lucia et~al. 2006; Kang et~al. 2005; 
Nagashima et~al. 2005c; Bower et~al. 2006; Lanzoni et~al. 2005). 
Helly et~al. (2003a) carried out a systematic study of the impact of 
using N-body merger trees on the predictions of the semi-analytic 
model. The calculation using N-body merger trees does not resolve 
galaxies down to the same luminosity as in the case with 
the Monte-Carlo trees. 
Helly et~al. found that the two approaches produced very similar 
predictions for the luminosity function for objects brighter than 
around one tenth of $L_{*}$; the mass resolution of the Millennium 
Simulation is around an order of magnitude better than that of the 
simulation used in the study of Helly et~al., so the discrepancy 
in the predictions would only become apparent at even fainter luminosities. 
The predicted star formation rates per unit volume are similar 
over the bulk of the history of the universe, diverging only at 
$z > 1 $ (the precise redshift depends upon the resolution of 
the simulation), beyond which the N-body trees underestimate the amount 
of star formation. 
This comparison can also be used to establish the accuracy 
of the Monte Carlo trees. Helly et~al. found that they could reproduce 
the results obtained with N-body trees if two adjustments were made 
to the Monte-Carlo trees. Firstly, they set the resolution of the Monte-Carlo 
trees to match the minimum halo mass available in the N-body trees.
Secondly, they applied an empirical correction to the distribution of 
progenitor masses expected in extended Press-Schechter model, in order 
to boost the number of massive halo progenitors (see Tormen 1998; 
Benson et~al. 2001). 

The advent of large volume, high resolution simulations such as 
the Millennium, with around twenty million dark haloes at the present 
day, means that careful studies can be carried out to reveal any 
environmental influences on the growth of dark matter haloes. 
The analysis of the correlation function of haloes in the Millennium  
has revealed a dependence of the clustering strength on the formation time 
of the halo (Gao et~al. 2005b; Harker et~al. 2006; see also 
Wechsler et~al. 2005; Reed et~al. 2006b; Zhu et~al. 2006). 

\subsection{Galaxy mergers} 
\label{sec:simmerge}

Semi-analytical models use dynamical friction arguments to compute the 
time needed for the orbit of a satellite galaxy to decay, causing it 
to merge with the central galaxy within a dark matter halo, as 
discussed in Section~\ref{sec:galmerge} 
(e.g. Kauffmann et~al. 1993; Cole et~al. 1994). 
More detailed analytical techniques have also been incorporated into the 
models, which include the stripping of mass from the satellites and other 
effects which promote the loss of orbital energy (Benson et~al. 2002b). 

Modern high resolution N-body simulations are able to resolve substructure 
within dark matter haloes. These substructures are the high density cores 
of the progenitors of the halo, which retain their identity after the more 
diffuse outer parts of their haloes have been tidally stripped. The 
resolution of substructure therefore offers an alternative to the 
analytic schemes to track galaxy mergers. A galaxy is assigned to the 
most bound particle in the halo in which it first forms. When this host 
halo merges with a more massive halo, the galaxy is assumed to track 
the most bound particle in the progenitor halo, which itself is part of 
the substructure that persists after the halo merger. As the substructure 
orbits within the more massive halo its mass is stripped until the point 
is reached where it can no longer be identified as a substructure, which 
occurs when the number of particles associated with the structure falls 
below the resolution limit of the group finder. At this point, a dynamical 
friction clock is started to monitor the final stages of the galaxy merger. 
Several such ``hybrid'' merger schemes have been implemented, either in 
high resolution simulations of individual dark haloes, or more 
recently, within cosmological volumes (Springel et~al. 2001, 
Kang et~al. 2005; Croton et~al. 2006; de Lucia et~al. 2006;  
Bower et~al. 2006).

\subsection{Gas cooling} 
\label{sec:simcool}

\begin{figure}
{\epsfxsize=16.truecm
\epsfbox[-90 500 562 800]{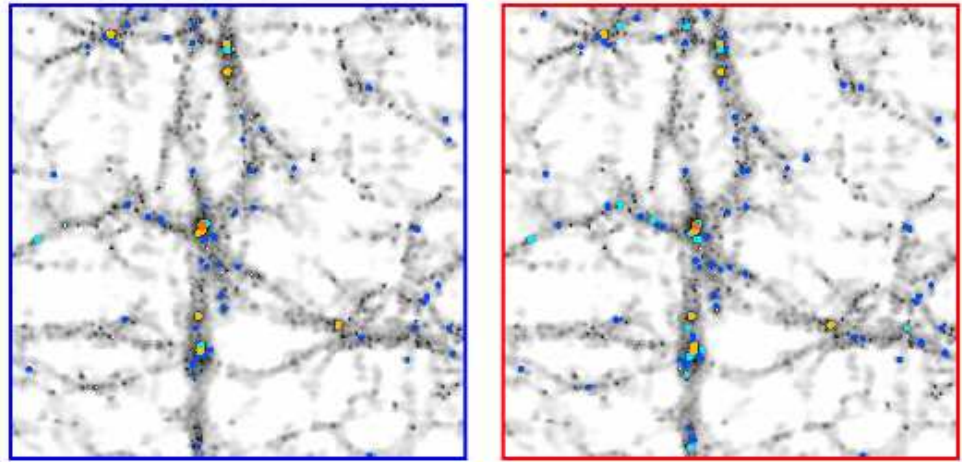}}
\caption{The distribution of cold gas and dark matter in two ``stripped-down'' 
calculations of gas cooling (Helly et~al. 2003b). The image on the left shows 
the results from an SPH simulation and the one on the right the predictions 
of a semi-analytical model which used the same dark matter haloes and merger 
trees. The dark matter is shown in grey. The different 
colour circles indicate different gas masses, with red indicating the highest 
mass. Based on a similar figure by Helly et~al. (2003b). 
}
\label{fig:cooltest}
\end{figure}

The recipe for the cooling and accretion of gas lies at the very heart of 
semi-analytical modelling of galaxy formation. The calculation of 
the cooling rate is carried out under specialized conditions and depends 
upon a number of assumptions and approximations, as set out in 
Section~\ref{sec:cooling}. It is therefore essential to test the accuracy 
of the cooling model against the results of gas dynamics simulations, which 
treat the problem under less restrictive conditions. 

The sceptical reader may question if it is wise to calibrate a simple 
recipe against the results of numerical simulations, which themselves 
may not give an accurate description of how gas cools and accumulates 
in galaxies. One could argue that since the gas simulations tend to 
overproduce massive galaxies, they must be considered to be incorrect 
at some level. However, the problem of ``super-sized'' galaxies is intimately 
connected to a number of processes in addition to gas cooling, such as 
star formation and feedback, so it is not clear that the treatment of gas 
cooling in the simulations is to blame. On the contrary, the infall of 
baryons into the gravitational potential wells of dark haloes and the 
radiative cooling of the gas is the one process which different 
numerical simulators, using different codes and algorithms, seem to 
agree upon.

The semi-analytic cooling model can be tested by designing a numerical 
simulation in which the less well understood phenomena, namely star 
formation and feedback, are simply omitted. These conditions are easily 
replicated in a semi-analytical model, due to its modular nature. The 
resulting calculation will not produce a galaxy mass function that agrees 
with observations. However, this is not the aim of the experiment. 
Such a test has been carried out by the Durham and Munich semi-analytic 
modellers. Benson et~al. (2001) compared the cold gas mass functions in 
a SPH simulation with radiative cooling and a ``stripped-down'' 
semi-analytical model using Monte-Carlo merger trees. Yoshida et~al. (2002) 
and Helly et~al. (2003b) extended this work to an object-by-object 
comparison, using halo merger trees drawn from an N-body simulation in 
the semi-analytical model. The level of agreement found by these studies 
is reassuring. The semi-analytical models have the flexibility 
to explore different density profiles for the hot gas halo. Good agreement 
with the simulation results was found for a particular choice of gas 
profile, which is not necessarily the one used in the ``bells and whistles'' 
semi-analytic models in which the goal is to reproduce 
the observed luminosity function.  

Despite these encouraging results, there has been much debate in the 
literature regarding the validity of the cooling recipe used in 
semi-analytical models. Tracking the thermal history of particles in SPH 
simulations reveals that a significant fraction of the gas never 
reaches the virial temperatures typical of galactic haloes, 
$\approx 10^{5}--10^{6}$K (Katz \& Gunn 1991; Navarro \& White 1994; 
Kay et~al. 2000; Keres et~al. 2005). 
Kay et~al. (2000) found that only $11\%$ of the particles found in 
galaxies in their simulations experience temperatures in excess 
of $10^{5}$K. Binney (2004) remarks that shocks would not be efficient 
enough to heat the gas at a sufficient rate to reach the virial temperature 
whilst competing with radiative cooling, pointing out that his earlier 
work on this topic (Binney 1977) had been widely misquoted as supporting 
the standard cooling picture set out in Section~\ref{sec:cooling}. 
Birnboim \& Dekel (2003) reached a similar conclusion using spherically 
symmetric analytic calculations to compute the halo mass below which a 
quasi-static hot halo does not form. The picture could be even more 
complicated. Maller \& Bullock (2004) argued that the hot gas halo 
is prone to thermal instabilities which can result in fragmentation of 
the hot halo into warm, pressure-supported clouds, similar 
to high velocity clouds. This process leads to a reduction of the 
cooling rate in massive haloes, offering another route to the suppression 
of the formation of massive galaxies (see also Sommer-Larsen 2006; Fukugita \& Peebles 2006; Kaufmann et~al. 2006)

Keres et~al. (2005) addressed the question of how galaxies acquire their 
gas in detail using SPH simulations. These authors characterize their results 
in terms of two cooling regimes: a cold mode in which gas is funnelled 
down filaments onto galaxies and a hot mode in which gas cools from a 
quasi-static halo. The cold mode is found to dominate in low mass haloes 
($ 3 \times 10^{11} M_{\odot}$) and at high redshift ($z > 3$). Keres et~al. 
present an interesting discussion of the implications of their results, 
speculating that the formation of disks could be linked to gas acquired 
during the cold accretion mode and that bulges could result from the 
``traditional'' hot accretion mode. In this scenario, most of the energy 
radiated by gas cooling to form galaxies is emitted at around $10^{5}$K 
and appears in Ly-$\alpha$ rather than in the X-rays expected if the gas was 
cooling from higher virial temperatures (Haiman, Spaans \& Quataert 2000; 
Fardal et~al. 2001). 
Thus, this model may explain the general 
lack of X-ray emission observed from galactic haloes which host spiral 
galaxies; such emission would be expected in the standard cooling model (Benson et~al. 2000; we note, however, that a 
recent detection of an extended X-ray halo around a spiral galaxy implying 
gas cooling has been reported by Pedersen et~al. 2005). 

Keres et~al. suggest that the cooling recipe used in semi-analytic models 
requires significant revision in light of their simulation results. 
Croton et~al. (2006) counter this argument by pointing out that the 
cooling model of White \& Frenk (1991), as set out in 
Section~\ref{sec:cooling}, contains two timescales which set the rate 
at which the gas is incorporated into a galactic disk; the cooling 
time ($t_{\rm cool}$) and the free-fall time ($t_{\rm ff}$). 
At high redshift, 
the cooling times are short and it is the free fall time that 
determines how quickly gas can accrue in the disk. There is 
little change in the amount of gas 
added to the disk even if the cooling time is arbitrarily set to zero in 
this regime. At later times, the cooling time lengthens as the density of 
the gas drops. The first scenario, where $t_{\rm cool} \ll t_{\rm ff}$ can 
be identified with the cold accretion mode and the second, in which 
$t_{\rm cool} \gg t_{\rm ff}$, can be associated with the hot accretion mode. 
Keres et~al. argue that the halo mass marking the transition between these 
regimes in the semi-analytical models is different from that found 
in their simulations. 
However, Croton et~al. call into question the density estimation used by 
Keres et~al. to compute their gas cooling rates. One outstanding difference, 
however, is the geometry of the accretion in the two modes. It will be 
interesting to investigate if this really does alter the amount of angular 
momentum added to the disk by the gas accumulated in the cold accretion mode. 
It will also be instructive to track the thermal history of halo gas in 
an AMR calculation (e.g. Quilis 2004), which treats the shocks in a 
different way to that used by SPH.  

Numerical and semi-analytical simulations suffer 
from an overcooling problem, if there is no attempt to 
include some form of feedback process in the calculation. 
A good example of this is given by the comparison of the galaxy mass 
functions predicted by a semi-analytical model and a SPH simulation run 
using the same dark matter merger trees, as shown by Fig.~1 of 
Berlind et~al. (2003). Part of this problem can be traced to the way in 
which the local density is estimated once gas starts to cool within a halo. 
If hot and cold particles are in close proximity, the presence of the 
dense cold gas can lead to an overestimate of the hot gas density, 
thus enhancing the cooling rate. Pearce et~al. (1999) used a ``decoupling'' 
technique to estimate the hot gas density, in which particles with 
temperatures below $12\,000$K are ignored when computing the density of 
gas particles with temperatures in excess of $T=10^{5}$K. 
Scannapieco et~al. (2006) describe a SPH code with a co-spatial 
treatment of the hot and cold phases of the interstellar medium. 

Finally, we note that, in the interests of simplicity and speed, 
older numerical simulations tended to adopt either a fixed metallicity 
when computing the rate at which gas cools (e.g. a primordial gas 
composition or some fraction of solar metallicity) or, in some cases, 
used a fixed global metallicity with some ad-hoc time evolution. 
We saw in Section~\ref{sec:cooling} that the metallicity of the 
gas can make a significant impact upon the cooling time. Scannapieco 
et~al. (2005) demonstrated that adopting a self-consistent calculation 
of the timescale for gas cooling in a model in which the gas metallicity 
evolves according to a chemical evolution model produces galaxies which 
are 25\% bigger than if a fixed, primordial metallicity is assumed.    

\subsection{Star formation}
\label{sec:stars}

Kay et~al. (2002) present a useful review of the various prescriptions 
used in gas dynamical simulations to turn cold gas into stars (see also 
Thacker \& Couchman 2000). 
The simulators have more information at their disposal than the 
semi-analytical modellers, which they can incorporate into the star 
formation recipe, such as the overdensity of cold gas, the motion 
of the gas and its temperature. Some simulators adopt a rule based 
upon the Schmidt law, which is closer to the approach taken in the 
semi-analytical models (e.g. Okamoto et~al. 2005). However, the 
prescriptions used in the simulations are still nevertheless recipes, 
because of the lack of a detailed theory of star formation and the 
impossibility of achieving the resolution necessary to follow the full 
range of physics that such a theory is likely to involve. 
Typically, once a packet of cold gas satisfies the conditions for star 
formation, the SPH particle spawns a star particle, and the mass of the 
SPH particle is reduced accordingly. The star particle is collisionless, 
i.e. it only responds to the gravity of the other particles and not to 
pressure forces. Kay et~al. tested the various implementations of star 
formation and found that, in the absence of some form of feedback to 
regulate the star formation rate, all of the algorithms produce too many 
bright (or massive) galaxies. When the parameters of the various schemes 
were adjusted to produce the same global stellar mass density, Kay et~al. 
found little difference between the results obtained with a Schmidt law 
prescription and those from the more complicated recipes.   

\subsection{Feedback}
\label{sec:simfeedback}

Kay et~al. (2002) review and test different 
implementations of feedback used in numerical simulations. 
Two classes of feedback are considered, thermal and kinetic, which differ 
in the way in which the energy released by supernova explosions 
influences the interstellar medium. In thermal feedback, the energy 
from the supernova heats the interstellar medium (Katz 1992). At 
high redshift when gas densities are very high, such heating has 
little effect upon the ISM, which simply radiates the additional energy 
away very rapidly. The impact of thermal feedback can be increased if 
the heated gas is only allowed to cool after a delay corresponding to 
the timescale of stellar associations  ($\sim 30$Myr;  Gerritsen 1997) 
or if the supernova heating causes turbulence that adds to the pressure 
of the gas (Springel 2000). Navarro \& White (1993) considered both 
thermal and kinetic feedback, parameterising the fraction of the supernova 
energy put into each mode. In kinetic feedback, the energy released by 
the supernova is used to boost the kinetic energy of the ISM.  
Navarro \& White found that simulations with a higher proportion of 
kinetic feedback resulted in lower star formation rates, whereas the 
star formation rates were unaffected in the runs where thermal feedback 
dominated. Springel \& Hernquist (2003) used an even more extreme form 
of kinetic feedback, in the same spirit as the superwinds described in 
Section~\ref{sec:feedback}.   

Semi-analytical models effectively use kinetic feedback, as they 
assume that energy injection results in cold gas being moved out of 
galactic disks. Cole et~al. (1994) used a feedback recipe based upon 
the simulation results of Navarro \& White (1993).

\subsection{The angular momentum of galactic disks}
\label{sec:disk}

The production of realistic disk galaxies has been 
a long standing problem for numerical simulations. 
Due to the high efficiency with which gas can cool at 
the high densities encountered at early times, disks tend 
to form by the coalescence of lumps of cold gas. As these 
lumps merge together, dynamical friction drains them of 
angular momentum, which is transferred to the outer parts 
of the dark matter halo (Navarro \& White 1994). 
The resulting disk retains an order of magnitude less angular momentum 
than originally possessed by the gas, resulting in a linear size 
for the disk that is much smaller than is observed for typical 
disk galaxies (e.g. Sommer-Larsen, Gelato \& Vedel 1999; 
Abadi et~al. 2003). 

Understanding and preventing the loss of angular momentum is 
the key to obtaining disks that look like real spiral galaxies. 
In semi-analytical modelling, the assumption that the angular 
momentum of the cooling gas is conserved leads to disk scalelengths 
that agree well with observations (Mo, Mao \& White 1998; 
de Jong \& Lacey 2000; Cole et~al. 2000; Firmani \& Avila-Reese 2000). 
Cole et~al. demonstrate how changing the strength of supernova 
feedback causes a change in the scalelength distribution of 
disks. If gas is allowed to cool in low mass haloes, then the 
predicted median sizes are too small.

A range of possible solutions have been proposed to the angular 
momentum problem in numerical simulations, ranging from numerical 
to astrophysical. 
Limited mass or force resolution and other numerical artefacts 
in the SPH method itself may contribute to the loss of angular momentum 
(Sommer-Larsen, Gelato \& Vedel 1999; Okamoto et~al. 2003; 
Governato et~al. 2004).  Governato et~al. (2004) point out that 
gas dynamic simulations, due to the added computational overhead, 
have traditionally employed far fewer particles to study the 
formation of disks than are used to attack the far simpler problem 
of establishing the internal structure of dark matter haloes. 
Sommer-Larsen \& Dolgov (2001) simulated the formation of disk 
galaxies in a warm dark matter universe and obtained galaxies 
whose properties were in better agreement with those of observed 
galaxies. Replacing cold dark matter by warm dark matter reduces the 
amount of small scale power which means that there are fewer low mass 
haloes in which gas may cool at high redshift and there are also fewer 
satellite galaxies to heat any disks that do form (see also Governato 
et~al. 2004). 

Most attention, however, has been directed towards astrophysical 
solutions which involve putting some form of feedback into the 
simulations to prevent gas from cooling in low mass haloes. 
Weil, Eke \& Efstathiou (1998) prevented the cooling of gas by hand 
until the bulk of the mass of the galactic halo had been assembled 
($z \approx 1$) to obtain disk galaxies which have the required specific 
angular momentum and which are not too centrally concentrated.
Sommer-Larsen, Gelato \& Vedel (1999) found that including strong 
feedback resulted in disks with scale lengths approaching those observed. 
Thacker \& Couchman (2001) used a feedback scheme in which the energy injected 
by supernovae persists in the interstellar medium for around $30$Myr, before 
the gas is allowed to radiate the energy away.  Sommer-Larsen, 
Gotz \& Portinari (2003) found that simulations with a strong early 
burst of star formation responsible for blowing away much of the cold 
gas in the halo before it is turned into stars produced disks with 
only a factor of two less specific angular momentum than observed. 
Okamoto et~al. (2005) used the top-heavy IMF in starbursts proposed 
in the semi-analytical model of Baugh et~al. (2005) as a solution 
to the problem of matching the number counts of galaxies detected 
at sub-millimetre wavelengths to produce disk galaxies. 
Robertson et~al. (2004) use a ``subresolution'' (semi-analytical) 
model to implement star formation and feedback in a multiphase interstellar 
medium. They find that the pressure support provided to the ISM by the 
processes of star formation and feedback allows the gas to retain more of its 
angular momentum, leading to a realistic disk galaxy, without a 
bulge component. Robertson et~al. (2006) develop this idea further to propose 
a merger driven scenario in which disks are the remnants of mergers between 
gas rich progenitors. 

Governato et~al. (2004) have argued that much of the angular momentum 
loss can be traced to limited mass or force resolution in previous 
simulations. These authors are able to produce viable disk galaxies 
without resorting to a significant injection of energy through feedback. 
However, the gas in the Governato et~al. simulations has a primordial 
composition, which reduces the rate at which it cools by almost an 
order of magnitude in haloes with circular velocities around 
$60 {\rm kms}^{-1}$.

\section{Successes and failures of hierarchical galaxy formation: 
areas for future improvement}

To complete this review, we discuss some of the areas in which the 
semi-analytical approach has enjoyed some notable successes and also 
those areas in which it has struggled or even failed to match observations. 
The continued evolution of the models is driven by the datasets for 
which the model predictions are at odds with the observations. Hence, 
the topics below in which difficulties are highlighted are likely to be 
those which will lead to future improvements to the modelling of the 
physics of galaxy formation. 

\subsection{Small scale fluctuations: problems for the cold dark matter model?}

First, we assess the health of the background cosmological model in which the 
semi-analytical models discussed in this review are set. We have already seen 
in the introduction that the cold dark matter model gives an impressively 
good description of the large scale fluctuations in the universe, as measured 
in the CMB and the large scale clustering 
of galaxies (e.g. Sanchez et~al. 2006). How does the CDM model 
fare on smaller scales 
which are of direct relevance for galaxy formation? Over the past decade, 
two serious challenges have emerged to the cold dark matter model: the 
abundance of sub-haloes in galactic sized haloes (the ``satellite problem'') 
and the shape of the inner density profile of cold dark matter haloes 
(the ``cuspy core problem''). The satellite problem 
is the result that high resolution simulations of galactic sized dark matter 
haloes reveal a plethora of substructures, and look liked scaled down 
versions of cluster mass dark haloes (Moore et~al. 1999a). Clusters 
contain many galaxies, so should we expect galactic haloes to also contain 
numerous satellite galaxies in the cold dark matter model? The number of 
satellite galaxies detected to date in the local group is over an order of 
magnitude fewer than would be expected from simple predictions based on 
the mass and number of dark matter substructures.
The cuspy core problem relates to the steepness of the central density 
profiles of cold dark matter haloes (e.g. Power et~al. 2003). Simple 
predictions of the rotation curves of cold dark matter haloes fail to 
match measurements made for a sample of low surface brightness galaxies 
dominated by dark matter, implying that the CDM haloes are too cuspy 
(Moore et~al. 1999b).
These two challenges to the cold dark matter model have generated a huge 
amount of interest. Two types of solution have been proposed. The first 
class focuses on modifying the properties of the dark matter 
particle. The problem of the over-abundance of low mass haloes and the 
cuspy inner density profiles of dark matter haloes can both be addressed 
by reducing the amount of small scale power below that predicted by 
the cold dark matter model. Possible ways to achieve this which have been 
put forwards include self-interacting dark matter (Spergel \& Steinhardt 2000) 
and replacing the cold dark matter by warm dark matter (e.g. Bode, 
Ostriker \& Turok 2001). Changing the nature of the dark matter has the 
drawback that it will be harder to account for the reionisation of the 
universe at high redshift (e.g. Spergel et~al. 2003); if the level of 
density fluctuations is reduced on small scales, then there will be fewer 
low mass haloes at high redshift, delaying the formation of the more massive 
structures which host the galaxies that produce large quantities of ionising 
photons. The second type of solution is to retain the cold 
dark matter framework whilst improving the treatment of the astrophysical 
phenomena which may influence the predictions of the model on these 
scales. For example, the number of luminous satellites within galactic 
haloes can be strongly affected by a number of processes: supernova 
feedback, the impact of a photoionising background on galaxy cooling and 
dynamical processes such as ram pressure stripping (e.g. 
Benson et~al. 2002a, 2002b; Somerville 2002).

\subsection{The present day galaxy luminosity function} 

The luminosity function is the most fundamental description 
of the galaxy population. Current determinations are able to 
measure the luminosity function to exquisite accuracy, with, 
for the most part, random errors which are smaller than the 
remaining systematic errors (Norberg et~al. 2002b; Blanton 
et~al. 2003). This observational breakthrough has led to a 
vigorous overhaul of the cooling and feedback prescriptions 
used in semi-analytical models. The faint end of the luminosity 
function can be matched with a combination of supernova feedback 
and the suppression of gas cooling in low mass haloes as a result 
of the presence of a background of photo-ionising radiation 
(e.g. Benson et~al. 2002a). Matching the bright end of the luminosity 
function has proven to be a more challenging problem, with a variety 
of phenomena proposed, such as the injection of energy into the hot 
gas halo to change its density, the heating of the hot gas to 
balance cooling and superwinds, which drive cold gas out of quite 
large galaxies (see Benson et~al. 2003 for a comparison of the 
effectiveness of different prescriptions; see also Kauffmann et~al. 1999; 
Somerville \& Primack 1999; Bower et~al. 2001; Hatton et~al. 2003). 
Any phenomenon which leads to a model successfully matching 
the observed break in the luminosity function uses up a dangerously 
high fraction of the energy released by supernova explosions. 
This led to the consideration of other possible energy sources 
and the implementation of AGN feedback into the models (Croton et~al. 2006; 
Bower et~al. 2006; see also Granato et~al. 2004). These models yield 
exceptionally faithful reproductions of the observed luminosity function. 
In view of the success of these first attempts to explicitly 
include the impact of AGN on galaxy formation, the energetics 
and the mechanics of the feedback required need to be considered in 
more detail. Do we actually observe the amount of AGN activity implied 
by the models and does this activity occur in the same type of objects 
that are observed to have AGN? One feature of the new models which gives 
cause for concern is the need for extremely strong supernova feedback 
(e.g. the model of Bower et~al. requires effective supernova feedback 
in galaxies with large circular velocities). Future work could focus on 
developing the cooling model further by looking at the response of the 
gas profile to entropy floors (Balogh, Babul \& Patton 1999; McCarthy et~al. 
2004). The next generation of the implementation of AGN feedback in 
semi-analytical models will be guided by detailed, high resolution 
simulations of the effects of energy input from AGN (e.g. Quilis, 
Bower \& Balogh 2001; Dalla Vecchia et~al. 2004; Ruszkowski, 
Bruggen \& Begelman 2004; Springel, di Matteo \& Hernquist 2005).

\subsection{The scaling relations of galaxies} 

A number of scaling relations are observed for 
spiral and elliptical galaxies. Some of these reveal common properties of 
the stellar populations of galaxies across a wide range of galaxy masses 
whilst others reveal connections between the stellar populations and the 
structural properties of galaxies. The existence and remarkable tightness 
of these relations point to clues about the star formation history and 
assembly of galaxies. 

\subsubsection{Matching the zero point of the Tully Fisher relation 
and the normalisation of the luminosity function} 

A long standing problem for hierarchical models has been to match the 
zero-point of the Tully-Fisher relation, the observed correlation between 
the rotation speed and luminosity of spiral disks (Tully \& Fisher 1977), 
at the same time as reproducing the luminosity function (Heyl et~al. 1995). 
We caution the reader to check carefully, when presented with model 
predictions for the Tully-Fisher relation, as to how the circular velocity of 
the model galaxy has been calculated. Some models use the effective rotation 
speed of the dark matter halo at the virial radius as a proxy for the 
rotation speed of the galaxy. Other models attempt to model the rotation 
curve of the galaxy, including the gravity of the baryons and dark matter. 
In the concordance $\Lambda$CDM model, the use of the effective circular 
velocity of the host dark matter halo measured at the virial radius tends 
to produce a very good match to the observed Tully-Fisher zero-point. To 
date, no model with a realistic calculation of galaxy size has been able to 
match the zero-point using the circular velocity of the disk measured at 
the half mass radius. This could imply that the model galaxies are either 
too condensed or contain to much mass. The first possibility points to a 
problem with the concentrations predicted for dark matter haloes in the 
cold dark matter model or calls into question some of the approximations 
used in the size calculation, such as adiabatic contraction of the baryons 
and dark matter (although this prescription has been tested successfully 
against numerical simulations e.g. Choi et~al. 2006). 
 
\subsubsection{The fundamental plane of elliptical galaxies} 

The radii, velocity dispersions and luminosities of elliptical galaxies 
show tight correlations which can be described by the so-called ``fundamental 
plane'' (Djorgovski \& Davies 1987; Dressler et~al. 1987).  Semi-analytical 
models have been able to produce a reasonable match to the local 
fundamental plane (Hatton et~al. 2003; Almeida et~al. 2006). However, 
the models do not reproduce the evolution seen in the zero-point of the plane, 
despite matching the evolution inferred in the mass-to-light ratios of 
ellipticals (Almeida et~al. 2006). This implies that another property of 
ellipticals is evolving at a rate which is at odds with observations to 
cancel the evolution in the mass-to-light ratios. 
Furthermore, one of the projections of the fundamental plane, between radius 
and luminosity is too shallow in the models (Almeida et~al. 2006). This could point to a problem in the way in which the size of merger remnants is computed 
in the models. 

\subsubsection{The sizes of galaxies} 
The scaling relations discussed in the preceeding two subsections rely upon 
the calculation of the scalelengths of galaxies. Although sophisticated 
alogrithms are now in use to compute sizes in some models, the problems 
experienced in matching the Tully-Fisher relation for spirals or the radius 
luminosity relation for ellipticals suggest the need for 
further improvements to these calculations. One area in which the 
semi-analytical recipe is undoubtedly oversimplified is in the tracking 
of the angular momentum vector or spin of a galaxy during mergers. 
In the models, the total angular momentum of the gas is conserved 
as it cools to form a disk, but no attention is paid to the direction 
of the angular momentum vector. In numerical simulations, the angular 
momentum vector of a galaxy is seen to change orientation after merger 
events, which could explain some of the difficulties in finding galaxies 
of the observed radius (Okamoto et~al. 2005).  
The first steps towards a more accurate modelling of the accretion 
of angular momentum have already been taken (van den Bosch 2001; 
Bullock et~al. 2001a; van den Bosch 2002; Chen \& Jing 2002; Maller \& Dekel 
2002).

\subsection{The metal enrichment of the intracluster medium and 
the abundance ratios of elliptical galaxies} 

Semi-analytical models can now follow the delayed enrichment 
of the ISM and IGM due to type Ia supernova explosions (Nagashima 
\& Okamoto 2006). By explicitly tracking the ejecta of type Ia and type 
II supernovae, predictions can be made for the production of $\alpha$ elements 
(e.g. O, Mg) and for iron-peak elements. The gas in hot, X-ray emitting 
clusters shows an $\alpha$ to iron ratio in excess of the solar value. 
The metal content of the ICM depends upon the star formation 
histories of the galaxies in the clusters and the way in which metals 
are expelled from the galaxies. The semi-analytical models do not reproduce 
the $\alpha$-enhancement of the ICM when star formation takes place with a 
standard IMF. However, if a top-heavy IMF is adopted in star bursts, the 
models can match the observed abundances of O, Mg, Si and Fe (Nagashima 
et~al. 2005a). The same model with a top-heavy IMF can also match the 
abundance of $\alpha$-elements observed in elliptical galaxies 
(Nagashima et~al. 2005b). However, this model does not reproduce the observed 
trend of $\alpha/$Fe increasing with the velocity dispersion of the 
galaxy; if anything, the models display a decrease of $\alpha$-enhancement 
with increasing velocity dispersion. 

\subsection{Tracing galaxy formation and evolution over the history 
of the universe} 

The first observations of significant numbers of high redshift galaxies 
provided an important challenge to galaxy formation modellers. Could the 
models, whose parameters were set with reference to the local universe, 
produce predictions which also matched the high redshift universe? 
Baugh et~al. (1998) showed that the Cole et~al. (1994) model gave a 
remarkably close match to the star formation history of the universe 
inferred by Madau et~al. (1996) and the luminosity function of Lyman 
break galaxies at $z \sim 3$ (Steidel et~al. 1999; see also the model of 
Somerville, Primack \& Faber 2001). As the models become more sophisticated, 
it is unlikely that the predictions for the high redshift universe are unique, 
once the model parameters have been set against a local reference point. 
Nevertheless, useful conclusions can still be reached, provided that the 
high redshift observations are not treated in isolation, but instead in the 
context of a model which still aims to explain the local universe. 
For example, 
using the superwind version of the Durham model, Baugh et~al. (2005) 
were able to reproduce the number counts of sub-millimetre selected 
galaxies, the luminosity function of Lyman-break galaxies at $z=3$ 
and $z=4$, whilst retaining a fair match to the present day optical 
and far infrared luminosity functions. However, this was only possible with 
the controversial assumption of a top-heavy initial mass function 
for star formation in merger driven starbursts. This model can also 
account for the metallicity of the intra-cluster medium and 
elliptical galaxies (Nagashima et~al. 2005a,b). 
Granato et~al. (2004) are able to account for the abundance of dusty 
galaxies without resort to changing the stellar initial mass function, but 
only considered the formation of spheroids. 
It will be interesting to see the predictions of the Croton et~al. (2006) 
and Bower et~al. (2006) models for the number of sub-millimetre emitting 
galaxies.  

\subsection{Galaxy downsizing: which galaxies are actively forming stars?} 

Observations indicate that the bulk of star formation since $z\sim1$ 
has taken place in intermediate mass galaxies, rather than in the 
most massive galaxies (e.g. Cowie et~al. 1996; Kodama et~al. 2004). 
Coupled with estimates of the stellar mass function which appear to 
show substantial numbers of massive galaxies in place at $z > 1$ 
(e.g. Drory et~al. 2005), this has been interpreted by some as 
evidence against hierarchical models and in favour of a monolithic 
collapse scenario. The naive expectation is that since the more massive 
haloes are assembled at relatively recent epochs in hierarchical models, 
then the most massive galaxies should also be acquiring their stellar 
mass at the same time. In practice this is not the case. Baugh et~al. 
(1999) showed that the star formation history of galaxies in clusters 
is shifted to earlier epochs relative to the field. A galactic halo which 
will ultimately become part of a cluster will begin to collapse at an 
earlier epoch than the same mass halo in a more average density environment, 
which in turn means that stars form earlier in the progenitor of the cluster 
halo. A significant fraction of the mass of massive galaxies is brought in 
through galaxy mergers. Typically, these are gas-poor mergers that occur 
since $z \sim 1$, which simply reassemble pre-existing stars into a spheroid 
(Baugh et~al. 1996b; an excellent discussion of how brightest cluster galaxies 
acquire their mass can be found in De Lucia \& Blaizot 2006). 
Observational evidence exists for these gas-poor or ``dry'' mergers 
(van Dokkum 2005; Bell et~al. 2006). 
It is true however, that without some means of 
suppressing the cooling of gas in massive haloes, there is  
too much star formation activity in massive galaxies at the present 
day in the models, which could obscure the observational signatures 
of ``downsizing''. The AGN feedback models help to redress this (Croton 
et~al. 2006; de Lucia et~al. 2006, Bower et~al. 2006). Croton et~al. and 
de Lucia et~al. show that in the model with AGN feedback, 
high velocity dispersion galaxies have older stellar populations. 
Bower et~al. demonstrate that, for massive galaxies, the star formation 
rate in objects of a given mass declines to the present day, 
giving a reasonable match to observational estimates (Juneau et~al. 
2005; Bauer, Drory \& Hill 2005). 
There still appears to be additional recent suppression of star formation 
observed in massive galaxies over and above that predicted by the models, 
which suggests than the onset of AGN feedback may need to be felt more keenly 
at earlier epochs, perhaps by more rapid growth of the black hole mass. 

\subsection{Massive galaxies at high redshift} 

There is now a significant amount of deep, multiband 
photometry which has been exploited to estimate the stellar 
mass function of galaxies across a wide range of redshifts 
(e.g. Fontana et~al. 2004; Glazebrook et~al. 2004; Drory et~al. 2005). 
Several models are on the market which attempt to explain 
these observations of the high redshift universe. 
Nagamine et~al. (2004, 2005) use numerical simulations to argue that hierarchical models do not 
have a problem in accounting for massive red galaxies at 
high redshift. However, these authors do not present predictions 
for the luminosity function at the present day, which, as we have 
argued is an important constraint on the gas cooling and feedback 
prescriptions used in the semi-analytical models.  
Granato et~al. (2004) reproduce the high redshift evolution of 
the K-band luminosity function and the abundance of sources 
detected at sub-millimetre wavelengths. However, their model 
does not treat the disk population and does not extend to 
the present day, so the high redshift successes may come at 
a high price. Bower et~al. (2006) find good agreement with the 
estimated stellar mass function up to $z \sim 5$, with more 
massive objects in place at earlier times than is predicted 
by previous versions of the Durham model.  
Again, caution should be exercised when comparing model predictions 
for the stellar mass function with observational estimates. 
Drory, Bender \& Hopp (2004) showed how estimates of the stellar 
mass from broad band photometry at low redshift compare with 
dynamical estimates. Eke et~al. (2006) repeated this test for 
the Durham semi-analytic model, in which case the true stellar 
mass is known a priori. Both studies show that there is a scatter 
and systematic differences between the mass estimated 
from photometry and the dynamical (Drory, Bender \& Hopp 2004) or 
true stellar mass (Eke et~al. 2006). 
Both effects are likely to increase in the case of optical and 
near infra-red photometry of high redshift galaxies, as a result of 
the wavelength coverage shifting towards the rest frame ultra-violet 
as the redshift at which the galaxy is observed increases.
The model predictions should be convolved with the estimated errors 
in the stellar mass determination, which could have a significant 
impact upon the high mass end of the mass function, before comparing 
with the observational estimates.

\section{Summary}

The goal of this review was to provide an overview of the physics of 
hierarchical galaxy formation and, in doing so, to convince the reader that  
semi-analytical models provide a powerful, complementary approach to studying 
galaxy formation to that offered by numerical simulations. The semi-analytical 
models currently provide the most detailed and complete predictions for the 
properties of the galaxy population in cold dark matter universes. The 
modular framework of the models means that it is straightforward to 
revise the description of the various phenomena, as required to reproduce 
the results of more detailed (and expensive) numerical simulations or as 
motivated by the need to reproduce new observations. In the future, the 
semi-analytical models will become more intimately linked with dark matter 
only N-body simulations, either the next generation of post-Millennium, 
high resolution simulations of cosmologically significant volumes or 
ultra-high resolution resimulations of individual dark matter structures. 
There is still a long way to go, but armed with this theoretical machinery 
and the ever increasing data on the high redshift data universe, these are 
promising times for advancing our understanding of how galaxies are made. 

\section*{Acknowledgements} 

I am indebted to current and past members of the Durham semi-analytical 
modelling group for shaping my ideas about galaxy formation: Shaun Cole, 
Cedric Lacey, Carlos Frenk, Andrew Benson, Richard Bower, John Helly, 
Rowena Malbon and Cesario Almeida. I have also benefitted from discussions 
at various times with Guinevere Kauffmann, Simon White, Vince Eke, 
Vladimir Alvia-Reese, Darren Croton and Rachel Somerville. I would like to thank 
Masahiro Nagashima, Rowena Malbon and the three anonymous referees 
for carefully reading the original manuscript and for making many helpful 
suggestions, and for correcting omissions, which helped to improve 
the article. I am indebted to Chris Power and Ariel Sanchez for supplying 
me with figures to use in the article. 
The Smithsonian/NASA Astrophysics Data System and the astro-ph 
preprint server were invaluable resources in preparing this review. 
Finally, I would also like to thank the editorial team at Reports on 
Progress in Physics for their considerable patience and indulgence 
whilst waiting for this article to be completed. The author is supported 
by the Royal Society through the award of a University Research Fellowship.

\References

\item[] Abadi M G, Navarro J F, Steinmetz M and Eke V R, 2003, 
{\it Astrop. J.}, {\bf 591}, 499-514

\item[] Abel T, Bryan G L and Norman M L, 2002, {\it Science}, 93-98

\item[] Adelberger K L, Steidel C C, Shapley A E and Pettini M, 
2003, {\it Astrop. J.}, {\bf 584}, 45-75

\item[] Alexander D M, Bauer F E, Chapman S C, Smail I, Blain A W, Brandt W N and Ivison R J, 2005, {\it Astrop. J.}, {\bf 632}, 736-750

\item[] Almeida C, Baugh C M and Lacey C G, 2006, {\it preprint}, astro-ph/0608544

\item[] Avila-Reese V, 2006, {\it Lectures given at the IV Mexican School of Astrophysics}, astro-ph/0605212. 

\item[] Babul A and Rees M J, 1992, {\it Mon. Not. Roy. Ast. Soc.}, {\bf 255}, 346-350

\item[] Baldry I K, Glazebrook K, Brinchmann J, Ivezic Z, Lupton R H, Nichol R C and Szalay A S, 2004,  {\it Astrop. J.}, {\bf 600}, 681-694

\item[] Balogh M L, Babul A and Patton D R, 1999, {\it Mon. Not. Roy. Ast. Soc.}, {\bf 307}, 463-479

\item[] Balogh M L, Pearce F R, Bower R G and Kay S T, 2001, 
{\it Mon. Not. Roy. Ast. Soc.}, {\bf 326}, 1228-1234

\item[] Balogh M L, Baldry I K, Nichol R, Miller C, Bower R and Glazebrook K, 
2004a, {\it Astrop. J.}, {\bf 615}, L101-L104

\item[] Balogh M L, et~al., 2004, {\it Mon. Not. Roy. Ast. Soc}, {\bf 348}, 1355-1372

\item[] Bardeen J M, Bond J R, Kaiser N and Szalay A S, 1986, 
{\it Astrop. J.}, {\bf 304}, 15-61

\item[] Barger A J, Cowie L L, Sanders D B, Fulton E, Taniguchi Y, Sato Y, 
Kawara K and Okuda H, 1998, {\it Nature}, {\bf 394}, 248-251 

\item[] Barkana R and Loeb A, 2001, {\it Phys. Rep.}, {\bf 349}, 125-238

\item[] Barnes J and Efstathiou G, 1987, {\it Astrop. J.}, {\bf 319}, 575-600

\item[] Barnes J E and Hernquist L E, 1991, {\it Astrop. J.}, {\bf 370}, L65-L68

\item[] Barnes J E and Hernquist L E, 1992, {\it Ann. Rev. Astron. Astrop.}, 
{\bf 30}, 705-742

\item[] Bauer A E, Drory N and Hill G J, 2005, {\it Astrop. J.}, {\bf 621}, L89-L92

\item[] Baugh C M, Cole S and Frenk C S, 1996a, {\it Mon. Not. Roy. Ast. Soc.}, {\bf 282}, L27-L32

\item[] Baugh C M, Cole S and Frenk C S, 1996b, {\it Mon. Not. Roy. Ast. Soc.}, {\bf 283}, 1361-1378

\item[] Baugh C M, Cole S, Frenk C S and Lacey C G, 1998, {\it Astrop. J.}, 
{\bf 498}, 504

\item[] Baugh C M, Cole S, Frenk C S, Benson A J and Lacey C G, 1999, Star Formation in Early Type Galaxies, ASP Conference Series 163, ed. P. Carral \& J. Cepa(Astronomy Society of the Pacific) 

\item[] Baugh C M, Lacey C G, Frenk C S, Granato G L, Silva L, Bressan A 
and Cole S, 2005, {\it Mon. Not. Roy. Ast. Soc.}, {\bf 356}, 1191-1200

\item[] Bell E F, Baugh C M, Cole S, Frenk C S and Lacey C G, 
2003, {\it Mon. Not. Roy. Ast. Soc.}, {\bf 343}, 367-384

\item[] Bell E F, McIntosh D H, katz N and Weinberg M D, 2003, {\it Astrop. J. Supp.}, {\bf 149}, 289-312

\item[] Bell E F, et~al., 2006, {\it Astrop. J.}, {\bf 640}, 241-251

\item[] Bender R M, Burstein D and Faber S M, 1993, {\it Astrop. J.}, {\bf 411}, 153

\item[] Benson A J, Bower R G, Frenk C S and White S D M, 2000, {\it Mon. Not. Roy. Ast. Soc.}, {\bf 314}, 557-565

\item[] Benson A J, Pearce F R, Frenk C S, Baugh C M and Jenkins A, 
2001, {\it Mon. Not. Roy. Ast. Soc.}, {\bf 320}, 261-280

\item[] Benson A J, Lacey C G, Baugh C M, Cole S and Frenk C S, 
2002a, {\it Mon. Not. Roy. Ast. Soc.}, {\bf 333}, 156-176

\item[] Benson A J, Frenk C S, Lacey C G, Baugh C M and Cole S, 
2002b, {\it Mon. Not. Roy. Ast. Soc.}, {\bf 333}, 177-190

\item[] Benson A J, Bower R G, Frenk C S Lacey C G, Baugh C M and Cole S, 
2003, {\it Astrop. J.}, {\bf 599}, 38-49

\item[] Benson A J, Lacey C G, Frenk C S, Baugh C M and Cole S, 2004, 
{\it Mon. Not. Roy. Ast. Soc.}, {\bf 351}, 1215-1236

\item[] Benson A J, Kamionkowski M and Hassani S H, 2005, {\it Mon. Not. Roy. Ast. Soc.}, {\bf 357}, 847-858

\item[] Berlind A A, et~al., 2003, {\it Astrop. J.}, {\bf 593}, 1-25

\item[] Bergstrom L., 2000, {\it Rep. Prog. Phys.}, {\bf 63}, 793-841

\item[] Binney J, 1977, {\it Astrop. J.}, {\bf 215}, 483-491

\item[] Binney J, 2004, {\it Mon. Not. Roy. Ast. Soc.}, {\bf 347}, 1093-1096

\item[] Binney J and Merrifield M, 1998, Galactic Astronomy, Princeton 
University Press, Princeton, New Jersey. 

\item[] Binney J and Tremaine S, 1987, Galactic Dynamics, Princeton 
University Press, Princeton, New Jersey.

\item[] Birnboim Y and Dekel A, 2003, {\it Mon. Not. Roy. Ast. Soc.}, {\bf 345}, 349-364

\item[] Blain A W, Smail I, Ivison R J, Kneib J P and Frayer D T, 
2002, {\it Phys. Rep.}, {\bf 369}, 111-176

\item[] Blanchard A, Douspis M, Rowan-Robinson M and Sarkar S, 2003, 
{\it Astron. \& Astrop.}, {\bf 412}, 35-44

\item[] Blanton M, Cen R, Ostriker J P and Strauss M A, 1999, {\it Astrop. J.}, {\bf 522}, 590-603

\item[] Blanton M R, et~al., 2003, {\it Astrop. J.}, {\bf 592}, 819-838

\item[] Blumenthal G R, Faber S M, Primack J R and Rees M J, 1984, 
{\it Nature}, {\bf 311}, 517-525

\item[] Bode P, Ostriker J P and Turok N, 2001, {\it Astrop. J.}, {\bf 556}, 93-107

\item[] Bond J R, Cole S, Efstathiou G and Kaiser N, 
1991, {\it Astrop. J.}, {\bf 379}, 440-460

\item[] Bonnell I A, Bate M R, Clarke C J and Pringle J E, 1997, 
{\it Mon. Not. Roy. Ast. Soc.}, {\bf 285}, 201-208

\item[] Bower R G, 1991, {\it Mon. Not. Roy. Ast. Soc.}, {\bf 248}, 332-352

\item[] Bower R G, Benson A J, Lacey C G, Baugh C M, Cole S and Frenk C S, 
2001, {\it Mon. Not. Roy. Ast. Soc.}, {\bf 325}, 497-508

\item[] Bower R G, Benson A J, Malbon R, Helly J C, Frenk C S, Baugh C M, 
Cole S and Lacey C G, 2006, {\it Mon. Not. Roy. Ast. Soc.}, {\bf 370}, 
645-655

\item[] Bromm V and Larson R B, 2004, {\it Ann. Rev. Astron. Astrop.}, 
{\bf 42}, 79-118

\item[] Bruzual G and Charlot S, 2003, {\it Mon. Not. Roy. Ast. Soc.}, {\bf 344}, 1000-1028

\item[] Bryan G L and Norman M L, 1997, {\it in Computational Astrophysics, 
ASP Conference Series 123}, eds. D A Clarke and M J West.  

\item[] Bruzual A G and Charlot S, 1993, {\it Astrop. J.}, {\bf 405}, 538-553

\item[] Bruzual A G and Charlot S, 2003, {\it Mon. Not. Roy. Ast. Soc.}, {\bf 344}, 1000-1028

\item[] Bullock J S, Dekel A, Kolatt T S, Kravtsov A V, Klypin A A, Porciani C and Primack J R, 2001a, {\it Astrop. J.}, {\bf 555}, 240-257

\item[] Bullock J S, Kolatt T S, Sigad Y, Somerville R S, Kravtsov A V, Klypin A A, Primack J R and Dekel A, 2001b, {\it Mon. Not. Roy. Ast. Soc.}, {\bf 321}, 
559-575.

\item[] Carroll S M, 2004, {\it New Cosmology}, {\bf 743}, 16-32

\item[] Carroll S M, Sawicki I, Silvestri A and Trodden M, 2006, 
{\it preprint}, astro-ph/0607458

\item[] Cattaneo A, Blaizot J, Devriendt J and Guiderdoni B, 2005, {\it Mon. Not. Roy. Ast. Soc.}, {\bf 364}, 407-423

\item[] Cattaneo A, Dekel A, Devriendt J, Guiderdoni B and Blaizot J, 
2006, {\it Mon. Not. Roy. Ast. Soc.}, {\bf 370}, 1651-1665

\item[] Cen R and Ostriker J P, 1999, {\it Astrop. J.}, {\bf 514}, 1-6

\item[] Chandrasekhar S, 1943, {\it Astrop. J.}, {\bf 97}, 255

\item[] Charlot S, Worthey G and Bressan A, 1996, {\it Astrop. J.}, {\bf 457}, 625

\item[] Chen D N and  Jing Y P, 2002, {\it Mon. Not. Roy. Ast. Soc.}, {\bf 336}, 55-65

\item[] Choi J H, Lu Y, Mo H J and Weinberg M D, 2006, {\it preprint}, astro-ph/0604587

\item[] Ciardi B and Ferrara A, 2005, {\it Space Sci. Rev.}, {\bf 116}, 625-705

\item[] Cole S, 1991, {\it Astrop. J.}, {\bf 367}, 45-53

\item[] Cole S and Lacey C, 1996, {\it Mon. Not. Roy. Ast. Soc.}, {\bf 281}, 716
\item[] Cole S, Aragon-Salamanca A, Frenk C S, Navarro J F and Zepf S E, 
1994, {\it Mon. Not. Roy. Ast. Soc.}, {\bf 271}, 781

\item[] Cole S, Lacey C G, Baugh C M and Frenk C S, 2000, {\it Mon. Not. Roy. Ast. Soc.}, {\bf 319}, 168-204

\item[] Cole S, et~al., 2001, {\it Mon. Not. Roy. Ast. Soc.}, {\bf 326}, 255-273

\item[] Cole S, et~al., 2005, {\it Mon. Not. Roy. Ast. Soc.}, {\bf 362}, 505-534
\item[] Coles P and Lucchin F 2002, {\it Cosmology: The Origin and Evolution of Cosmic Structure}, John Wiley ansd Sons Ltd

\item[] Colin P, Avila-Reese V and Valenzuela O, 2000, {\it Astrop. J.}, {\bf 542}, 622-630

\item[] Colless M, et~al., 2001, {\it Mon. Not. Roy. Ast. Soc.}, 
{\bf 328}, 1039

\item[] Cowie L L, Songaila A, Hu E M and Cohen J G, 1996, 
{\it Astron. J}, {\bf 112}, 839

\item[] Couchman H M P and Rees M J, 1986, {\it Mon. Not. Roy. Ast. Soc.}, 
{\bf 221}, 53-62

\item[] Couchman H M P, Thomas P A and Pearce F R, 1995, {\it Astrop. J.}, 
{\bf 452}, 797

\item[] Croton D J, Springel V, White S D M, De Lucia G, Frenk C S, Gao L, 
Jenkins A, Kauffmann G, Navarro J F and Yoshida N, 2006, {\it Mon. Not. Roy. 
Ast. Soc.}, {\bf 365}, 11-28 

\item[] Dalcanton J J, Spergel D N and Summers F J, 1997, {\it Astrop. J.}, 
{\bf 482}, 659

\item[] Dalla Vecchia C, Bower R G, Theuns T, Balogh M L, Mazzotta P and Frenk 
C S, 2004, {\it Mon. Not. Roy. Ast. Soc.}, {\bf 355}, 995-1004

\item[] Davis M, Efstathiou G, Frenk C S and White S D M, 1985, {\it Astrop. J.}, {\bf 292}, 371-394

\item[] Dawson S, Spinrad H, Stern D, Dey A, van Breugel W, de Vries W and 
Reuland M, 2002, {\it Astrop. J.}, {\bf 570}, 92-99

\item[] de Bernardis P et~al., 2000, {\it Nature}, {\bf 404}, 955-959

\item[] De Jong R S and Lacey C G, 2000, {\it Astrop. J.}, {\bf 545}, 781-797

\item[] De Lucia G, Kauffmann G and White S D M, 2004, {\it Mon. Not. Roy. Ast. Soc.}, {\bf 349}, 1101-1116

\item[] De Lucia G, Springel V, White S D M, Croton D and Kauffmann G, 
2006, {\it Mon. Not. Roy. Ast. Soc.}, {\bf 366}, 499-509

\item[] De Lucia G and Blaizot J, 2006, {\it Preprint}, astro-ph/0606519

\item[] Deffayet C, Dvali G and Gabadadze G, 2002, {\it Phys. Rev. D.}, 
{\bf 65} 04023

\item[] Dekel A and Silk J, 1986, {\it Astrop. J.} {\bf 303}, 39-55

\item[] Devriendt J E G, Guiderdoni B and Sadat R, 1999, {\it Astron. \& Astrop.}, {\bf 350}, 381-398 

\item[] Devriendt J E G and Guiderdoni B, 2000, {\it Astron. Astrop.}, {\bf 363}, 851-862

\item[] Diemand J, Moore B and Stadel J, 2005, {\it Nature}, {\bf 433}, 389-391

\item[] Dickinson C, et~al., 2004, {\it Mon. Not. Roy. Ast. Soc.}, {\bf 353}, 732

\item[] Di Matteo T, Springel V and Hernquist L, 2005, {\it Nature}, {\bf 433}, 604-607

\item[] Djorgovski S and Davis M, 1987, {\it Astrop. J.}, {\bf 313}, 59 

\item[] Dressler A, 1980, {\it Astrop. J.}, {\bf 236}, 351-365 

\item[] Dressler A, Lynden-Bell D, Burstein D, Davies R L, Faber S M, Terlevich R J and Jackson R, 1987, {\it Astrop. J.}, {\bf 313}, 42

\item[] Drory N, Bender R and Hopp U, 2004, {\it Astrop. J.} {\bf 616}, L103-L106
\item[] Drory N, Salvato M, Gabasch A, Bender R, Hopp U, Feulner G and 
Pannella M, 2005, {\it Astrop. J.} {\bf 619}, L131-L134

\item[] Dubinski J and Carlberg R G, 1991, {\it Astrop. J}, {\bf 478}, 496-503

\item[] Efstathiou G, 1992, {\it Mon. Not. Roy. Ast. Soc.}, {\bf 256}, 43P-47P

\item[] Efstathiou G, 2000, {\it Mon. Not. Roy. Ast. Soc.}, {\bf 317}, 697-719

\item[] Efstathiou G, 2003, preprint, astro-ph/0303623

\item[] Efstathiou G and Jones B J T, 1979, {\it Mon. Not. Roy. Ast. Soc.}, {\bf 186}, 133-144

\item[] Efstathiou G, Frenk C S, White S D M and Davis M, 1988, {\it Mon. Not. Roy. Ast. Soc.}, {\bf 235}, 715-748

\item[] Efstathiou G, Sutherland W J and Maddox S J, 1990, {\it Nature}, {\bf 348}, 705-707

\item[] Efstathiou G, et~al., 2002, {\it Mon. Not. Roy. Ast. Soc.}, {\bf 330}, L29-L35

\item[] Eisenstein D J and Hut P, 1998, {\it Astrop. J.}, {\bf 498}, 137

\item[] Eisenstein D J, et~al., 2005, {\it Astrop. J.}, {\bf 633}, 560-574

\item[] Eke V R, Cole S and Frenk C S, 1996, {\it Mon. Not. Roy. Ast. Soc.}, 
{\bf 282}, 263-280

\item[] Eke V R, Navarro J F and Steinmetz M, 2001, {\it Astrop. J.}, {\bf 554}, 114-125

\item[] Eke V R et~al., 2004a, {\it Mon. Not. Roy. Ast. Soc}, {\bf 348}, 866-878

\item[] Eke V R et~al. 2004b, {\it Mon. Not. Roy. Ast. Soc}, {\bf 355}, 769-784 

\item[] Eke V R, Baugh C M, Cole S, Frenk C S and Navarro J F, 2006, {\it Mon. Not. Roy. Ast. Soc}, {\bf 370}, 1147-1158 

\item[] Ellis R S, 1997, {\it Ann. Rev. Astron. Astrop.}, {\bf 35}, 389-443

\item[] Elmegreen B G and Scalo J, 2004, {\it Ann. Rev. Astron. Astrop.}, 
{\bf 42}, 211-273. 

\item[] Enoki M, Nagashima M and Gouda N, 2003, {\it Pub. Astron. Soc. Jap.}, 
{\bf 55}, 133-142

\item[] Faber S M, 1973, {\it Astrop. J.}, {\bf 179}, 731

\item[] Faber S M and Jackson R, 1976, {\it Astrop. J.}, {\bf 204}, 668

\item[] Fabian A C, Mushotzky R F, Nulsen P E J and Peterson J R, 2001, {\it Mon. Not. Roy. Ast. Soc}, {\bf 321}, L20-L24

\item[] Fabian A C, Voigt L M and Morris R G, 2002, {\it Mon. Not. Roy. Ast. Soc}, {\bf 335}, L71-L74

\item[] Fabian A C, Reynolds C S, Taylor G B and Dunn R J H, 2005, 
{\it Mon. Not. Roy. Ast. Soc}, in press, astro-ph/0501222

\item[] Fall S M and Efstathiou G, 1980, {\it Mon. Not. Roy. Ast. Soc}, {\bf 193}, 189-206

\item[] Fardal M A, Katz N, Gardner J P, Hernquist L, Weinberg D H and 
Dave R, 2001, {\it Astrop. J.} {\bf 539}, L9-L12

\item[] Ferguson H C, Dickinson M and Williams R, 2000, {\it 
Ann. Rev. Astron. Astrop.}, {\bf 38}, 667-715

\item[] Ferrara A, Bianchi S, Cimatti A and Giovanardi C, 1999, 
{\it Astrop. J. Supp.} {\bf 123}, 437-445

\item[] Ferrarese L and Merritt D, 2000, {\it Astrop. J.} {\bf 539}, L9-L12

\item[] Fioc M and Rocca-Volmerange B, 1999, {\it Astron. \& Astrop.}, {\bf 351} 869-882

\item[] Firmani C and Avila-Reese V, 2000, {\it Mon. Not. Roy. Ast. Soc}, {\bf 315}, 457-472

\item[] Fontana A, et~al. 2004, {\it Astron. \& Astrop.}, {\bf 424}, 23-42

\item[] Fontanot F, Monaco P, Cristiani S and Tozzi P, 2006, {\it Preprint}, 
astro-ph/0609823

\item[] Frenk C S, White S D M, Davis M and Efstathiou G, 1988, {\it Astrop. J.} {\bf 327}, 507-525

\item[] Frenk C S, Evrard A E, White S D M and Summers F J, 1996, {\it Astrop. J.} {\bf 472}, 460

\item[] Frenk C S, et~al., 1999,  {\it Astrop. J.}, {\bf 525}, 554-582

\item[] Fukigita M, Hogan C J and Peebles P J E, 1998, 
{\it Astrop. J.} {\bf 503}, 518-530

\item[] Fukugita M, Yamashita K, Takahara F and Yoshii Y, 1990, 
{\it Astrop. J.}, {\bf 361}, L1-L4

\item[] Fukugita M and Peebles P J E, 2006, {\it Astrop. J.}, {\bf 639}, 590-599

\item[] Fukushige T and Makino J, 1997, {\it Astrop. J.}, {\bf 477}, L9-L12

\item[] Fukushige T and Makino J, 2001, {\it Astrop. J.}, {\bf 557}, 533-545

\item[] Gao L, White S D M, Jenkins A, Frenk C S and Springel V, 
2005a, {\it Mon. Not. Roy. Ast. Soc.}, {\bf 363}, 379-392

\item[] Gao L, Springel V and White S D M, 2005b, {\it Mon. Not. Roy. Ast. Soc.}, {\bf 363}, L66-L70

\item[] Gebhardt K, et~al., 2000, {\it Astrop. J.}, {\bf 539}, L13-L16

\item[] Gelb J M and Bertschinger E, 1994, {\it Astrop. J.}, {\bf 436}, 467-490

\item[] Gerritsen J P E, 1997, PhD Thesis, Groningen University

\item[] Ghigna S, Moore B, Governato F, Lake G, Quinn T and 
Stadel J, 1998, {\it Mon. Not. Roy. Ast. Soc.}, {\bf 300}, 146-162

\item[] Ghigna S,  Moore B, Governato F, Lake G, Quinn T and Stadel J, 2000, 
{\it Astrop. J.}, {\bf 544}, 616-628

\item[] Girardi L, Bressan A, Bertelli G and Chiosi C, 2000, {\it Astron. \& Astrop. Suppl.}, {\bf 141}, 371-383. 

\item[] Glazebrook K, et~al., 2004, {\it Nature}, {\bf 430}, 181-184

\item[] Gnedin N Y, 1995, {\it Astrop. J. Suppl.}, {\bf 97}, 231-257

\item[] Gnedin N Y, 2000, {\it Astrop. J.}, {\bf 542}, 535-541

\item[] Gondolo P, 2004, {\it preprint}, astro-ph/0403064

\item[] Governato F, Moore B, Cen R, Stadel J, Lake G and Quinn T, 
1997, {\it New Astron.}, {\bf 2}, 91-106

\item[] Governato F, Babul A, Quinn T, Tozzi P, Baugh C M, Katz N and 
Lake G, 1999, {\it Mon. Not. Roy. Ast. Soc.}, {\bf 307}, 949-966

\item[] Governato F, Mayer L, Wadsley J, Gardner J P, Willman B, Hayashi E, 
Quinn T, Stadel J and Lake G, 2004, {\it Astrop. J.}, {\bf 607}, 688-696

\item[] Granato G L, Lacey C G, Silva L, Bressan A, Baugh C M, Cole S and 
Frenk C S, 2000, {\it Astrop. J.}, {\bf 542}, 710-730

\item[] Granato G L, De Zotti G, Silva L, Bressan A and Danese L, 
2004, {\it Astrop. J.}, {\bf 600}, 580-594

\item[] Gross M A K, Somerville R S, Primack J R, Holtzman J and Klypin A, 
1998, {\it Mon. Not. Roy. Ast. SOc.}, {\bf 301}, 81-94 

\item[] Guiderdoni B, Hivon E, Bouchet F R and Maffei B, 1998, {\it Mon. Not. Roy. Ast. Soc.}, {\bf 295}, 877-898

\item[] Gunn J E and Gott J R, 1972, {\it Astrop. J.}, {\bf 176}, 1

\item[] Haardt F and Madau P, 1996, {\it Astrop. J.}, {\bf 461}, 20

\item[] Haiman Z, Spaans M and Quataert E, 2000, {\it Astrop. J.}, {\bf 537}, L5-L8

\item[] Hanany S, et~al., 2000, {\it Astrop. J.}, {\bf 545}, L5-L9

\item[] Harker G, Cole S, Helly J, Frenk C S and Jenkins A, 
2006,  {\it Mon. Not. Roy. Ast. Soc.}, {\bf 367}, 1039-1049

\item[] Hatton S, Devriendt J E G, Ninin S, Bouchet F R, Guiderdoni B and 
Vibert D, 2003, {\it Mon. Not. Roy. Ast. Soc.}, {\bf 343}, 75-106

\item[] Hayashi E, Navarro J F, Taylor J E, Stadel J and Quinn T, 2003, 
{\it Astrop. J}, {\bf 584}, 541-558

\item[] Hayashi E, Navarro J F, Power C, Jenkins A, Frenk C S, White S D M, 
Springel V, Stadel J and Quinn T R, 2004, {\it Mon. Not. Roy. Ast. Soc.}, {\bf 355}, 794-812

\item[] Helly J C, Cole S, Frenk C S, Baugh C M, Benson A and 
Lacey C, 2003a, {\it Mon. Not. Roy. Ast. Soc.}, {\bf 338}, 903-912

\item[] Helly J C, Cole S, Frenk C S, Baugh C M, Benson A, 
Lacey C and Pearce F R, 2003b, {\it Mon. Not. Roy. Ast. Soc.}, {\bf 338}, 913-925

\item[] Heyl J S, Cole S, Frenk C S and Navarro J F, 1995, {\it Mon. Not. Roy. Ast. Soc.}, {\bf 274}, 755-768

\item[] Hinshaw G, et~al., 2003, {\it Astrop. J. Supp.}, {\bf 148}, 135-159 

\item[] Hinshaw G, et~al., 2006, {\it preprint}, astro-ph/0603451

\item[] Hoyle F, 1949, in 'Problems of Cosmical Aerodynamics', 
published by Central air documents office, Ohio, 195-197

\item[] Hoyle F, 1953, {\it Astrop. J.}, {\bf 118}, 513-528 

\item[] Huang J S, Glazebrook K, Cowie L L and Tinney C, 
2003, {\it Astrop. J.}, {\bf 584}, 203-209

\item[] Hughes D H, Serjeant S, Dunlop J, Rowan-Robinson M, Blain A, Mann R G, 
Ivison R, Peacock J, Efstathiou A, Gear W, Oliver S, Lawrence A, Longair M, 
Goldschmidt P and Jenness T, 1998, {\it Nature}, {\bf 394}, 241-247

\item[] Jarrett T H, Chester T, Cutri R, Schneider S, Skrutskie M 
and Huchra J P, 2000, {\it Astron. J.}, {\bf 119}, 2498-2531

\item[] Jedamzik K, 1995, {\it Astrop. J.}, {\bf 448}, 1-7

\item[] Jenkins A, Frenk C S, White S D M, Colberg J M, Cole S, Evrard A E, 
Couchman H M P, Yoshida N, 2001, {\it Mon. Not. Roy. Ast. Soc.}, {\bf 321}, 372-384

\item[] Jing Y P, 2000, {\it Astrop. J.}, {\bf 535}, 30-36

\item[] Jones W C, et~al., 2006, {\it Astrop. J.}, {\bf 647}, 823-832

\item[] Juneau S, et~al. 2005, {\it Astrop. J.}, {\bf 619}, L135-L138

\item[] Kang X, Jing Y P, Mo H J and Borner G, 2005, {\it Astrop. J.}, 
{\bf 631}, 21-40

\item[] Kang X, Jing Y P and Silk J, 2006, {\it Astrop. J.}, {\bf 648}, 820-825

\item[] Kaastra J S, Ferrigno C, Tamura T, Paerels F B S, Peterson J R and 
Mittaz J P D, 2001, {\it Astron. \& Astrop.}, {\bf 365}, L99-L103

\item[] Katz N, 1992, {\it Astrop. J.}, {\bf 391}, 502-517

\item[] Katz N and Gunn J E, 1991, {\it Astrop. J.}, {\bf 377}, 365-381

\item[] Kauffmann G, 1996, {\it Mon. Not. Roy. Ast. Soc.}, {\bf 281}, 475-486

\item[] Kauffmann G, White S D M and Guiderdoni B, 1993, {\it Mon. Not. Roy. Ast. Soc.}, {\bf 264}, 201

\item[] Kauffmann G and White S D M, 1994, in The Formation and evolution of 
galaxies, Cambridge University Press, Cambridge. 

\item[] Kauffmann G and Charlot S, 1994, {\it Astrop. J.}, 
{\bf 430}, L97-L100

\item[] Kauffmann G, Colberg J M, Diaferio A and White S D M, 1999, {\it Mon. Not. Roy. Ast. Soc.}, {\bf 303}, 188-206

\item[] Kauffmann G and Haehnelt M, 2000, {\it Mon. Not. Roy. Ast. Soc.}, {\bf 311}, 576-588

\item[] Kauffmann G, et~al., 2003, {\it Mon. Not. Roy. Ast. Soc.}, {\bf 341}, 54-69

\item[] Kaufmann T, Mayer L, Wadsley J, Stadel J and Moore B, 2006, {\it Mon. Not. Roy. Ast. Soc.}, {\bf 370}, 1612-1622

\item[] Kaviani A, Haehnelt M G and Kauffmann G, 2003, {\it Mon. Not. Roy. Ast. Soc.}, {\bf 340}, 739-746

\item[] Kay S T, Pearce F R, Jenkins A, Frenk C S, White S D M, Thomas P A, 
and Couchman P A, 2000, {\it Mon. Not. Roy. Ast. Soc.}, {\bf 316}, 374

\item[] Kay S T, Pearce F R, Frenk C S and Jenkins A, 2002, {\it Mon. Not. Roy. Ast. Soc.}, {\bf 330}, 113-128

\item[] Kazantzidis S, Mayer L, Mastropietro C, Diemand J, Stadel J and Moore B, 2004, {\it Astrop. J.}, {\bf 608}, 663-679

\item[] Kennicutt R C, 1998a, {\it Astrop. J.}, {\bf 498}, 541

\item[] Kennicutt R C, 1998b, {\it Ann. Rev. Astron. Astrop.}, {\bf 36}, 189-232

\item[] Keres D, Katz N, Weinberg D H and Dave R, 2005, {\it Mon. Not. Roy. Ast. Soc.}, {\bf 363}, 2-28

\item[] Klessen R S, Heitsch F and Mac Low M, 2000, {\it Astrop. J.}, {\bf 535}, 887-906

\item[] Klypin A, Nolthenius R and Primack J, 1997, {\it Astrop. J.}, {\bf 474}, 533

\item[] Klypin A, Kravtsov A V, Valenzuela O and Prada F, 1999, {\it Astrop. J.}, {\bf 522}, 82-92

\item[] Klypin A, Kravtsov A V, Bullock J S and Primack J R, 2001, {\it Astrop. J}, {\bf 554}, 903-915

\item[] Kodama T et~al. 2004, {\it Mon. Not. Roy. Ast. Soc.}, {\bf 350}, 1005-1014

\item[] Kochanek C S, Pahre M A, Falco E E, Huchra J P, Mader J, Jarrett T H, 
Chester T, Cutri R and Schneider S E, 2001, {\it Astrop. J.}, {\bf 560}, 
566-579

\item[] Kormendy J, 1977, {\it Astrop. J.}, {\bf 218}, 333

\item[] Kravtsov A V, 2003, {\it Astrop. J.}, {\bf 590}, L1-L4

\item[] Kravtsov A V, Gnedin O Y and Klypin A A, 2004, {\it Astrop. J.}, 
{\bf 609}, 482-497

\item[] Krumholz M R, McKee C F and Klein R I, 2005, {\it Nature}, {\bf 438}, 332-334

\item[] Kuo C L, et~al., 2004, {\it Astrop. J.}, {\bf 600}, 32

\item[] Lacey C and Silk J, 1991, {\it Astrop. J.}, {\bf 381}, 14-32

\item[] Lacey C and Cole S, 1993, {\it Mon. Not. Roy. Ast. Soc.}, 
{\bf 262}, 627-649 

\item[] Lacey C and Cole S, 1994, {\it Mon. Not. Roy. Ast. Soc.}, 
{\bf 271}, 676 

\item[] Lacey C, Guiderdoni B, Rocca-Volmerange B and Silk J, 1993, 
{\it Astrop. J.}, {\bf 402}, 15-41

\item[] Lanzoni B, Guiderdoni B, Mamon G A, Devriendt J and Hatton S, 
2005, {\it Mon. Not. Roy. Ast. Soc.}, {\bf 361}, 369-384 

\item[] Larson R B, 1974, {\it Mon. Not. Roy. Ast. Soc.}, 
{\bf 169}, 229-246 

\item[] Larson R B, 1975, {\it Mon. Not. Roy. Ast. Soc.}, 
{\bf 173}, 671 

\item[] Le Borgne D, Rocca-Volmerange B, Prugniel P, Lancon A, Fioc M 
and Soubiran C, 2004, {\it Astron. \& Astroph.}, {\bf 425}, 881-897

\item[] Leitherer C, et~al., 1999, {\it Astrop. J. Supp.}, {\bf 123}, 3-40

\item[] Lemson G and Kauffmann G, 1999, {\it Mon. Not. Roy. Ast. Soc.}, 
{\bf 302}, 111-117

\item[] Li P S, Norman M L, Mac Low M and Heitsch F, 2004, 
{\it Astrop. J.}, {\bf 605}, 800-818

\item[] Loveday J, Peterson B A, Efstathiou G and Maddox S J, 1992, 
{\it Astrop. J.}, {\bf 390}, 338-344

\item[] Mac Low M and Klessen R S, 2004, {\it Rev. Mod. Phys.}, {\bf 76}, 
125-194

\item[] MacTavish C J, et~al., 2006, {\it Astrop. J.}, {\bf 647}, 799-812. 

\item[] Madau P, Ferguson H C, Dickinson M E, Giavalisco M, Steidel 
C C and Fruchter A, 1996, {\it Mon. Not. Roy. Ast. Soc.}, {\bf 283}, 1388-1404.

\item[] Magorrian J, et~al., 1998, {\it Astron. J.}, {\bf 115}, 2285-2305

\item[] Malbon R K, Baugh C M, Frenk C S and Lacey C G, 2006, {\it preprint}, 
astro-ph/0607424

\item[] Maller A H and Dekel A, 2002, {\it Mon. Not. Roy. Ast. Soc.}, 
{\bf 335}, 487-498

\item[] Maller A H and Bullock J S, 2004, {\it Mon. Not. Roy. Ast. Soc.}, 
{\bf 355}, 694-712

\item[] Maraston C, 1998, {\it Mon. Not. Roy. Ast. Soc.}, 
{\bf 300}, 872-892 

\item[] Maraston C, 2005, {\it Mon. Not. Roy. Ast. Soc.}, 
{\bf 362}, 799-825 

\item[] Martin C L, 1997, {\it Astrop. J.}, {\bf 491}, 561

\item[] Martin C L, 1998, {\it Astrop. J.}, {\bf 506}, 222-252

\item[] Martin C L, 1999, {\it Astrop. J.}, {\bf 513}, 156-160

\item[] Martin C L, 2005, {\it Astrop. J.}, {\bf 621}, 227-245

\item[] Mayer L, Mastropietro C, Wadsley J, Stadel J and Moore B, 
2006, {\it Mon. Not. Roy. Ast. Soc.}, {\bf 369}, 1021-1038

\item[] McCarthy I G, Balogh M L, Babul A, Poole G B and Horner D J, 2004, 
{\it Astrop. J.}, {\bf 613}, 811-830

\item[] McKee C F and Ostriker, 1977, {\it Astrop. J.}, {\bf 218}, 148-169

\item[] Menci N, Cavaliere A, Fontana A, Giallongo E and Poli F, 2002, 
{\it Astrop. J.}, {\bf 575}, 18-32

\item[] Menci N, Cavaliere A, Fontana A, Giallongo E, Poli F and Vittorini V, 
2004, {\it Astrop. J.}, {\bf 604}, 12-17

\item[] Menci N, Fontana A, Giallongo E, Grazian A and Salimbeni S, 2006, 
{\it Astrop. J.}, {\bf 647}, 753-762

\item[] Merritt D, Navarro J F, Ludlow A and Jenkins A, 2005a, {\it Astrop. J.},  {\bf 624}, L85-L88

\item[] Merritt D, Graham A W, Moore B, Dieman J and Terzic B, 2005b, 
{\it Preprint}, astro-ph/0509417

\item[] Mihos C J and Hernquist L, 1994, {\it Astrop. J.}, {\bf 425}, L13-L16

\item[] Mihos C J and Hernquist L, 1996, {\it Astrop. J.}, {\bf 464}, 641

\item[] Monaco P, 2004, {\it Mon. Not. Roy. Ast. Soc.}, {\bf 352}, 181-204.

\item[] Monaco P and Fontanot F, 2005, {\it Mon. Not. Roy. Ast. Soc.}, {\bf 359}, 283-294.

\item[] Monaco P, Murante G, Borgani S and Fontanot F, 2006, {\it Preprint}, 
astro-ph/0610045

\item[] Monaco P, Fontanot F and Taffoni G, 2006, {\it Preprint}, astro-ph/0610805

\item[] Monaghan J J, 1992, {\it Ann. Rev. Astron. Astrop.}, {\bf 30}, 
543-574

\item[] Mo H J, Mao S and White S D M, 1998, {\it Mon. Not. Roy. Ast. Soc.}, {\bf 295}, 319-336

\item[] Mo H J and White S D M, 2002, {\it Mon. Not. Roy. Ast. Soc.}, {\bf 336}, 112-118

\item[] Moore B, Katz N, Lake G, Dressler A and Oemler A, 1996, {\it Nature}, 
{\bf 389}, 613

\item[] Moore B, Governato F, Quinn T, Stadel J and Lake G, 
1998, {\it Astroph. J.}, {\bf 499}, L5

\item[] Moore B, Ghigna S, Governato F, Lake G, Quinn T, Stadel J and Tozzi P, 
1999a, {\it  Astroph. J.}, {\bf 524}, L19-L22

\item[] Moore B, Quinn T, Governato F, Stadel J and Lake G, 1999b, 
{\it Mon. Not. Roy. Ast. Soc.}, {\bf 310}, 1147-1152. 

\item[] Mushotzky R, Loewenstein M, Arnaud K A, Tamura T, Fukazawa Y, Matsushita K, Kikuchi K and Hatsukade L, 1996, {\it Astrop. J.}, {\bf 466}, 686

\item[] Nagai D and Kravtsov A V, 2005, {\it Astrop. J.}, {\bf 618}, 557-568

\item[] Nagamine K, Cen R, Hernquist L, Ostriker J P and Springel V, 
2004, {\it  Astroph. J.}, {\bf 610}, 45-50

\item[] Nagamine K, Cen R, Hernquist L, Ostriker J P and Springel V, 
2005, {\it  Astrop. J.}, {\bf 627}, 608-620

\item[] Nagashima M, 2001, {\it Astrop. J.}, {\bf 562}, 7-23

\item[] Nagashima M, Gouda N and Sugiura N, 1999, {\it Mon. Not. Roy. Ast. Soc.}, {\bf 305}, 449-456

\item[] Nagashima M and Okamoto T, 2006, {\it Astrop. J.}, {\bf 643}, 863-880

\item[] Nagashima M, Lacey C G, Baugh C M, Frenk C S and Cole S, 
2005a, {\it Mon. Not. Roy. Ast. Soc.}, {\bf 358}, 1247-1266

\item[] Nagashima M, Lacey C G, Okamoto T, Baugh C M, Frenk C S and Cole S, 
2005b, {\it Mon. Not. Roy. Ast. Soc.}, {\bf 363}, L31-L35

\item[] Nagashima M, Yahagi H, Enoki M, Yoshii Y and Gouda N, 
2005c, {\it  Astrop. J.}, {\bf 634}, 26-50

\item[] Navarro J F and White S D M, 1993, {\it Mon. Not. Roy. Ast. Soc.}, {\bf 265}, 271

\item[] Navarro J F and White S D M, 1994, {\it Mon. Not. Roy. Ast. Soc.}, {\bf 267}, 401-412

\item[] Navarro J F, Frenk C S and White S D M, 1996, {\it  Astrop. J.}, {\bf 462}, 563

\item[] Navarro J F, Frenk C S and White S D M, 1997, {\it  Astrop. J.}, {\bf 490}, 493

\item[] Navarro J F, Hayashi E, Power C, Jenkins A R, Frenk C S, White S D M, 
Springel V, Stadel J nd Quinn T R, 2004, {\it Mon. Not. Roy. Ast. Soc.}, {\bf 349}, 1039-1051

\item[] Norberg P, et~al., 2001, {\it Mon. Not. Roy. Ast. Soc.}, {\bf 328}, 64-70

\item[] Norberg P, et~al., 2002a, {\it Mon. Not. Roy. Ast. Soc.}, {\bf 332}, 827-838

\item[] Norberg P, et~al., 2002b, {\it Mon. Not. Roy. Ast. Soc.}, {\bf 336}, 907-931

\item[] Okamoto T and Nagashima M, 2001, {\it Astroph. J.}, {\bf 547}, 109-116 

\item[] Okamoto T and Nagashima M, 2003, {\it Astroph. J.}, {\bf 587}, 500-513 

\item[] Okamoto T, Jenkins A, Eke V R, Quilis V and Frenk C S, 2003, {\it Mon. Not. Roy. Ast. Soc.}, {\bf 345}, 429-446

\item[] Okamoto T, Eke V R, Frenk C S and Jenkins A, 2005, {\it Mon. Not. Roy. Ast. Soc.}, {\bf 363}, 1299-1314

\item[] O'Shea B W, Bryan G, Bordner J, Norman M L, Abel T, Harkness R and Kritsuk A, 2004, {\it preprint}, astro-ph/0403044

\item[] O'Shea B W, Nagamine K, Springel V, Hernquist L and Norman M L, 2005 
{\it Astrop. J. Suppl.}, {\bf 160}, 1-27

\item[] Ott J, Walter F and Brinks E, 2005, {\it Mon. Not. Roy. Ast. Soc.}, {\bf 358}, 1453-1471

\item[] Padmanabhan T, 1993, {\it Structure Formation in the Universe}, Cambridge University Press, Cambridge UK. 

\item[] Padmanabhan N, et~al., 2006, {\it preprint}, astro-ph/0605302

\item[] Peacock J A, 1999, Cosmological Physics, Cambridge

\item[] Peacock J A and Heavens A F, 1990, {\it Mon. Not. Roy. Ast. Soc.}, {\bf 243}, 133-143

\item[] Pearce F R, Jenkins A, Frenk C S, Colberg J M, White S D M, Thomas P A, Couchman H M P, Peacock J A and Efstathiou G, 1999, {\it Astroph. J.}, {\bf 521}, L99-L102

\item[] Pedersen K, Ramussen J, Sommer-Larsen J, Toft S, Benson A J and 
Bower R G, 2005, preprint, astro-ph/0511682

\item[] Peebles P J E, 1969, {\it Astroph. J.}, {\bf 155}, 393

\item[] Peebles P J E, 1980, {\it The Large Scale Structure of the Universe}, Princeton University Press, Princeton NJ. 

\item[] Peebles P J E, 1982, {\it Astrop. J.}, {\bf 263}, L1-L5

\item[] Peebles P J E, 1993, {\it Physical Cosmology}, Princeton University Press, Princeton. 

\item[] Peterson J R, Paerels F B S, Kaastra J S, Arnaud M, Reiprich T H, Fabian A C, Mushotzky R F, Jernigan J G and Sakelliou I, 2001, {\it Astron. \& Astrop.}, {\bf 365}, L104-L109

\item[] Percival W J, et~al. 2001, {\it Mon. Not. Roy. Ast. Soc.}, 
{\bf 327}, 1297-1306

\item[] Percival W J, et~al. 2002, {\it Mon. Not. Roy. Ast. Soc.}, 
{\bf 337}, 1068-1080

\item[] Percival W J, Scott D, Peacock J A and Dunlop J S, 2003, {\it Mon. Not. Roy. Ast. Soc.}, {\bf 338}, L31-L35

\item[] Percival W J, et~al., 2006a, {\it preprint}, astro-ph/0608636

\item[] Percival W J, et~al., 2006b, {\it preprint}, astro-ph/0608635

\item[] Perlmutter S, Aldering G, Goldhaber G, et~al., 1999, 
{\it Astrop. J.}, {\bf 517}, 565

\item[] Persic M and Salucci P, 1992, {\it Mon. Not. Roy. Ast. Soc.}, 
{\bf 258}, 14P-18P

\item[] Pettini M, Shapley A E, Steidel C C, Cuby J-G, Dickinson M, 
Moorwood A F M, Adelberger K L and Giavalisco M, 2001, 
{\it Astrop. J.}, {\bf 554}, 981-1000

\item[] Pettini M, Rix S A, Steidel C C, Adelberger K L, Hunt M P and 
Shapley A E, 2002, {\it Astrop. J}, {\bf 569}, 742-757

\item[] Poli F, Giallongo E, Menci N, D'Odorico S and Fontana A,  1999, 
{\it Astrop. J}, {\bf 527}, 662-672

\item[] Pope A C, et~al., 2004, {\it Astrop. J}, {\bf 607}, 655-660 

\item[] Power C, Navarro J F, Jenkins A, Frenk C S, White S D M, Springel V,
Stadel J and Quinn T, 2003, {\it Mon. Not. Roy. Ast. Soc.}, 
{\bf 338}, 14-34

\item[] Press W H and Schechter P, 1974, {\it Astrop. J}, {\bf 187}, 425-438 
 
\item[] Quilis V, 2004, {\it Mon. Not. Roy. Ast. Soc.}, 
{\bf 352}, 1426-1438

\item[] Quilis V, Ibanez J M and Saez D, 1994, {\it Astron. Astrop.}, {\bf 286}, 1-16

\item[] Quilis V, Bower R G and Balogh M L, 2001, {\it Mon. Not. Roy. Ast. Soc.}, {\bf 328}, 1091-1097

\item[] Readhead A C S, Mason B S, Contaldi C R et~al., 2004, 
{\it Astrop. J.}, {\bf 609}, 498

\item[] Reed D, Bower R, Frenk C S, Gao L, Jenkins A, Theuns T and White 
S D M, 2005, {\it Mon. Not. Roy. Ast. Soc.}, {\bf 363}, 393-404

\item[] Reed D S, Bower R G, Frenk C S, Jenkins A R and Theuns T, 
2006a, {\it preprint}, astro-ph/0607150

\item[] Reed D S, Governato F, Quinn T, Stadel J and Lake G, 2006b, 
{\it preprint}, astro-ph/0602003

\item[] Rees M J  and Ostriker J P, 1977, {\it Mon. Not. Roy. Ast. Soc.}, 
{\bf 179}, 541-559

\item[] Renzini A, Ciotti L, D'Ercole A and Pellegrini S, 1993, {\it Astrop. J.}, {\bf 419}, 52

\item[] Riess A G, et~al., 1998, {\it Astron. J.}, 
{\bf 116}, 1009-1038

\item[] Riess A G, Strolger L, Tonry J, et~al., 2004, 
{\it Astrop. J.}, {\bf 607}, 665

\item[] Robertson B, Yoshida N, Springel V and Hernquist L, 2004, 
{\it Astrop. J.}, {\bf 606}, 32-45

\item[] Robertson B, Bullock J S, Cox T J, Di Matteo T, Hernquist L, Springel V and Yoshida N, 2006, {\it Astrop. J.}, {\bf 645}, 986-1000

\item[] Roukema B F, Quinn P J, Peterson B A and Rocca-Volmerange B, 
1997, {\it Mon. Not. Roy. Ast. Soc.}, {\bf 292}, 835

\item[] Ruszkowski M, Bruggen M and Begelman M C, 2004, {\it Astrop. J.}, 
{\bf 615}, 675-680

\item[] Ryu D, Ostriker J P, Kang H and Cen R, 1993, {\it Astrop. J.}, 
{\bf 414}, 1-19

\item[] Sanchez A G, Baugh C M, Percival W J, Peacock J A, Padilla N, 
Cole S, Frenk C S and Norberg P, 2006, {\it Mon. Not. Roy. Ast. Soc.}, 
{\bf 366}, 189-207 

\item[] Scannapieco C, Tissera P B, White S D M and Springel V, 2005, 
{\it Mon. Not. Roy. Ast. Soc.}, 
{\bf 364}, 552-564 

\item[] Scannapieco C, Tissera P B, White S D M and Springel V, 2006, 
{\it Mon. Not. Roy. Ast. Soc.}, 
{\bf 371}, 1125-1139 

\item[] Schechter P, 1976, {\it Astrop. J.}, {\bf 203}, 297-306

\item[] Schmidt M, 1959, {\it Astrop. J.}, {\bf 129}, 243-259

\item[] Seljak U, et~al., 2005, {\it Phys. Rev. D.}, {\bf 71}, 103515

\item[] Shapley A E, Steidel C C, Pettini M and Adelberger K L, 2003, 
{\it Astrop. J.}, {\bf 588}, 65-89

\item[] Sheth R K, Mo H J and Tormen G, 2001, {\it Mon. Not. Roy. Ast. Soc.}, 
{\bf 323}, 1

\item[] Sheth R K and Tormen G, 2004, {\it Mon. Not. Roy. Ast. Soc.}, 
{\bf 350}, 1385

\item[] Silk J, 1977, {\it Astrop. J.}, {\bf 211}, 638-648

\item[] Silva L, Granato G L, Bressan A and Danese L, 1998, {\it Astrop. J.}, {\bf 509}, 103-117

\item[] Smail I, Ivison R J and Blain A W, 1997, {\it Astrop. J.}, 
{\bf 490}, L5

\item[] Smoot G F et~al., 1992, {\it Astrop. J.}, {\bf 396}, L1-L5

\item[] Somerville R S, 2002, {\it Astrop. J.}, {\bf 572}, L23-L26

\item[] Somerville R and Kolatt T S, 1999, {\it Mon. Not. Roy. Ast. Soc.}, 
{\bf 305}, 1-14

\item[] Somerville R S and Primack J R, 1999, {\it Mon. Not. Roy. Ast. Soc.}, 
{\bf 310}, 1087-1110

\item[] Somerville R S, Lemson G, Kolatt T S and Dekel A, 2000, {\it Mon. Not. Roy. Ast. Soc.},  {\bf 316}, 479-490

\item[] Somerville R S, Primack J R and Faber S M, 2001, {\it Mon. Not. Roy. Ast. Soc.}, {\bf 320}, 504-528

\item[] Sommer-Larsen J, 2006, {\it Astrop. J.}, {\bf 644}, L1-L4

\item[] Sommer-Larsen J, Gelato S and Vedel H, 1999, {\it Astrop. J.}, 
{\bf 519}, 501-512

\item[] Sommer-Larsen J and Dolgov A, 2001, {\it Astrop. J.}, 
{\bf 551}, 608-623

\item[] Sommer-Larsen J, Gotz M and Portinari L, 2003, {\it Astrop. J.}, 
{\bf 596}, 47-66

\item[] Spergel D N and Steinhardt P J, 2000, {\it Phys. Rev. Letts}, {\bf 84}, 3760-3763

\item[] Spergel D N, Verde L, Peiris H V, et~al., 2003, {\it Astrop. 
J. Supp.}, 148, 175

\item[] Spergel D N, et~al. 2006, {\it preprint}, astro-ph/0603449

\item[] Spitzer L, 1962, Physics of fully ionised gases (New York, Wiley)

\item[] Springel V, 2000, {\it Mon. Not. Roy. Ast. Soc.}, {\bf 312}, 859-879

\item[] Springel V, 2005, {\it Mon. Not. Roy. Ast. Soc.}, {\bf 364}, 1105-1134

\item[] Springel V, White S D M, Tormen G and Kauffmann G, 2001, 
{\it Mon. Not. Roy. Ast. Soc.}, {\bf 328}, 726-750

\item[] Springel V and Hernquist L, 2002, {\it Mon. Not. Roy. Ast. Soc.}, 
{\bf 333}, 649-664

\item[] Springel V and Hernquist L, 2003, {\it Mon. Not. Roy. Ast. Soc.}, 
{\bf 339}, 312-334

\item[] Springel V, Di Matteo T and Hernquist L, 2005, {\it Mon. Not. Roy. Ast. Soc.}, {\bf 361}, 776-794

\item[] Springel V, et~al. 2005, {\it Nature}, {\bf 435}, 629-636

\item[] Springel V, Frenk C S and White S D M, 2006, {\it Nature}, {\bf 440}, 1137-1144

\item[] Sutherland R S and Dopita M A, 1993, {\it Astrop. J. Supp.}, 
{\bf 88}, 253-327

\item[] Summers F J, Davis M and Evrard A E, 1995, {\it Astrop. J.}, {\bf 454}, L1

\item[] Steidel C C, Giavalisco M, Pettini M, Dickinson M and Adelberger K L, 
1996, {\it Astrop. J.}, {\bf 462}, L17

\item[] Steidel C C, Adelberger K L, Giavalisco M, Dickinson M and Pettini M, 
1999, {\it Astrop. J.}, {\bf 519}, 1

\item[] Tan J C, 2000, {\it Astrop. J.}, {\bf 536}, 173-184

\item[] Tassis K and Mouschovias T Ch, 2004, {\it Astrop. J.}, {\bf 616}, 283-287

\item[] Taylor J E and Babul A, 2001, {\it Astrop. J.}, {\bf 559}, 716-735

\item[] Tegmark M, Silk J, Rees M J, Blanchard A, Abel T and Palla F, 
1997, {\it Astrop. J.}, {\bf 474}, 1

\item[] Tegmark M, et~al., 2004a, {\it Astrop. J.} {\bf 606}, 702

\item[] Tegmark M, et~al., 2004b, {\it Phys. Rev. D.}, {\bf 69},  103501

\item[] Tegmark M, et~al., 2006, {\it preprint}, astro-ph/0608632

\item[] Teyssier R, 2002, {\it Astron. Astrop.}, {\bf 385}, 337-364

\item[] Thacker R J and Couchman H M P, 2000, {\it Astrop. J.}, {\bf 545}, 728-752

\item[] Thacker R J and Couchman H M P, 2001, {\it Astrop. J.}, {\bf 555}, L17-L20

\item[] Thacker R J, Tittley E R, Pearce F R, Couchman H M P and Thomas P A, 
2000, {\it Mon. Not. Roy. Ast. Soc.}, {\bf 319}, 619-648

\item[] Thomas D, 1999, {\it Mon. Not. Roy. Ast. Soc.}, {\bf 306}, 655-661

\item[] Thomas D, Greggio L and Bender R, 1998, {\it Mon. Not. Roy. Ast. Soc.}, {\bf 296}, 119

\item[] Thomas D and Kauffmann G, 1999, Spectrophotometric Dating of Stars 
and Galaxies, ASP Conference Proceedings, {\bf 192}, 261. 
 
\item[] Thoul A A and Weinberg D H, 1996, {\it  Astrop. J.}, {\bf 465}, 608

\item[] Tinsley B M, 1980, {\it Fun. of Cosmic Phys.}, {\bf 5}, 287-288

\item[] Tormen G, 1998, {\it Mon. Not. Roy. Ast. Soc.}, {\bf 297}, 648-656

\item[] Tully R B and Fisher J R, 1977, {\it Astron. Astrop.}, {\bf 54}, 661-673

\item[] Tully R B, Somerville R S, Trentham N and Verheijen M A W, 2002, {\it  Astrop. J.}, {\bf 569}, 573-581

\item[] van den Bosch F C, 2001, {\it Mon. Not. Roy. Ast. Soc.}, {\bf 327}, 1334-1352

\item[] van den Bosch F C, 2002, {\it Mon. Not. Roy. Ast. Soc.}, {\bf 331}, 98-110

\item[] van Dokkum P G, 2005, {\it Astron. J.}, {\bf 130}, 2647-2665

\item[] van Kampen E, Jimenez R and Peacock J A, 1999, {\it Mon. Not. Roy. Ast. Soc.}, {\bf 310}, 43-56

\item[] Vazdekis A, 1999, {\it  Astrop. J.}, {\bf 513}, 224-241

\item[] Wadsley J W, Stadel J and Quinn T, 2004, {\it New Astronomy}, {\bf 9}, 
137-158

\item[] Walker I R, Mihos J C and Hernquist L, 1996, {\it  Astrop. J.}, {\bf 460}, 121

\item[] Warren M S, Quinn P J, Salmon J K and Wojciech H, 1992, {\it Astrop. J.}, {\bf 399}, 405-425

\item[] Warren M S, Abazajian K, Holz D E and Teodoro L, 2006, {\it Astrop. J.}, {\bf 646}, 881-885

\item[] Weil M L, Eke V R and Efstathiou G, 1998, {\it Mon. Not. Roy. Ast. Soc.}, 
{\bf 300}, 773-789

\item[] Wechsler R H, Bullock J S, Primack J R, Kravtsov A V and Dekel A, 
2002, {\bf 568}, 52-70 

\item[] Wechsler R H, Zentner A R, Bullock J S, Kravtsov A V and Allgood B, 
2005, {\it preprint}, astro-ph/0512416

\item[] White M, 2002, {\it Astrop. J. Supp.}, {\bf 143}, 241-255

\item[] White S D M, 1984, {\it Astrop. J.}, {\bf 286}, 38-41

\item[] White S D M, 1994, {\it Formation and Evolution of Galaxies}, Les Houches Lectures, astro-ph/9410043

\item[] White S D M and Rees M J, 1978,  {\it Mon. Not. Roy. Ast. Soc.}, 
{\bf 183}, 341-358

\item[] White S D M and Frenk C S, 1991, {\it Astrop. J.}, {\bf 379}, 52-79 

\item[] Williams R E et~al. 1996, {\it Astron. J.}, {\bf 112}, 1335

\item[] Wilman D J, Balogh M L, Bower R G, Mulchaey J S, Oemler A, Carlberg R G, Morris S L and Whitaker R J, 2005a,  {\it Mon. Not. Roy. Ast. Soc.}, 
{\bf 358}, 71-87

\item[] Wilman R J, Gerssen J, Bower R G, Morris S L, Bacon R, de Zeeuw P T 
and Davies R L, 2005b, {\it Nature}, {\bf 436}, 227-229

\item[] Worthey G, 1994, {\it Astrop. J. Supp.}, {\bf 95} 107-149

\item[] Wu K K S, Fabian A C and Nulsen P E J, 1998, {\it Mon. Not. Roy. Ast. Soc.}, {\bf 301}, L20-L24

\item[] Wu K K S, Fabian A C and Nulsen P E J, 2000, {\it Mon. Not. Roy. Ast. Soc.}, {\bf 318}, 889-912

\item[] Yahagi H, Nagashima M and Yoshii Y, 2004, {\it Astrop. J}, {\bf 605}, 709-713

\item[] Yang X, Mo H J and van den Bosch F C, 2003, {\it Mon. Not. Roy. Ast. Soc.}, {\bf 339}, 1057-1080

\item[] Yang X, Mo H J, van den Bosch and Jing Y P, 2005, {\it Mon. Not. Roy. Ast. Soc.}, {\bf 356}, 1293-1307

\item[] Yano T, Nagashima M and Gouda N, 1996, {\it Astrop. J.} {\bf 466}, 1-12

\item[] York D, et~al., 2000, {\it Astron. J.}, {\bf 120}, 1579

\item[] Yoshii Y and Takahara F, 1988, {\it Astrop. J.}, {\bf 326}, 1-18

\item[] Yoshida N, Abel T, Hernquist L and Sugiyama N, 2003, {\it Astrop. J.}, 
{\bf 592}, 645-663

\item[] Yoshida N, Stoehr F, Springel V and White S D M, 2002, {\it Mon. Not. Roy. Ast. Soc.}, {\bf 335}, 767-772

\item[] Zehavi I, et~al., 2002, {\it Astrop. J.}, {\bf 571}, 172-190

\item[] Zentner A R and Bullock J S, 2003, {\it Astrop. J.}, {\bf 598}, 49-72

\item[] Zhu G, Zheng Z, Lin W P, Jing Y P, Kang X and Gao L, 2006, 
{\it Astrop. J.}, {\bf 639}, L5-L8

\item[] Zwaan M A, et~al., 2003, {\it Astron. J.}, {\bf 125}, 2842-2858

\endrefs

\end{document}